\documentclass[letter,scriptaddress,twocolumn, showkeys,aps]{revtex4}
\usepackage{ amssymb }
    \usepackage{amsmath}%,amssymb} 
    \usepackage{makeidx}
    \usepackage{amsfonts}
        \usepackage{bm}
    \usepackage[ansinew]{inputenc}
    \usepackage[usenames,dvipsnames]{pstricks}
    \usepackage{subfigure}
    \usepackage{epsfig}
    \usepackage{pst-grad} % For gradients
    \usepackage{pst-plot} % For axes
        \usepackage{amsthm}
    \usepackage{makeidx}
    \usepackage{listings}
    \usepackage{mathtools}
    \usepackage{float}
    \usepackage{tabularx}
    
    \usepackage[colorlinks,hyperindex]{hyperref}
    \hypersetup
    {
        colorlinks,%
        citecolor=blue,%
        linkcolor=blue,%
        urlcolor=blue,%
    }
    \usepackage{tikz}

\usepackage{amssymb}
\usepackage{graphicx}
\usepackage{xcolor}
\usepackage{float,graphicx}
\usepackage{placeins}
\usepackage[colorinlistoftodos]{todonotes}
\usepackage{cases}

 \usepackage{subfigure}
\theoremstyle{definition}
    \newtheorem{definition}{Definition}
    \newtheorem{example}{Example}
    
    \theoremstyle{plain}

\makeindex
\begin{document}
\def\definitionautorefname{Definition}
\def\sectionautorefname{Section}
\def\subsectionautorefname{Subsection}
\def\subsubsectionautorefname{Subsubsection}
\def\lemmaautorefname{Lemma}
\def\propautorefname{Proposition}
\def\theoremautorefname{Theorem}
\def\remarkautorefname{Remark}

%\title{Infection Spread and Hotspot Growth in Networked Bi-Pathogen Dynamics}

\title{Dynamics of Infection Spread and Hotspot Growth in Bi-Pathogen Networks}

\author{Alyssa Yu$^{a}$, Laura P. Schaposnik$^{b}$} 

\begin{abstract}
Understanding the spatio-temporal evolution of epidemics with multiple pathogens requires not only new theoretical models but also careful analysis of their practical consequences. Building on the Multiplex Bi-Virus Reaction-Diffusion framework (MBRD) introduced in our companion paper, we investigate how the super-infection model (MBRD-SI) and the co-infection model (MBRD-CI) behave under different epidemiological and network conditions. Through numerical experiments, we study the effects of pathogen virulence, transmission parameters, and source locations on epidemic hotspot formation, initial outbreaks, and long-term prevalence. Our results highlight the role of multiplex structure in amplifying or suppressing co-circulating infections, and provide quantitative insight into conditions that drive persistent epidemic patterns. Beyond epidemiology, these findings have broader implications for multiplex contagion processes such as information diffusion and malware propagation.
\end{abstract}

\maketitle
 
\section{Introduction}

Spatio-temporal epidemic modeling has emerged as a critical area in understanding infection spread. In many scenarios, the severity of infections in a region depends on infections in neighboring regions. As a result, factoring in spatial information leads to more accurate models. Moreover, spatial data and analysis are especially useful for identifying and targeting high-risk areas, particularly in the early stages of an epidemic~\cite{lin2022spatial}. Such approaches date back as early as 1798, when a mapping study was conducted on yellow fever~\cite{stevenson1965putting}. More recently, spatio-temporal-focused studies such as~\cite{castro2021spatio-temporal} have aided in the use of prior data to inform mitigation efforts during the COVID-19 pandemic.
Prior research on spatio-temporal infection spread includes methods such as cellular automata~\cite{white2007modeling}, stochastic modeling~\cite{arenas2020modeling}, Bayesian analysis of spatio-temporal data~\cite{adeoye2025bayesian}, and partial differential equations~\cite{bertuzzo2010spatially} including but not limited to reaction-diffusion dynamics. Reaction-diffusion dynamics have proven particularly versatile, with applications not only in epidemiology but also in vegetation pattern formation~\cite{yin2017cross} and blood clotting~\cite{lobanova2004running}.

In our companion work~\cite{yu2025spatial}, we introduced two reaction-diffusion models for bi-pathogen dynamics on multiplex metapopulation networks: the super-infection model (MBRD-SI) and the co-infection model (MBRD-CI), which together form the Multiplex Bi-Virus Reaction-Diffusion (MBRD) framework. That paper focused on the theoretical foundations of MBRD, establishing conditions for Turing and Turing-Hopf instabilities. In this paper, we build on that framework to analyze bi-pathogen dynamics from a simulation-based standpoint, with an emphasis on the types of spatio-temporal phenomena that arise under different network and epidemiological conditions, from  two complementary perspectives:
\begin{itemize}
\item {\bf Hotspot growth. } We investigate the formation and growth of stationary hotspots that emerge from perturbations to a steady state. Such hotspots, driven by Turing instability as shown in~\cite{yu2025spatial}, can lead to system collapse. For example, cholera is largely endemic in countries such as Bangladesh, yet hotspots often occur due to seasonal plankton blooms~\cite{emch2010local}. Recent research has also shown that interacting contagions can generate persistent Turing-type spatial heterogeneity, even under simple SIS dynamics, when mobility rates differ between susceptible and infected populations~\cite{chen2019persistent, Chen_2017}. While such prior work established the possibility of persistent spatial structures in interacting contagions, no previous study has examined this phenomenon in bi-pathogen dynamics through a simulation-based approach as we do here.

\item {\bf Point-source infections.} Infections often originate from localized events such as concerts, airports, or schools. Historical examples include \emph{Coxiella burnetii} spreading from a dairy-goat farm in the Netherlands~\cite{hackert2012q}, or a \emph{campylobacteriosis} outbreak from raw milk consumed on a school field trip~\cite{korlath1985point}. In the context of bi-pathogen spread, we investigate point-source infections where both pathogens originate independently in the network, and analyze how MBRD-SI and MBRD-CI respond under varying parameter regimes.
\end{itemize}

Through our analysis of the different settings, we have the following   main results:

\begin{itemize}
\item Stationary hotspots can grow in severity over time under both MBRD-SI and MBRD-CI dynamics, potentially leading to system collapse. For both superinfection and co-infection dynamics, this typically occurs within a close range and for relatively extreme parameter values. Varying network average degrees across the layers can inhibit pattern formation and growth, and the dominance of one pathogen can suppress hotspot growth.
\item A higher superinfection coefficient accelerates the spread of the more severe pathogen, while a higher co-transmission coefficient accelerates the spread of co-infections. Moreover, lower removal rates of co-infected individuals can increase endemicity of co-infections.
\item During the early stages of bi-pathogen spread, limiting migration of infected individuals is crucial to containment, verifying the importance of quarantine policies.
\item Network topology strongly influences pathogen spread, offering a possible explanation for seasonal surges in outbreaks - even under bi-pathogen dynamics.
\end{itemize}

The rest of this paper is organized as follows. Section~\ref{sec:prelim} provides an overview of the MBRD-SI and MBRD-CI models introduced in~\cite{yu2025spatial}. Section~\ref{sec:pattern-form} investigates how various factors influence pattern formation and hotspot growth. Section~\ref{sec:point_source} introduces metrics for analyzing spread from point sources and discusses the impact of model parameters and network layers. Section~\ref{sec:realworld} connects our simulation results with real-world co-infection data. Finally, Section~\ref{sec:conclusion} summarizes this work and outlines future directions.

\section{Preliminaries}\label{sec:prelim}

As mentioned before, in \cite{yu2025spatial} we introduced two reaction-diffusion models for two-pathogen dynamics on metapopulation networks: the {\bf super-infection model (MBRD-SI)} and the {\bf co-infection model (MBRD-CI)}. The {\bf MBRD-SI} model incorporates superinfection dynamics but does not include co-infection, while the {\bf MBRD-CI} model incorporates co-infection dynamics only.  

\begin{figure}[htbp]
\centering
\includegraphics[width=0.34\textwidth]{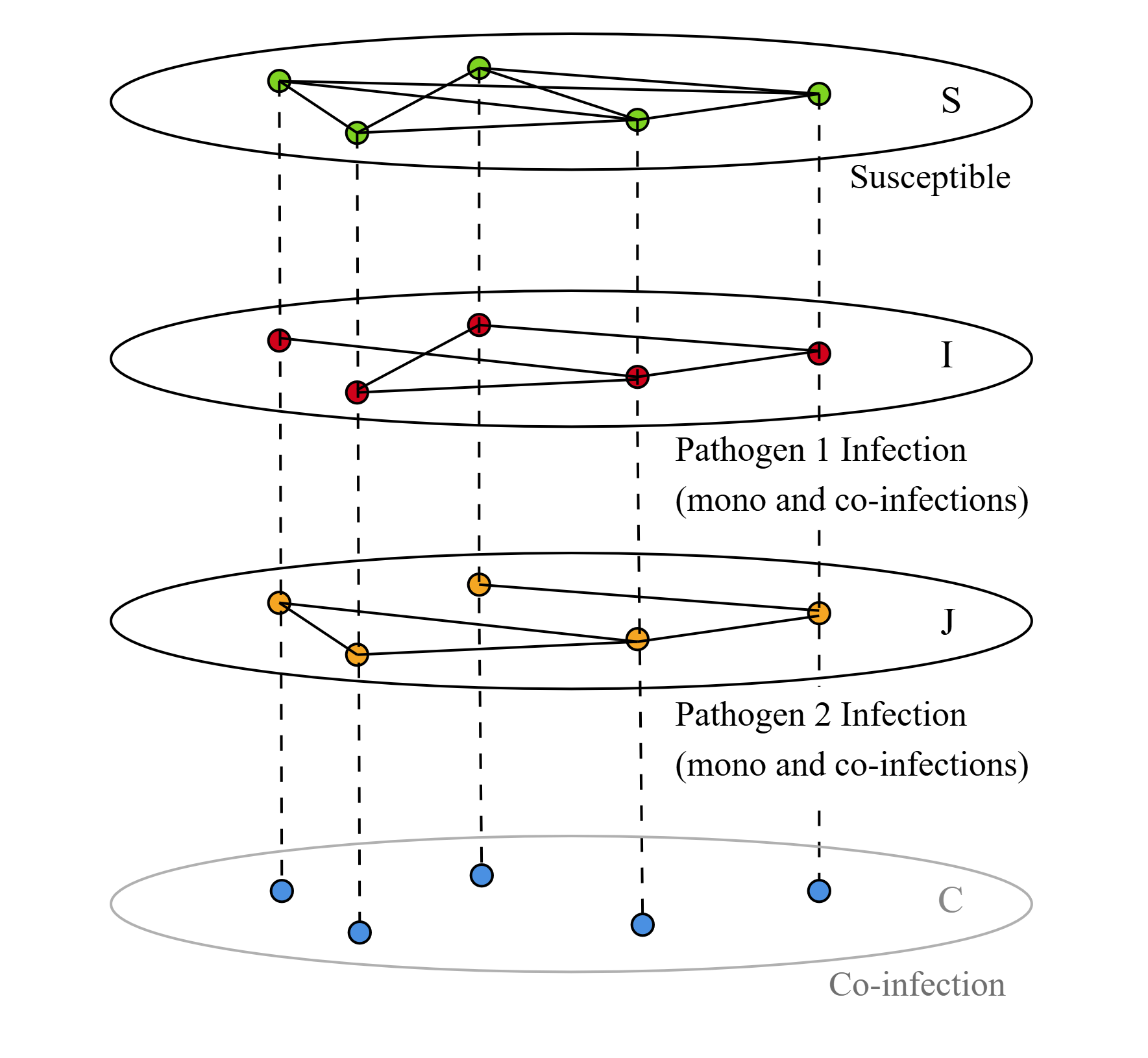}
\caption{Multiplex network for both models, inspired by figures in \cite{yu2025spatial}.}
\label{fig:multiplex_prelim}
\end{figure}

In what follows, we discuss the multiplex networks for {\it  MBRD-SI} and {\it MBRD-CI}. We model { \it MBRD-SI} dynamics on a three-layer multiplex network and { \it MBRD-CI} dynamics on a four-layer multiplex network. In both networks, the first layer describes the flow of the susceptible population density in response to the spatial distribution of susceptible individuals. The second (resp. third) layer describes the flow of pathogen~1-infections (resp. pathogen~2-infections), including both mono- and co-infections, across the network. The fourth layer is used only for modeling { \it MBRD-CI} and describes the density of co-infections at each region. Because these densities can be calculated from the densities of the other layers, this layer has no edges. Both networks are shown in \autoref{fig:multiplex_prelim}.

In the multiplex networks, the $i$-th region is represented by one node in each of the layers and the densities $S_i$, $I_i$, $J_i$, and $C_i$ (only for the {\it MBRD-CI} model) assigned to those distinctive nodes.
The {\it MBRD-SI} model incorporates superinfection dynamics but does not allow for co-infection. We have the following system of equations from \cite{yu2025spatial} defining the {\it MBRD-SI} model in Equation~\eqref{eq:superinfect-prelim}:
\begin{small}

\begin{equation}\label{eq:superinfect-prelim}
\begin{aligned}
    \frac{dS_i}{dt} &= rS_i\left(1-\frac{S_i}{K}\right)\left(\frac{S_i}{A}-1\right)-\frac{(\beta_1I_i+\beta_2 J_i)S_i}{S_i+I_i+J_i}\\
    &\quad+\gamma_1 I_i+\gamma_2 J_i-\mu S_i\\
    &\quad+d_{11}\sum_{j=1}^NL_{ij}^{(S)}S_j+d_{12}\sum_{j=1}^NL_{ij}^{(I)}I_j+d_{13}\sum_{j=1}^NL_{ij}^{(J)}J_j, \\
\frac{dI_i}{dt} &= I_i\left(\frac{\beta_1 S_i}{S_i+I_i+J_i}-\mu-\alpha_1-\gamma_1-\frac{\sigma\beta_2 J_i}{S_i+I_i+J_i}\right)\\
&\quad+d_{22}\sum_{j=1}^NL_{ij}^{(I)}I_j, \\
\frac{dJ_i}{dt} &= J_i\left(\frac{\beta_2S_i}{S_i+I_i+J_i}-\mu-\alpha_2-\gamma_2+\frac{\sigma\beta_2I_i}{S_i+I_i+J_i}\right)\\
&\quad+d_{33}\sum_{j=1}^NL_{ij}^{(J)}J_j.
\end{aligned}
\end{equation}
\end{small}
The {\it MBRD-CI} model, also proposed in \cite{yu2025spatial}, incorporates co-infection dynamics. It is given by the system in Equation~\eqref{eq:coinfect-prelim}:
\begin{small}
\begin{equation}
\label{eq:coinfect-prelim}
\begin{aligned}
    \frac{dS_i}{dt}&=rS_i\left(1-\frac{S_i}{K}\right)\left(\frac{S_i}{A}-1\right)-\frac{(\beta_1 I_i+\beta_2 J_i)S_i}{S_i+I_i+J_i-C_i}\\
    &\quad-\frac{(\beta_{10}+\beta_{02}+\beta_{12}-\beta_1-\beta_2)C_iS_i}{S_i+I_i+J_i-C_i}\\
    &\quad+\gamma_1I_i+\gamma_2J_i-(\gamma_{1}+\gamma_{2})C_i-\mu S_i\\
    &\quad+d_{11}\sum_{j=1}^NL_{ij}^{(S)}S_j+d_{12}\sum_{j=1}^NL_{ij}^{(I)}I_j+d_{13}\sum_{j=1}^NL_{ij}^{(J)}J_j,\\
    \frac{d I_i}{dt}&=[\beta_1I_i+(\beta_{10}+\beta_{12}-\beta_1)C_i]\cdot \frac{S_i+J_i-C_i}{S_i+I_i+J_i-C_i}\\
    &\quad-\gamma_1I_i-\alpha_1(I_i-C_i)-\alpha_{12}C_i-\mu I_i\\
    &\quad+d_{22}\sum_{j=1}^NL_{ij}^{(I)}I_j,\\
    \frac{d J_i}{dt}&=[\beta_2J_i+(\beta_{02}+\beta_{12}-\beta_2)C_i]\cdot\frac{S_i+I_i-C_i}{S_i+I_i+J_i-C_i}\\
    &\quad-\gamma_2J_i-\alpha_2(J_i-C_i)-\alpha_{12}C_i-\mu J_i\\
    &\quad+d_{33}\sum_{j=1}^NL_{ij}^{(J)}J_j,\\
    \frac{d C_i}{dt}&=\frac{\beta_{12}C_iS_i}{S_i+I_i+J_i-C_i}\\
    &\quad+\frac{[\beta_2J_i+(\beta_{02}+\beta_{12}-\beta_2)C_i](I_i-C_i)}{S_i+I_i+J_i-C_i}\\
    &\quad+\frac{[\beta_1I_i+(\beta_{10}+\beta_{12}-\beta_1)C_i](J_i-C_i)}{S_i+I_i+J_i-C_i}\\
    &\quad-(\gamma_{1}+\gamma_{2}+\alpha_{12})C_i-\mu C_i.
\end{aligned}
\end{equation}
\end{small}
Descriptions for the parameters of both models can be found in \cite{yu2025spatial}. Because {\it MBRD-SI} uses a three-layer multiplex network and {\it MBRD-CI} uses a four-layer network (with an additional co-infection density layer), their multiplex topologies differ slightly, but the same framework applies. We use the systems in Equations~\eqref{eq:superinfect-prelim} and~\eqref{eq:coinfect-prelim} to perform our simulations in the following sections.

\section{Pattern Formation}\label{sec:pattern-form}

Many epidemics are comprised of waves of infections. Here, we aim to simulate such waves by adding stochastic noise to an initial steady state, which creates spatial oscillations. In a physical sense, this noise can represent fluctuations in infections that are induced by changes in external dynamics such as weather. We shall dedicate this section to analyzing simulation results with the proposed {\bf MBRD-SI} and {\bf MBRD-CI} models. Particularly, in Subsection~\ref{sec:methods}, we describe the methodology for our experiments. In Subsection~\ref{sec:turing-pattern-simu}, we analyze three examples of pattern formation. In Subsection~\ref{sec:pattern-parameters}, we aim to understand the effect of different parameters on hotspot growth. Finally, in Subsection~\ref{sec:simu-deg}, we explore the effect of variations in layerwise network degrees on pattern formation and growth.

\subsection{Methodology}\label{sec:methods}

Because infection densities cannot be negative in the real world, we set $0$ as the minimum threshold for all densities. We conduct simulations on three types of networks:

\begin{itemize}
    \item \textbf{Lattice networks.} With simulations on lattice networks, we can more easily view patterns that form. We implement the LA4, LA12, and LA24 lattice networks, in which most nodes have degree $4$, $12$, and $24$, respectively. These are represented in \autoref{fig:la-networks}. In each of the lattices in the figure, the center red vertex is connected to the green vertices through edges. Lattice networks have a deterministic structure and give us identical solutions across trials.

\begin{figure}[htbp]
\centering
\includegraphics[width=0.48\textwidth]{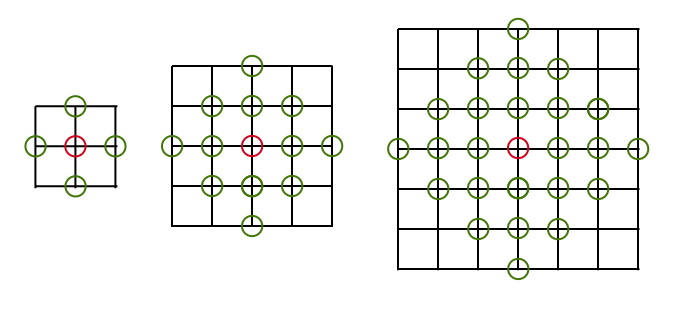}
\caption{The LA4 (left), LA12 (middle), and LA24 (right) networks.}
\label{fig:la-networks}
\end{figure}
    \item \textbf{Watts-Strogatz (WS) network.} The Watts-Strogatz model~\cite{watts1998collective} is characterized by short geodesic distances between nodes; this phenomenon is called the small-world effect. 
    \item \textbf{Barab\'asi-Albert (BA) network.} The BA network~\cite{barabasi2003scale} is a scale-free network and follows a power-law degree distribution. It also incorporates preferential attachment, meaning that each node is more likely to create new connections if it already is connected to many other nodes. 
\end{itemize}

\begin{figure*}[htbp]
    \centering

    \makebox[\textwidth]{%
        \subfigure[]{%
            \includegraphics[width=0.31\textwidth]{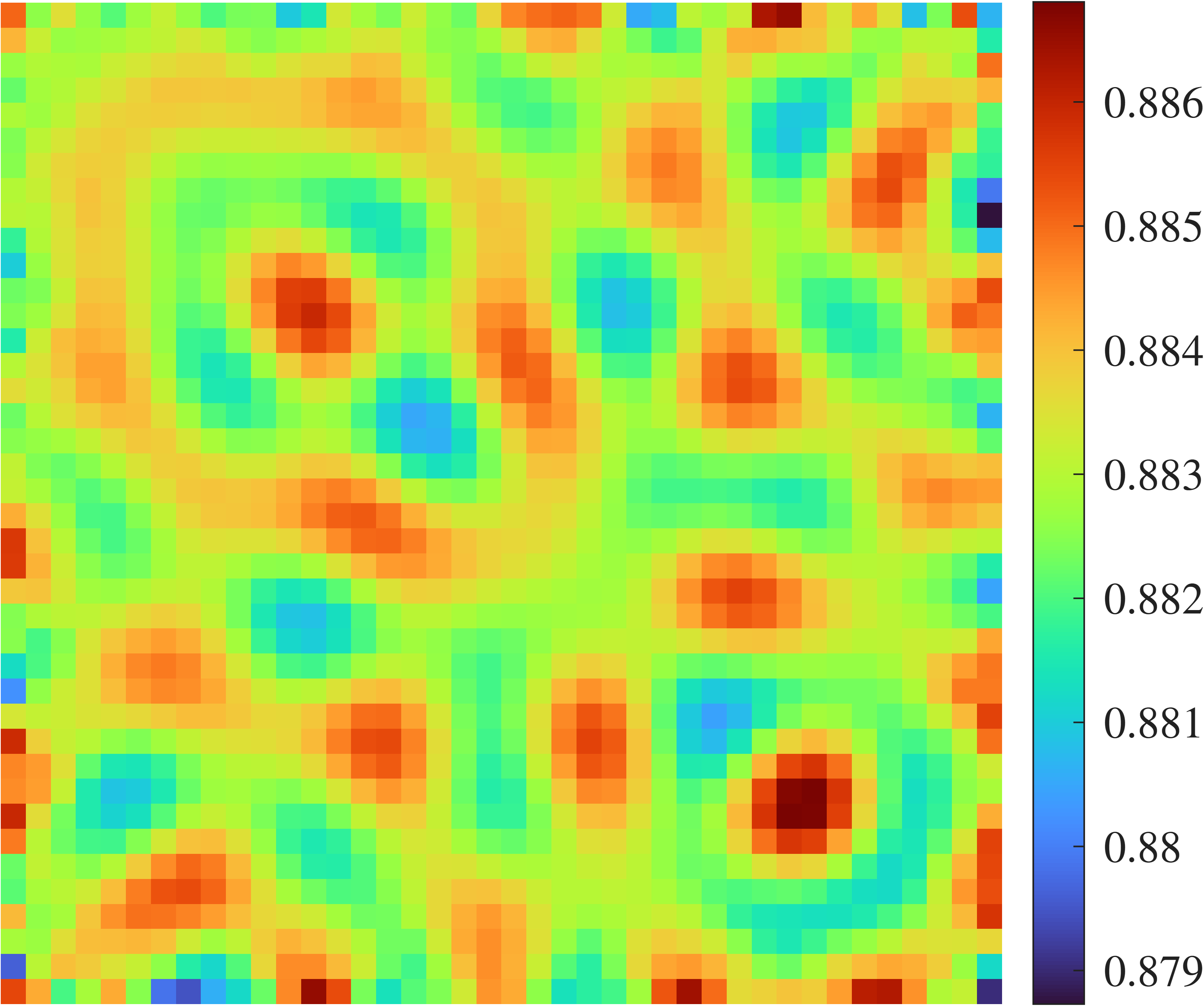}
            \label{}
        }
        \subfigure[]{%
            \includegraphics[width=0.31\textwidth]{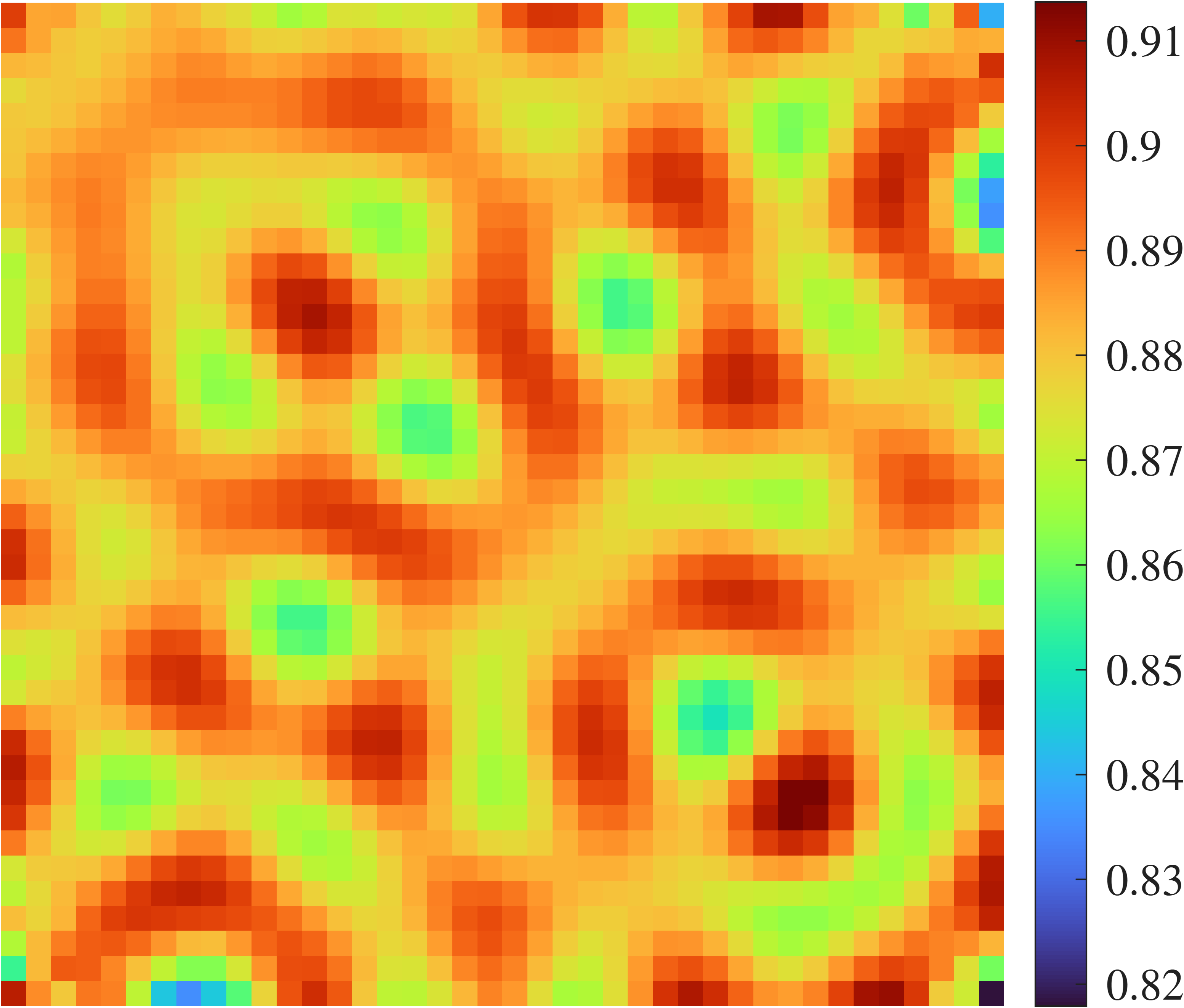}
            \label{}
        }
        \subfigure[]{%
            \includegraphics[width=0.30\textwidth]{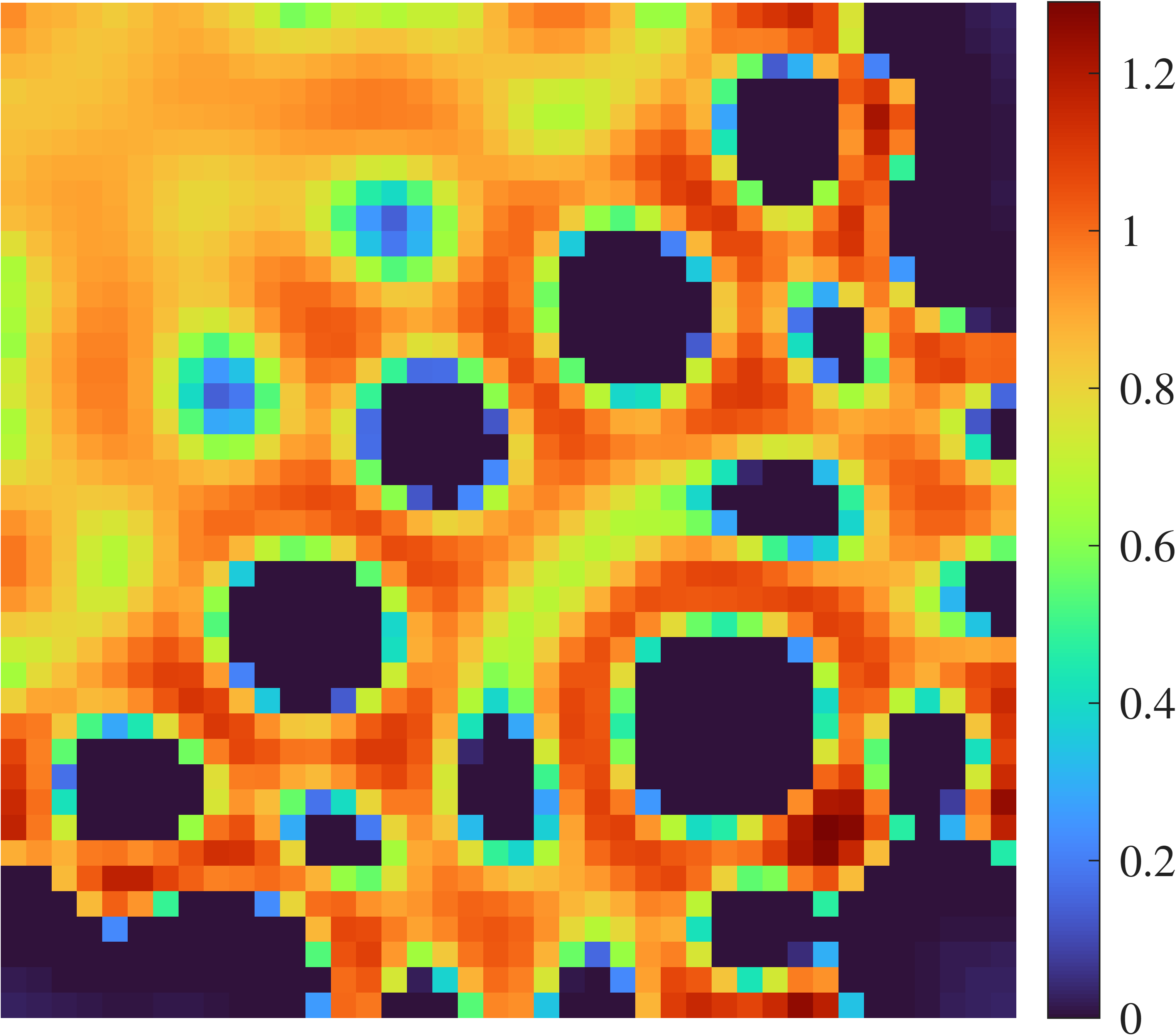}
            \label{}
        }
    }
    \makebox[\textwidth]{%
        \subfigure[]{%
            \includegraphics[width=0.31\textwidth]{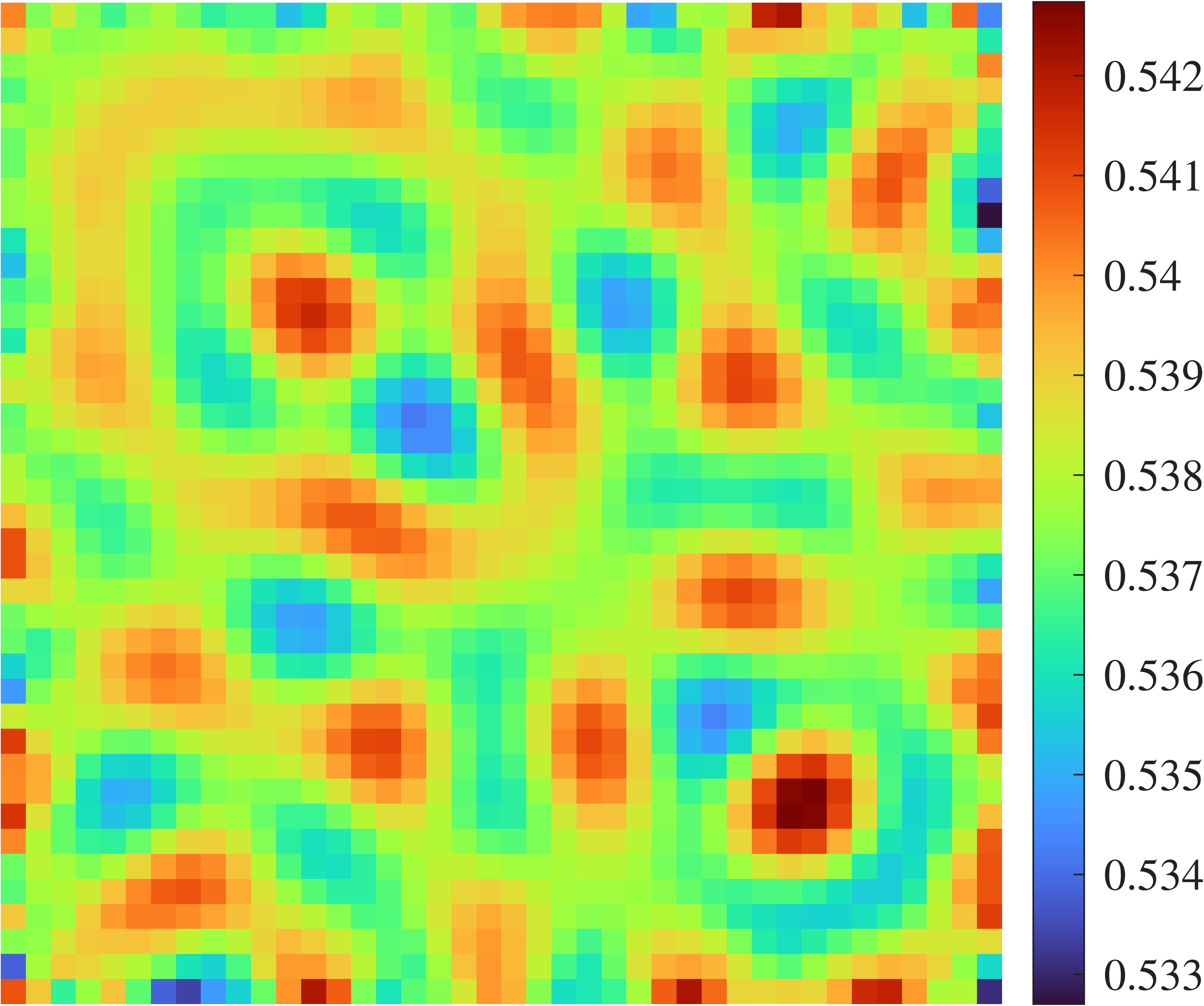}
            \label{}
        }
        \subfigure[]{%
            \includegraphics[width=0.31\textwidth]{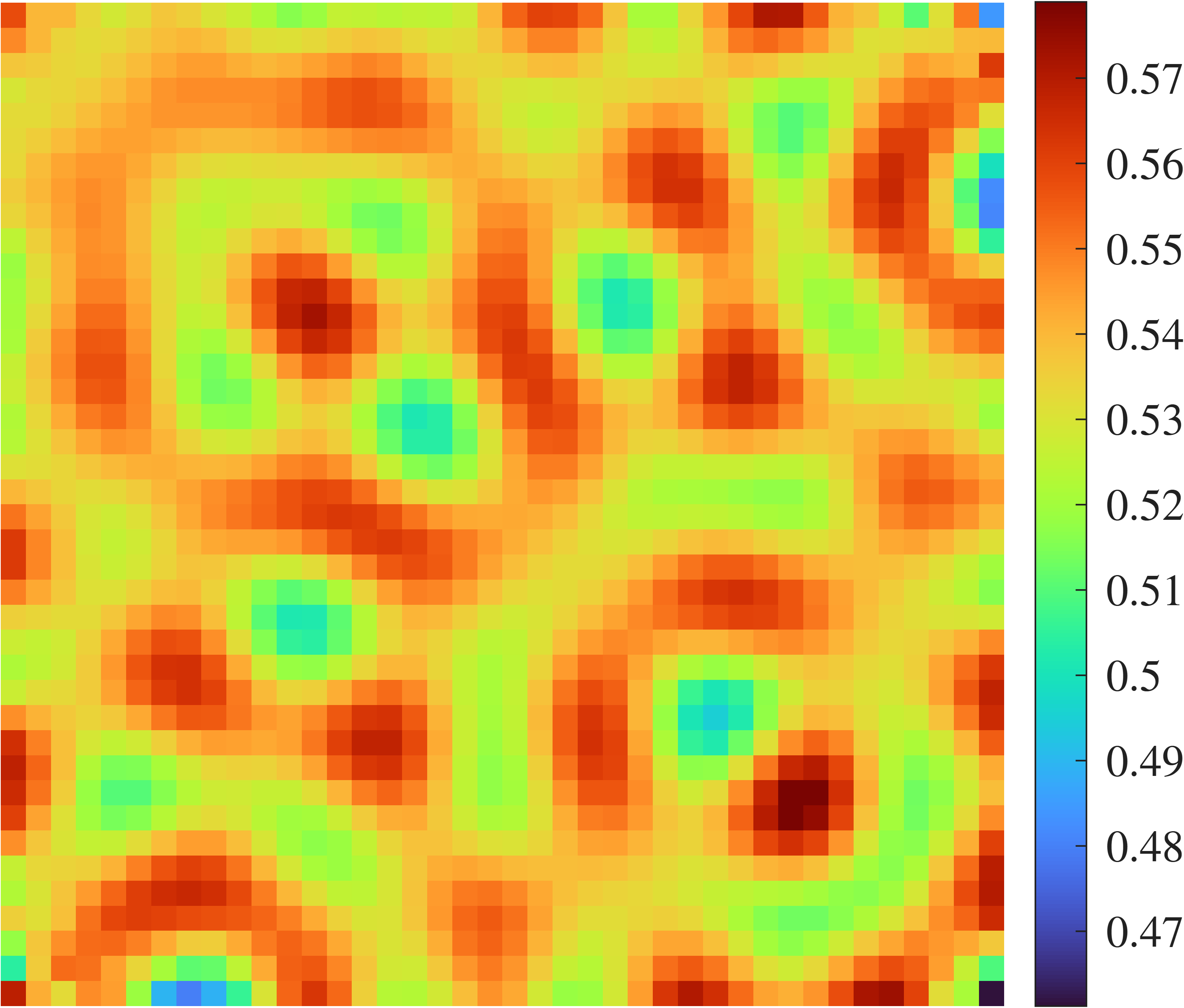}
            \label{}
        }
        \subfigure[]{%
            \includegraphics[width=0.30\textwidth]{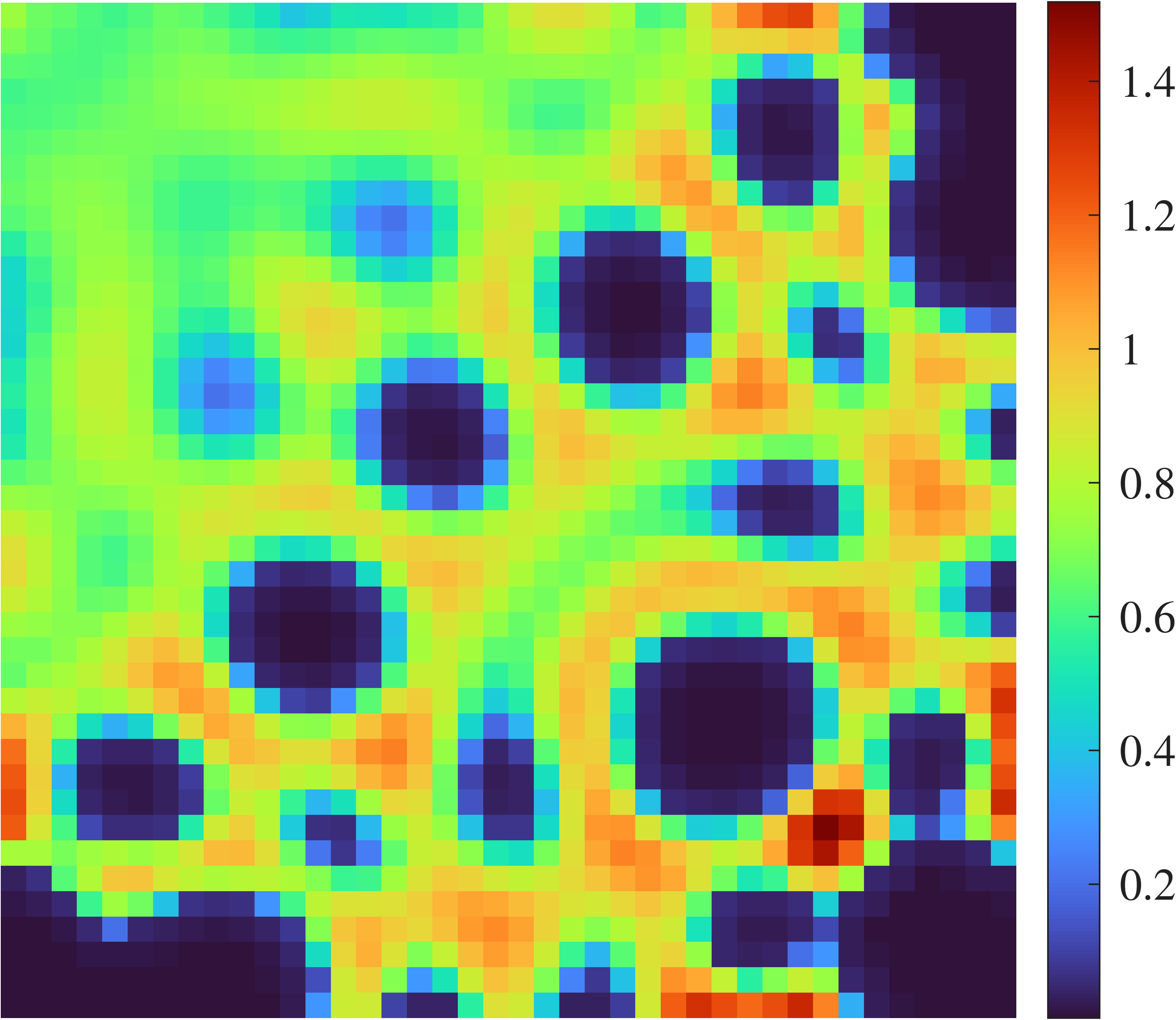}
            \label{}
        }
    }
    \makebox[\textwidth]{%
        \subfigure[]{%
            \includegraphics[width=0.31\textwidth]{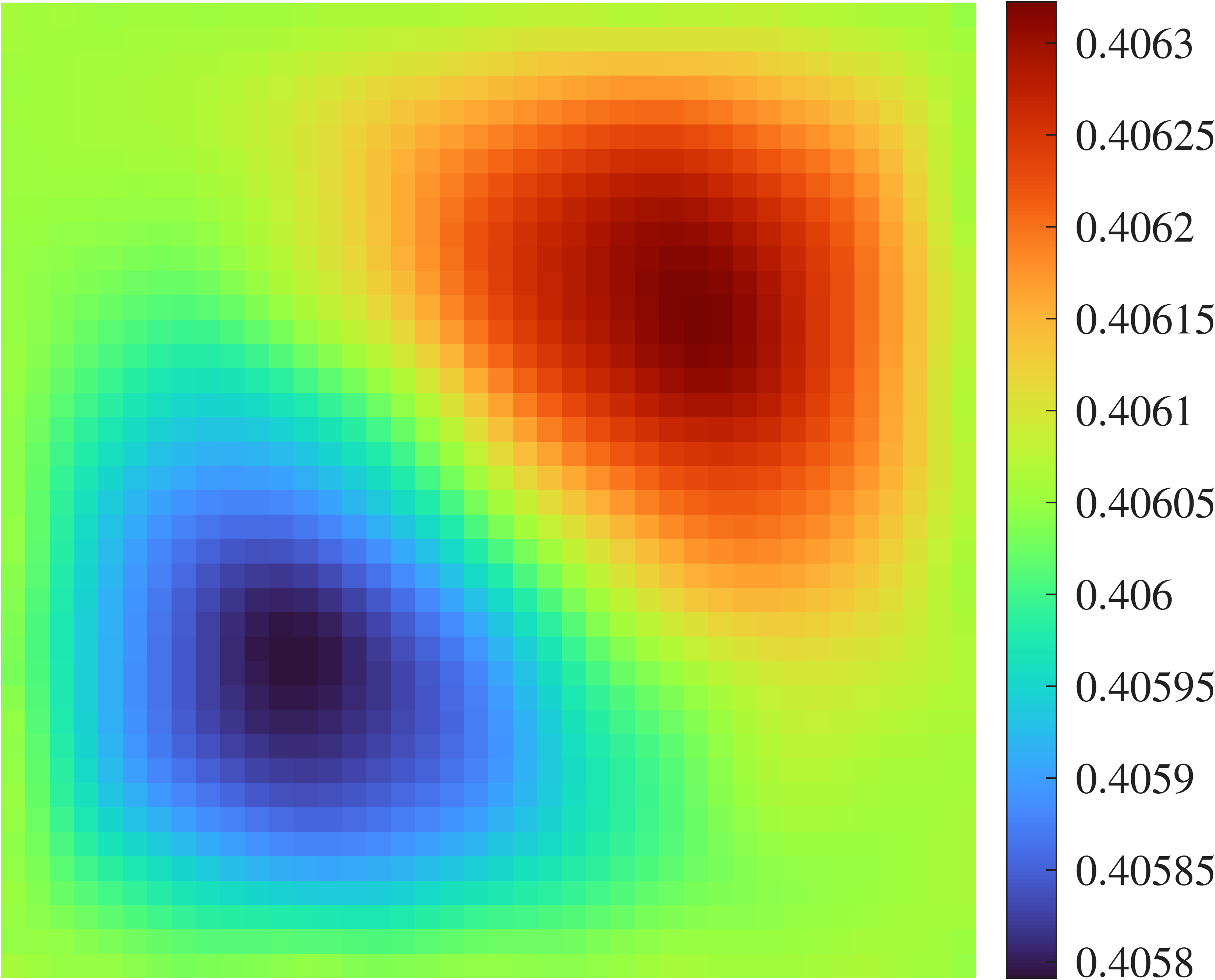}
            \label{}
        }
        \subfigure[]{%
            \includegraphics[width=0.31\textwidth]{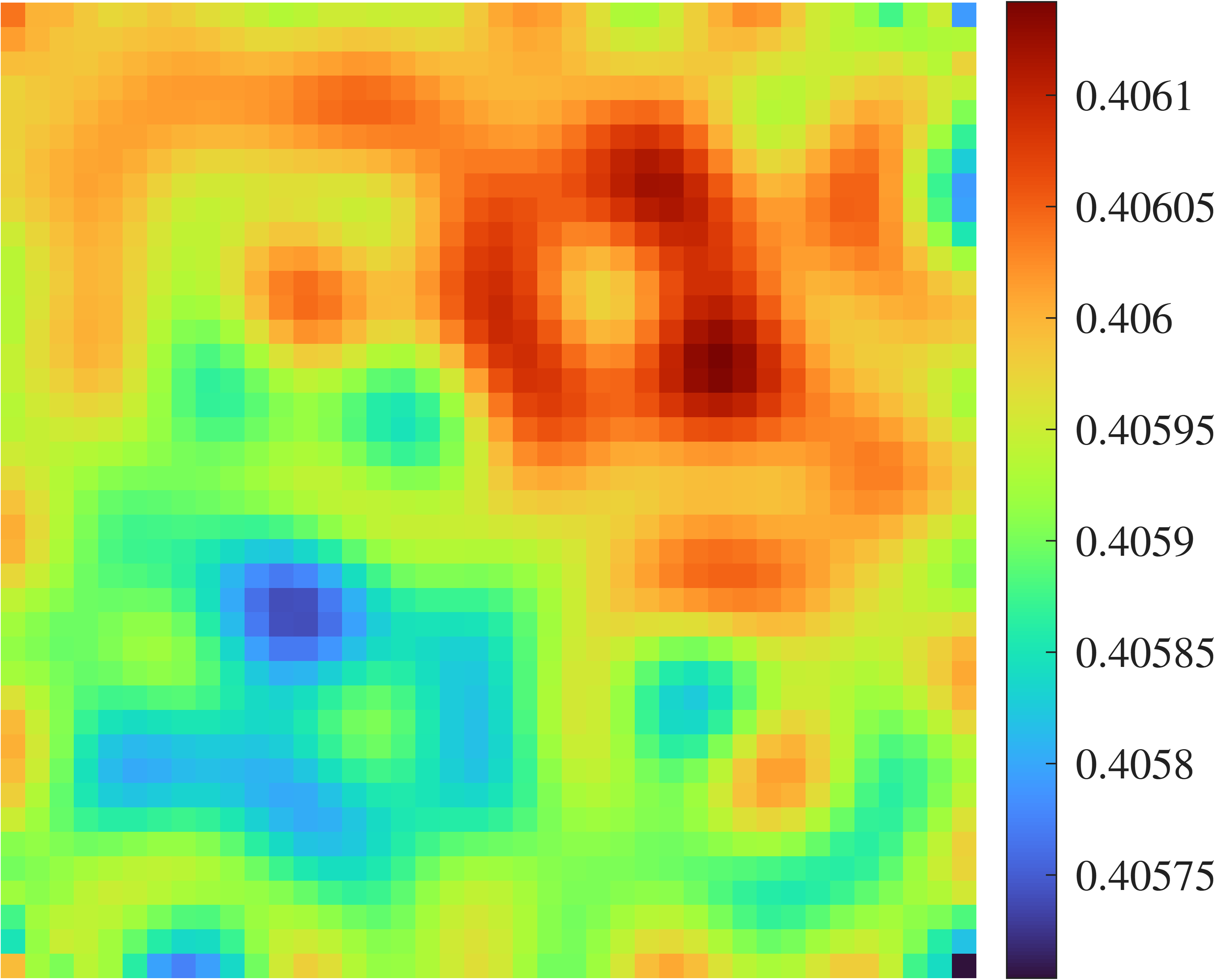}
            \label{}
        }
        \subfigure[]{%
            \includegraphics[width=0.30\textwidth]{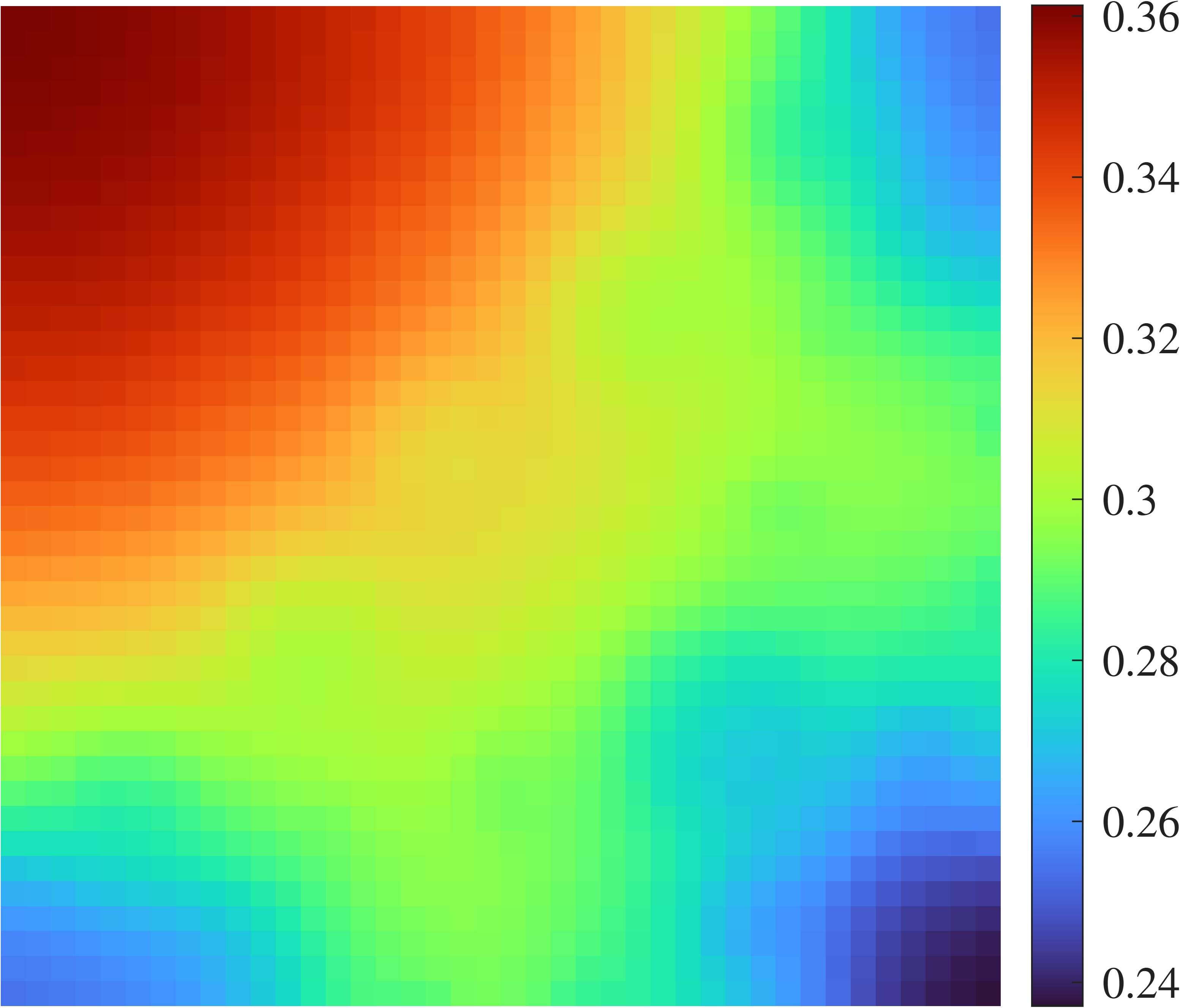}
            \label{}
        }
    }
    \caption{Spatial distribution of susceptible (first row), pathogen $1$ infection densities (second row), and pathogen $2$ infection densities (third row). For each, plots are displayed for $t=750$ (first column), $t=1450$ (second column), and $t=1900$ (third column).}
    \label{fig:superinfect_pattern}
\end{figure*}

\begin{figure*}[htbp]
    \centering

    \makebox[\textwidth]{%
        \subfigure[]{%
            \includegraphics[width=0.32\textwidth]{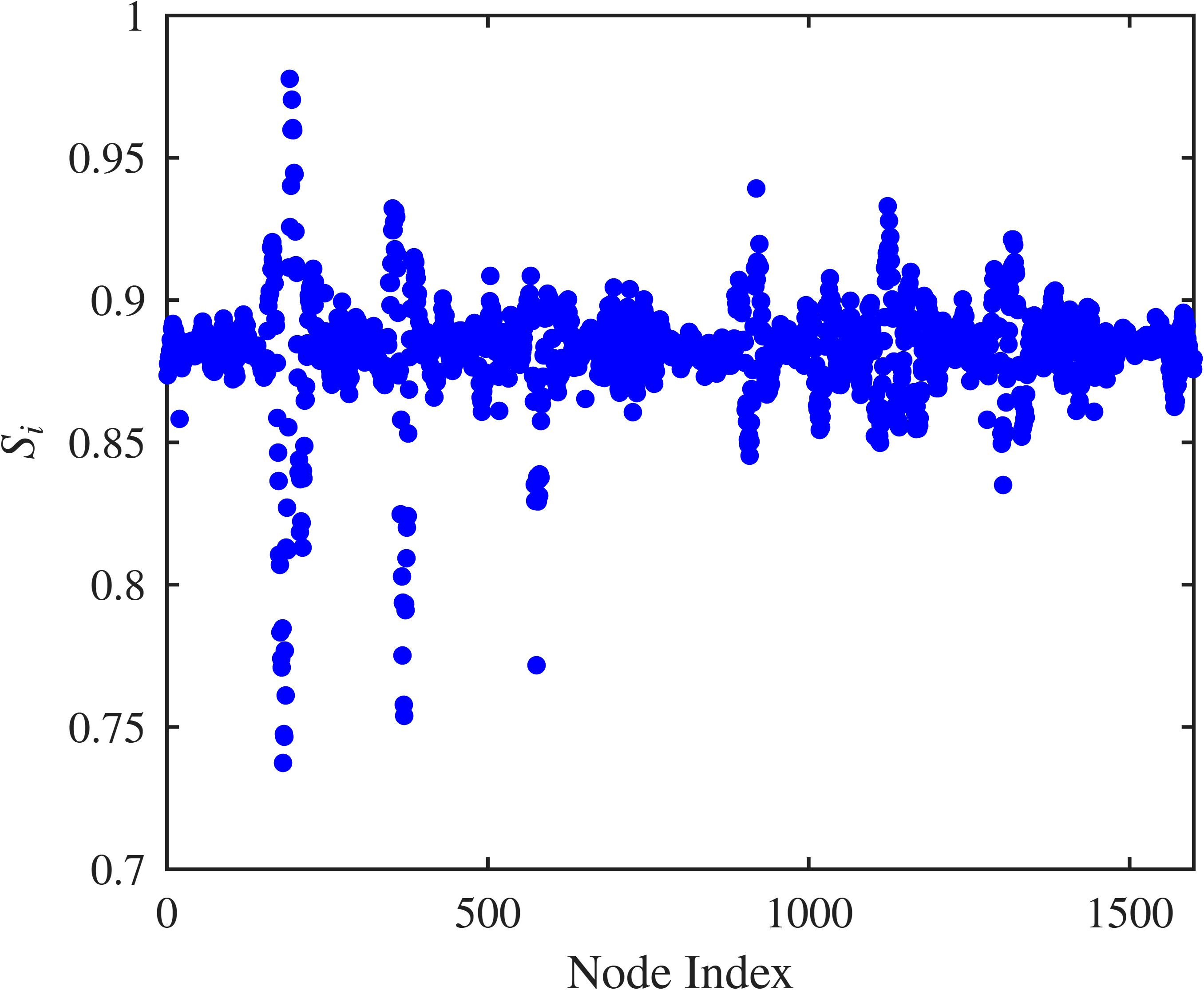}
            \label{}
        }
        \subfigure[]{%
            \includegraphics[width=0.32\textwidth]{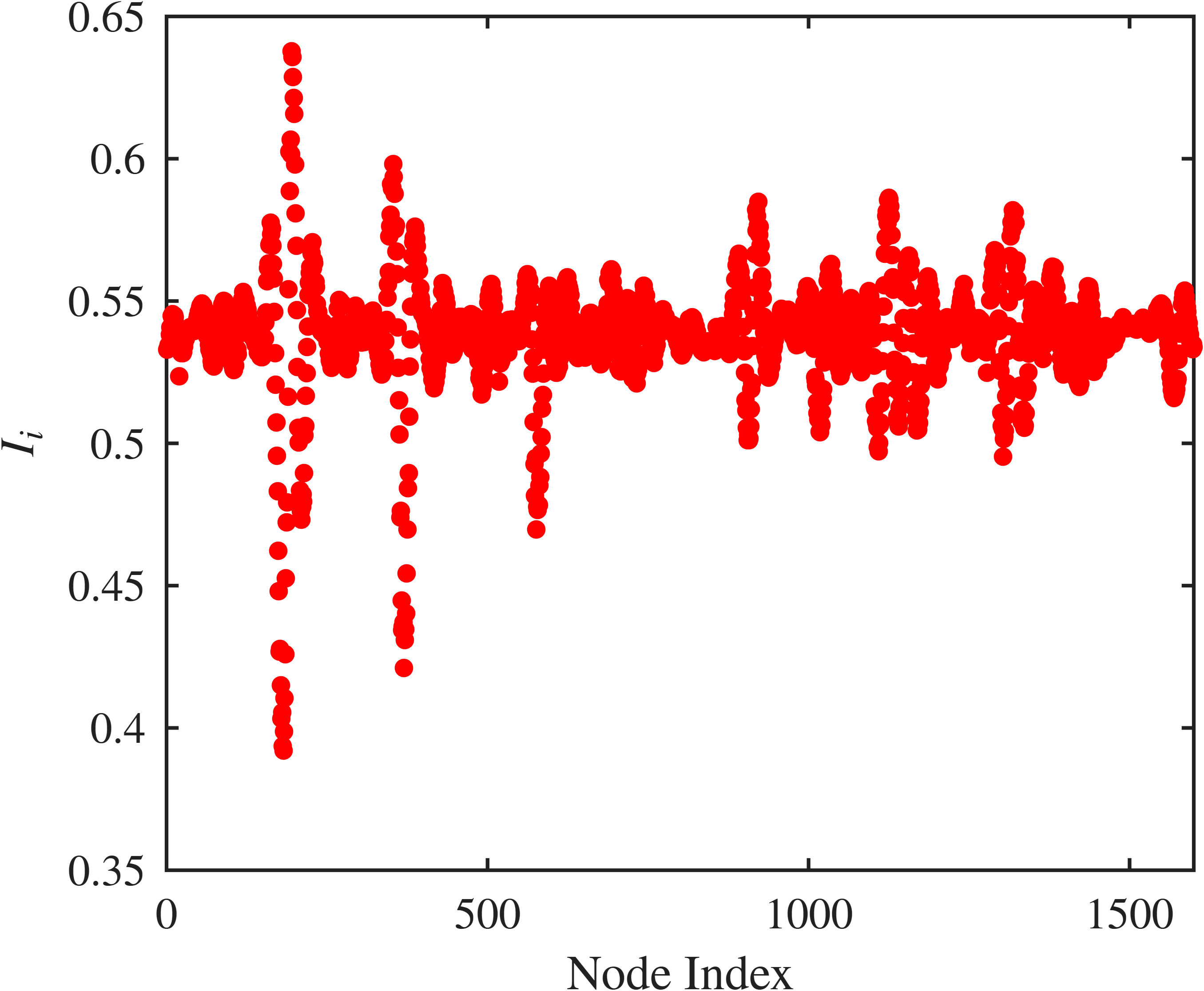}
            \label{}
        }
        \subfigure[]{%
            \includegraphics[width=0.32\textwidth]{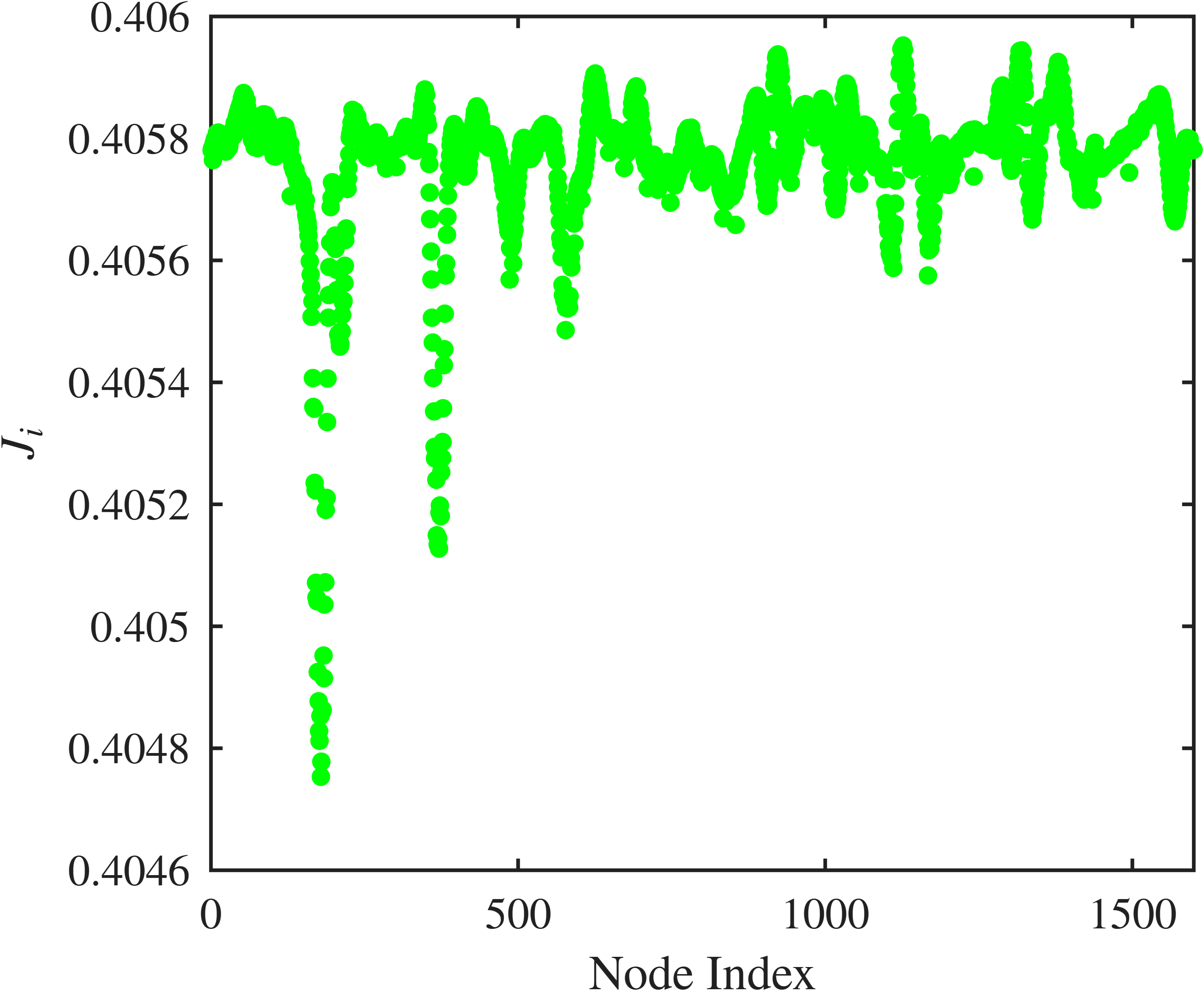}
            \label{}
        }
    }
    \caption{Spatial distribution of densities on each layer at time $t=700$ on an WS multiplex network with all layers having average degree of $12$.}
    \label{fig:superinfect_ws}
\end{figure*}

We simulate with the WS and BA networks because both the small-world phenomenon and scale-free properties are present in social networks~\cite{davidsen2002emergence, barabasi2003scale}, the World Wide Web~\cite{kleinberg2000small, barabasi2000scale}, and migration between physical human communities~\cite{peres2016community, levy2010scale}. We consider the following metrics throughout this section:
\begin{definition}[Pattern amplitude for superinfection dynamics]\label{def:amp_superinfect}
    Because some patterns shift from the original equilibrium, we define the amplitude of the densities among all layers of the multiplex network in superinfection dynamics to be
    \begin{equation}
        A\coloneqq\sqrt{\sum_{i=1}^N\left[(S_i-\bar{S})^2+(I_i-\bar{I})^2+(J_i-\bar{J})^2\right]},
    \label{eqn:def-A}
    \end{equation}
    where $\bar{S}$, $\bar{I}$, and $\bar{J}$ are the mean densities of all nodes in the network: $\bar{S}=\frac{1}{N}\sum_{i=1}^N S_i$, $\bar{I}=\frac{1}{N}\sum_{i=1}^N I_i$, and    
    $\bar{J}=\frac{1}{N}\sum_{i=1}^N J_i$.
\end{definition}

\begin{definition}[Pattern amplitude for co-infection dynamics]\label{def:amp_coinfect}
    Similarly, we use the following alternate definition of $A$ from Equation~(\ref{eqn:def-A}):
    \begin{equation}
        \sqrt{\sum_{i=1}^N\left[(S_i-\bar{S})^2+(I_i-\bar{I})^2+(J_i-\bar{J})^2+(C-\bar{C})^2\right]},
    \end{equation}
    where $\bar{S}$, $\bar{I}$, $\bar{J}$, and $\bar{C}$ are the mean densities of all nodes in the network: $\bar{S}=\frac{1}{N}\sum_{i=1}^N S_i$, $\bar{I}=\frac{1}{N}\sum_{i=1}^N I_i$,       
    $\bar{J}=\frac{1}{N}\sum_{i=1}^N J_i$ and $\bar{C}=\frac{1}{N}\sum_{i=1}^N C_i$.
\end{definition}

\subsection{Amplifying Hotspots}\label{sec:turing-pattern-simu}

Emerging spatial hotspots have been investigated in the spread of COVID-19~\cite{purwanto2021spatiotemporal}. Here, we focus on hotspots that arise from Turing instability. In SIS dynamics, prior research has shown that Turing patterns often become spatio-temporally stationary after a period of time ~\cite{zhao2025navigating}. Meanwhile, in in both the {\bf MBRD-SI} and {\bf MBRD-CI} dynamics discussed in this paper, we find that it is possible for these peaks to be both spatially stationary while also consistently growing over time until system collapse occurs. We illustrate this phenomena with Examples~\ref{ex:1} and~\ref{ex:2} for superinfection and co-infection dynamics, respectively. 

%We first analyze the set of parameters for the superinfection model below.

\begin{example}[{\bf MBRD-SI} model, Yu $\&$ Schaposnik \cite{yu2025spatial}] \label{ex:1} 
Consider the {\bf MBRD-SI} model in Equation~(\ref{eq:superinfect-prelim}) with the following parameters:
\begin{equation}\label{eq:p1-settings}
\begin{aligned}
\mu &= 0.005\quad &r&=0.1\quad & A&=0.1\quad & K&=1\\
\beta_1 &=0.3, \quad & \beta_2&=0.15,\quad & \sigma &=3\\
\gamma_1&=0.02,\quad &\gamma_2&=0.05,\quad & \alpha_1&=0.02,\quad & \alpha_2&=0.15\\
d_{11}&=0.1,\quad & d_{12}&=-0.2\quad & d_{13}&=-0.2\\
d_{22}&=0.01,\quad & d_{33}&=4.8.
\end{aligned}
\end{equation}
\end{example}

\begin{figure*}[htbp]
    \centering

    \makebox[\textwidth]{%
        \subfigure[]{%
            \includegraphics[width=0.32\textwidth]{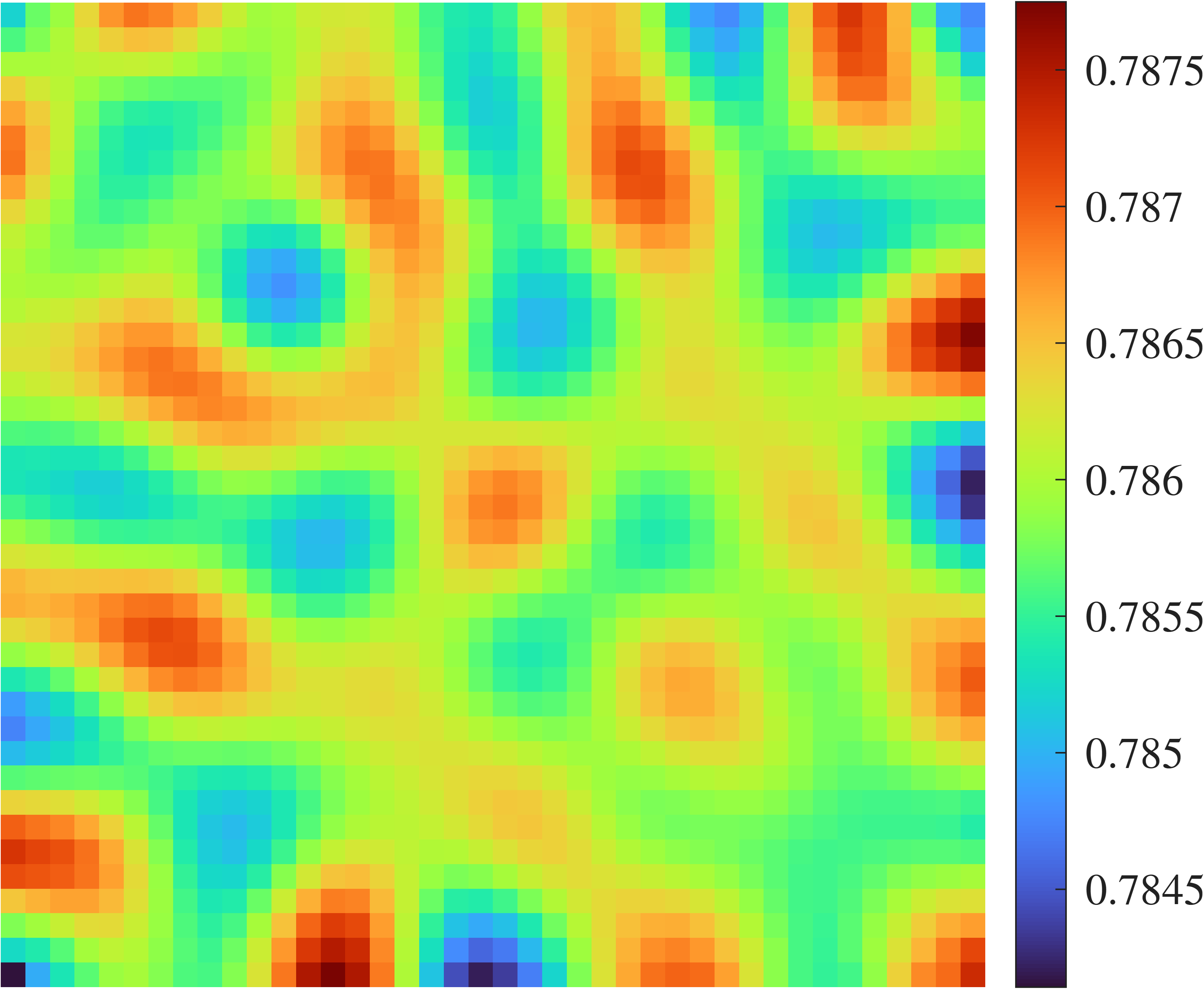}
            \label{}
        }
        \subfigure[]{%
            \includegraphics[width=0.31\textwidth]{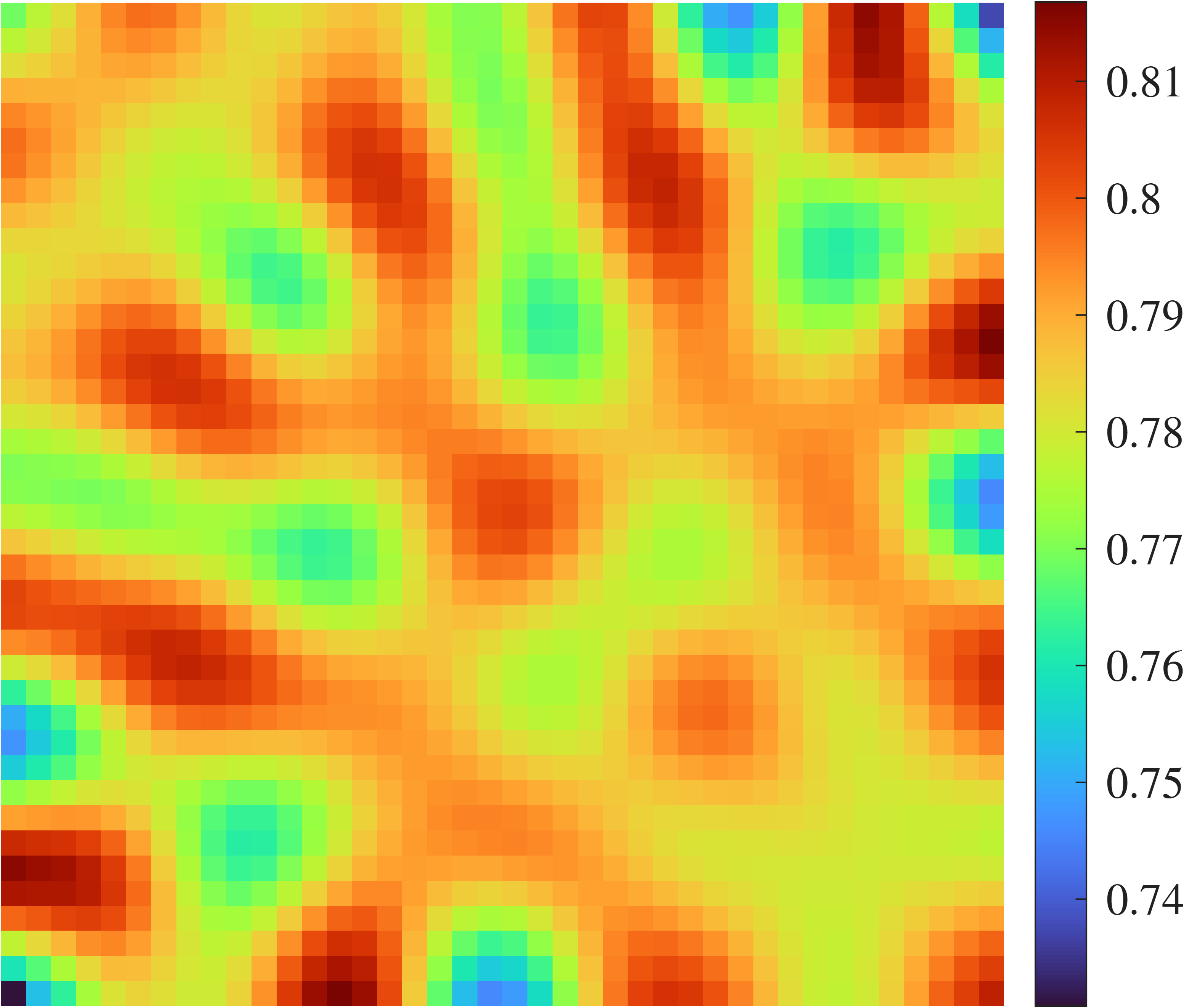}
            \label{}
        }
        \subfigure[]{%
            \includegraphics[width=0.30\textwidth]{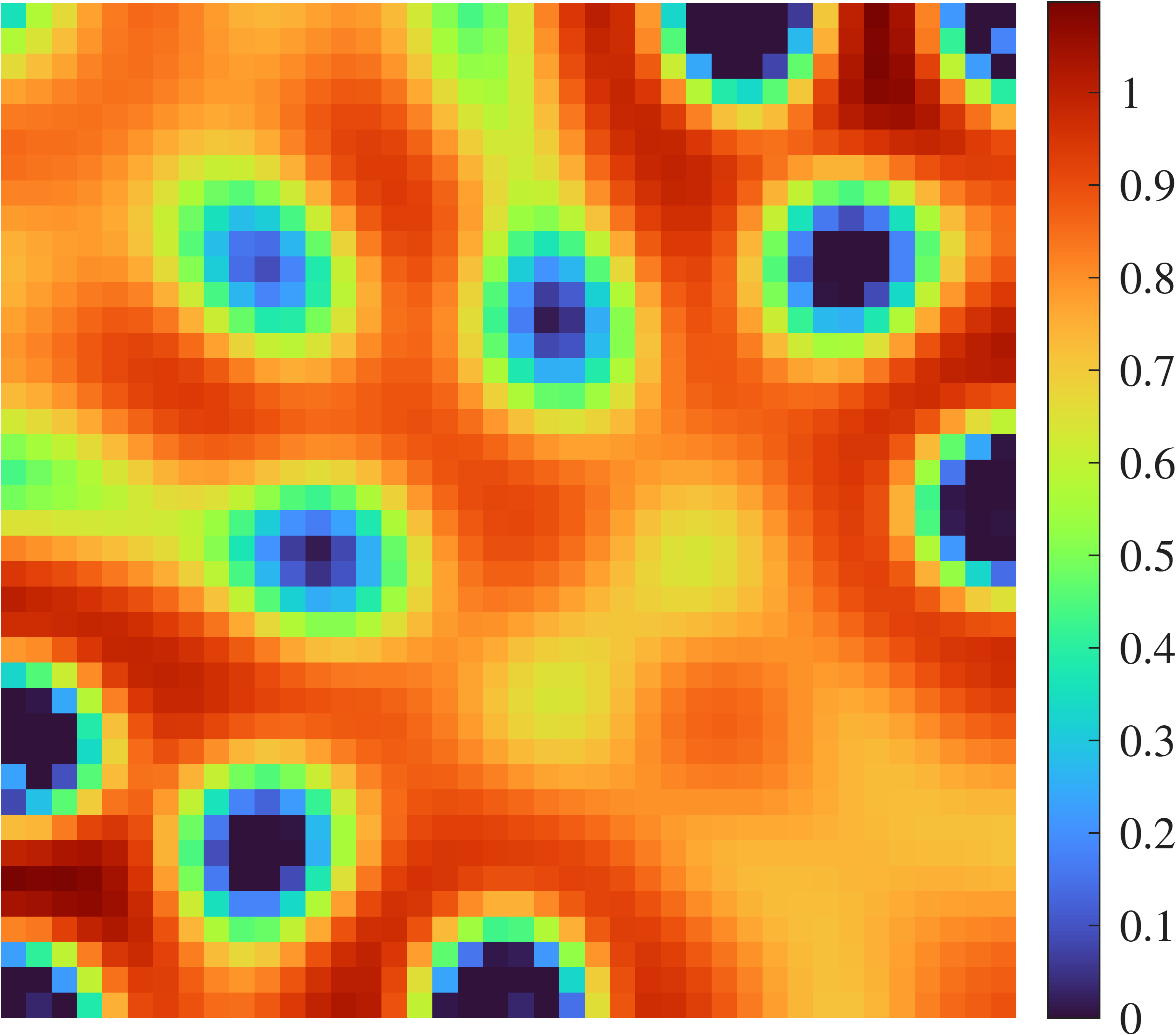}
            \label{}
        }
    }
    \makebox[\textwidth]{%
        \subfigure[]{%
            \includegraphics[width=0.315\textwidth]{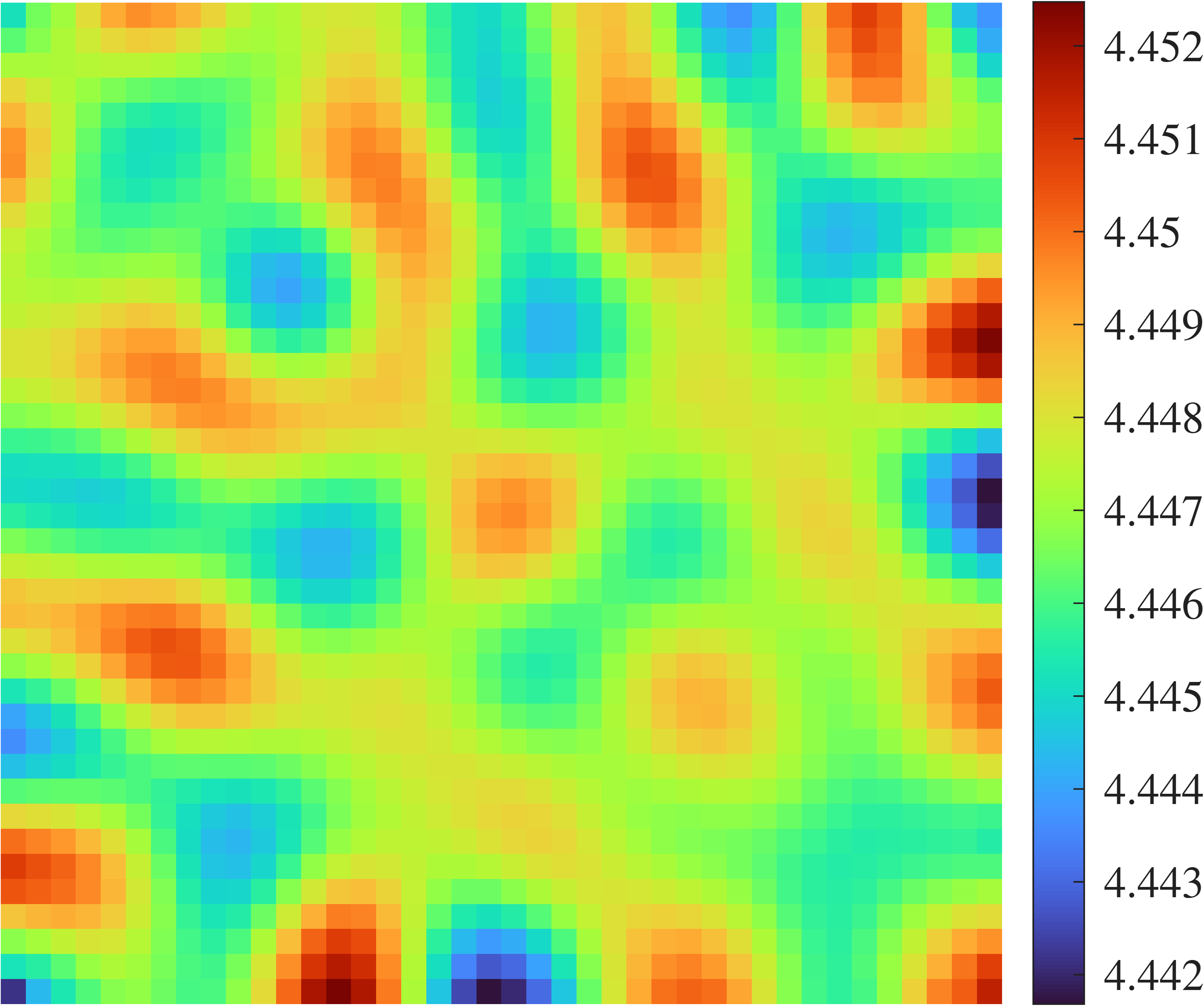}
            \label{}
        }
        \subfigure[]{%
            \includegraphics[width=0.31\textwidth]{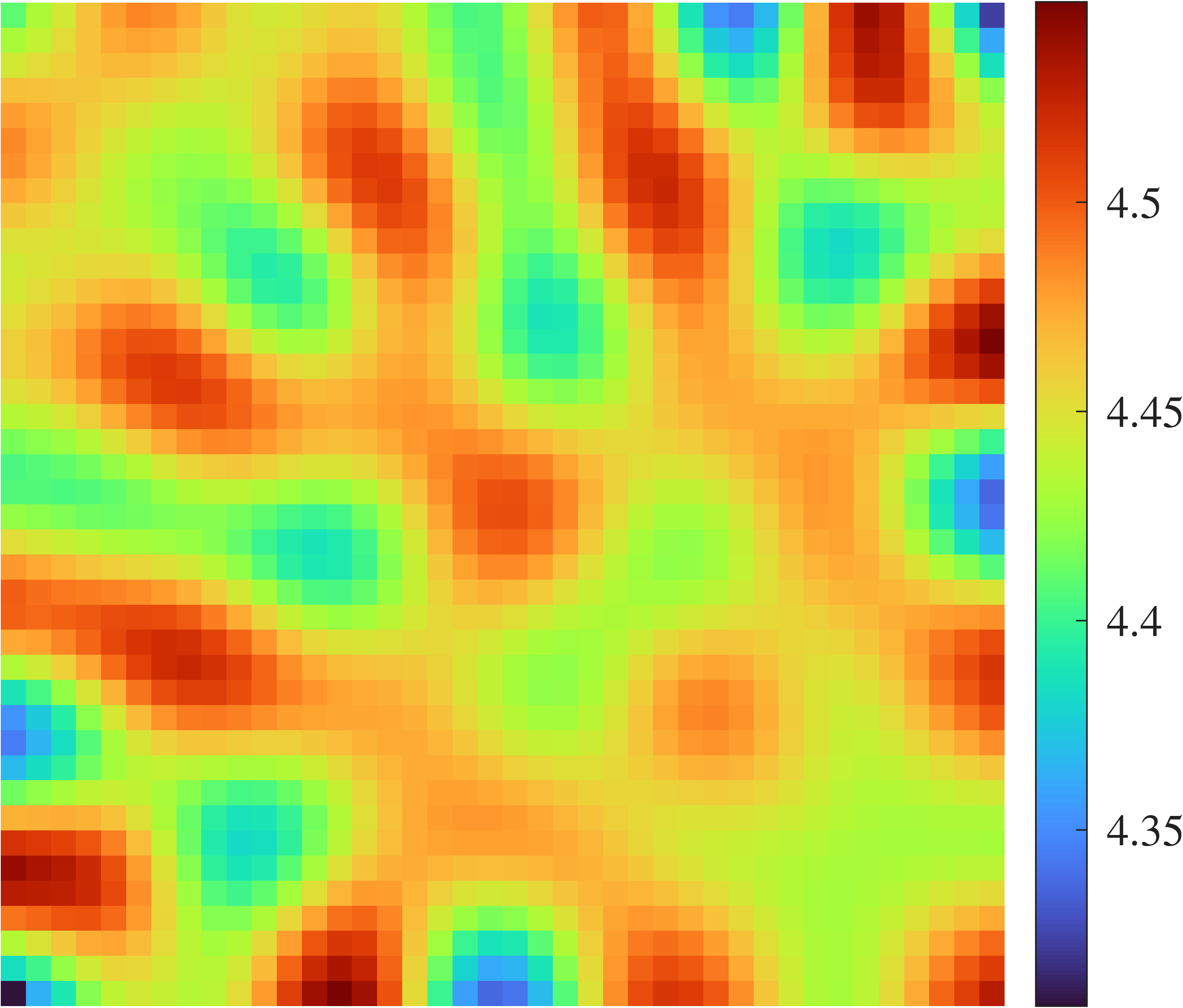}
            \label{}
        }
        \subfigure[]{%
            \includegraphics[width=0.305\textwidth]{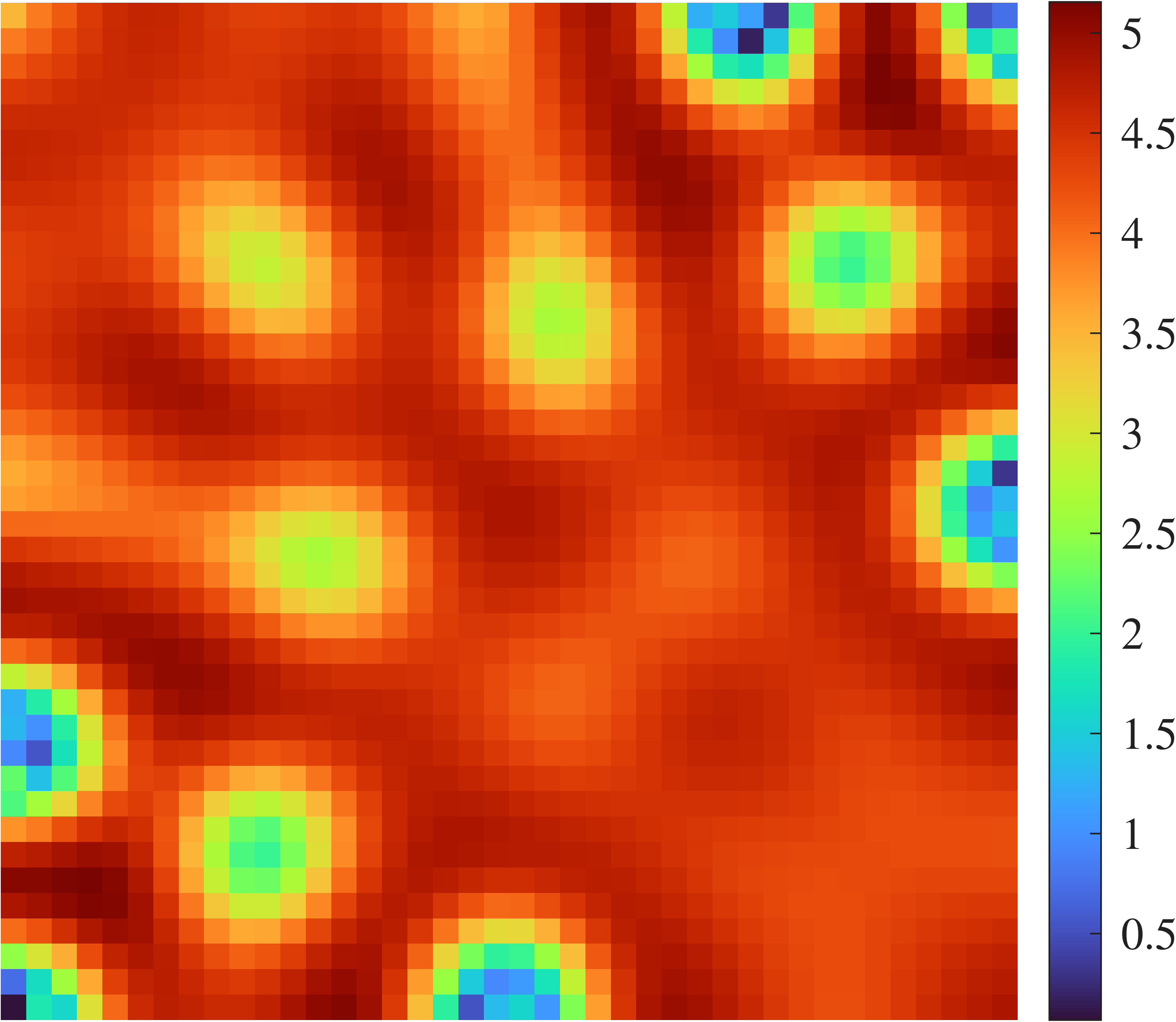}
            \label{}
        }
    }
    \makebox[\textwidth]{%
        \subfigure[]{%
            \includegraphics[width=0.32\textwidth]{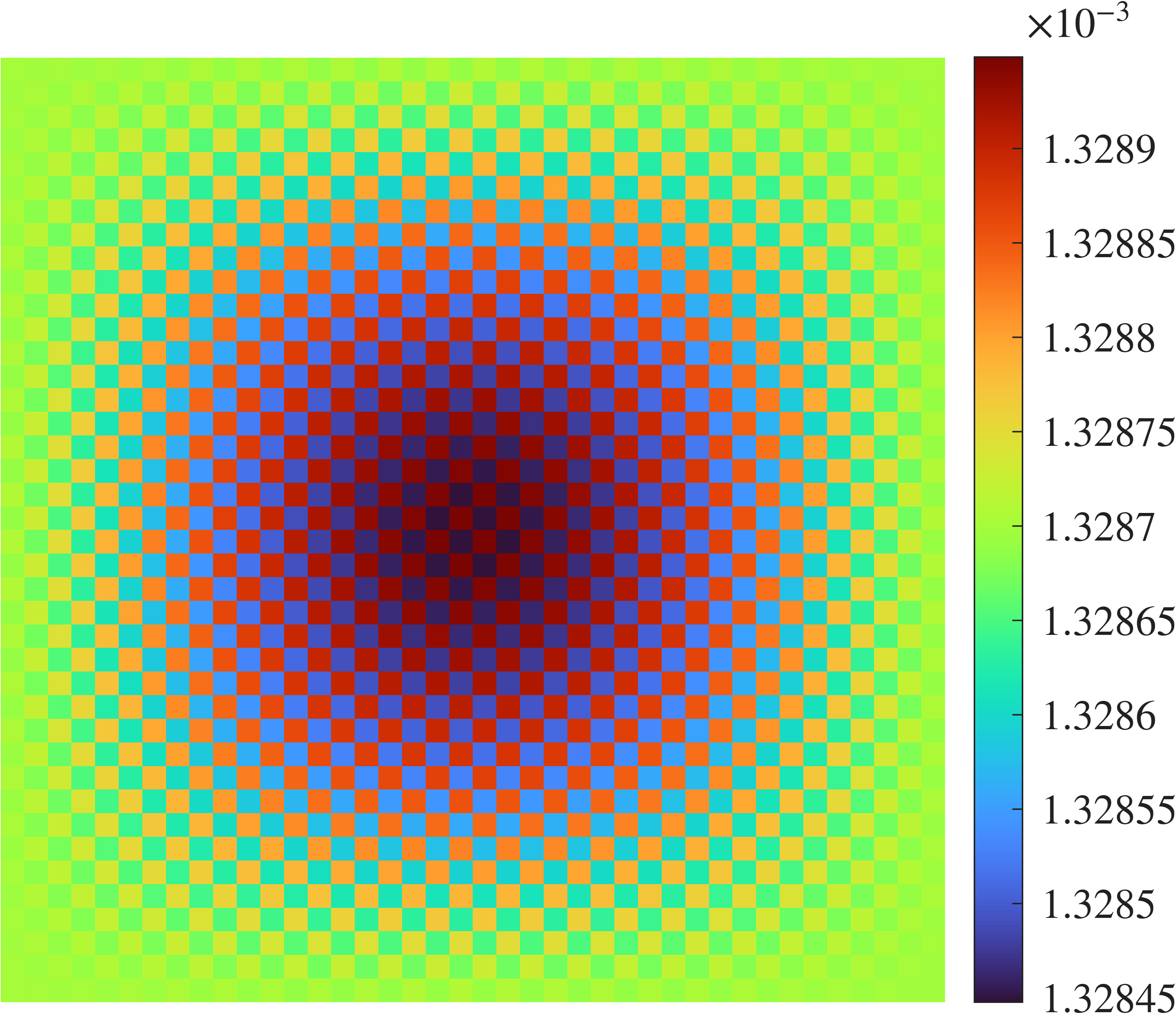}
            \label{}
        }
        \subfigure[]{%
            \includegraphics[width=0.32\textwidth]{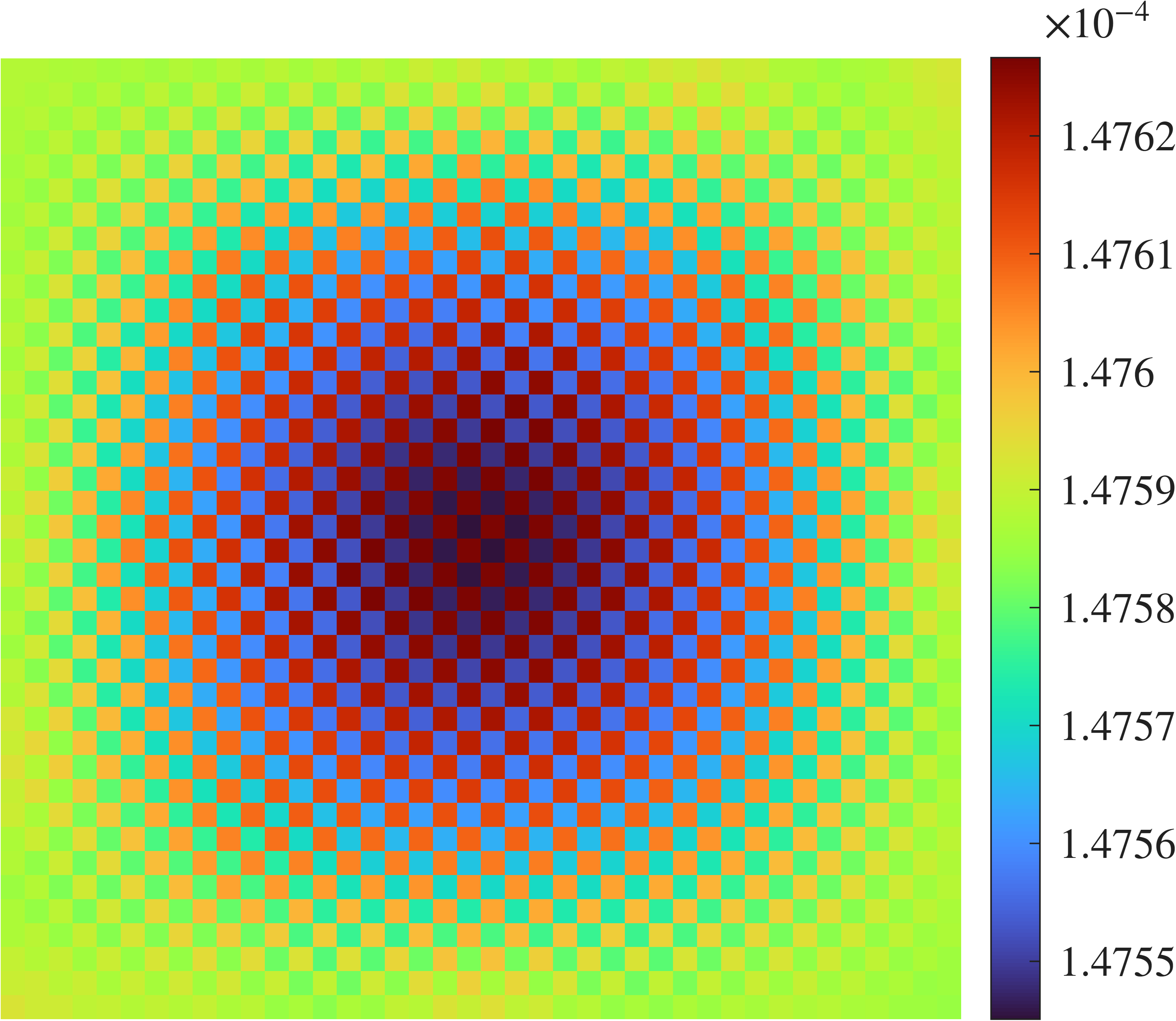}
            \label{}
        }
        \subfigure[]{%
            \includegraphics[width=0.31\textwidth]{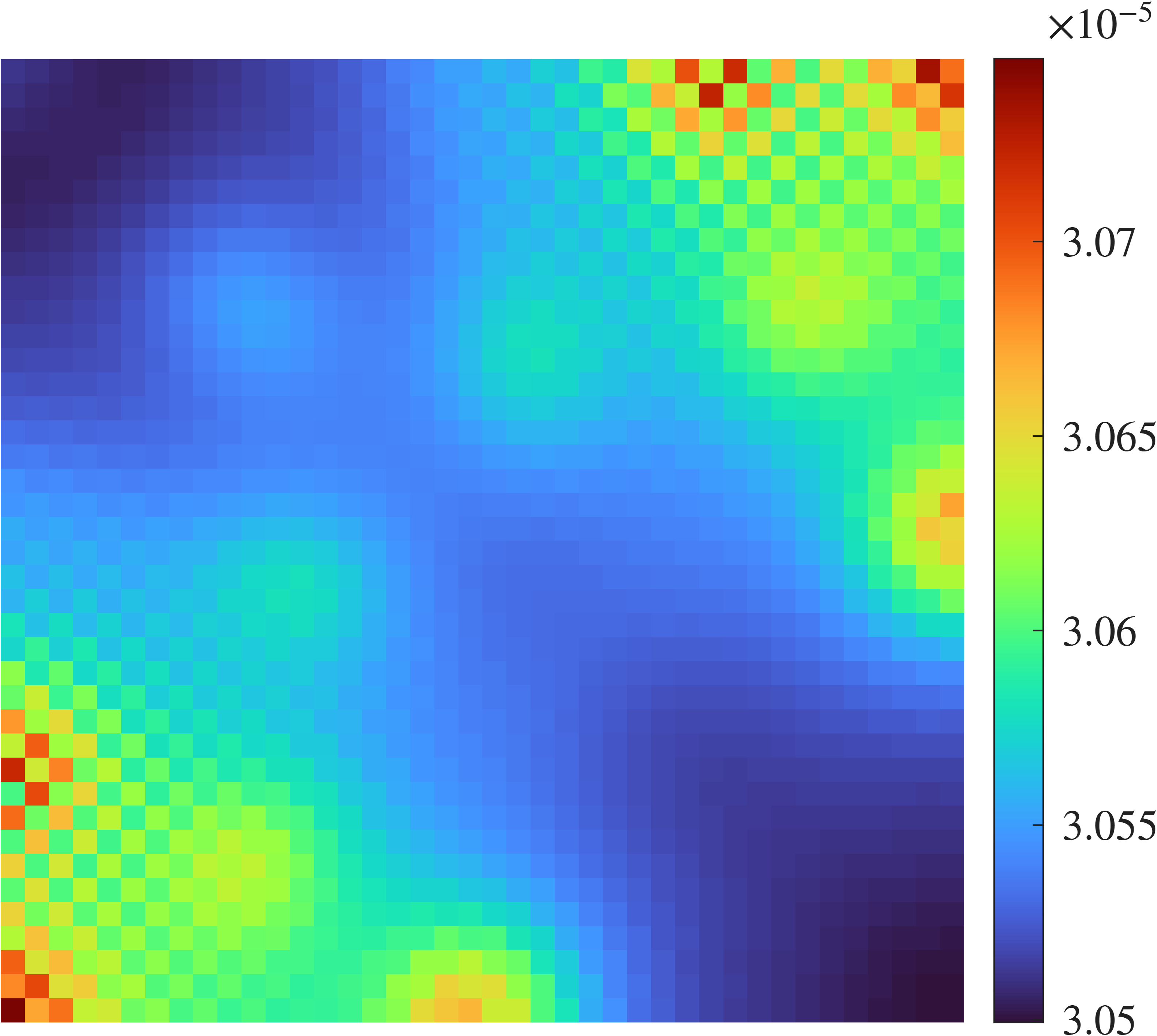}
            \label{}
        }
    }
    \makebox[\textwidth]{%
        \subfigure[]{%
            \includegraphics[width=0.32\textwidth]{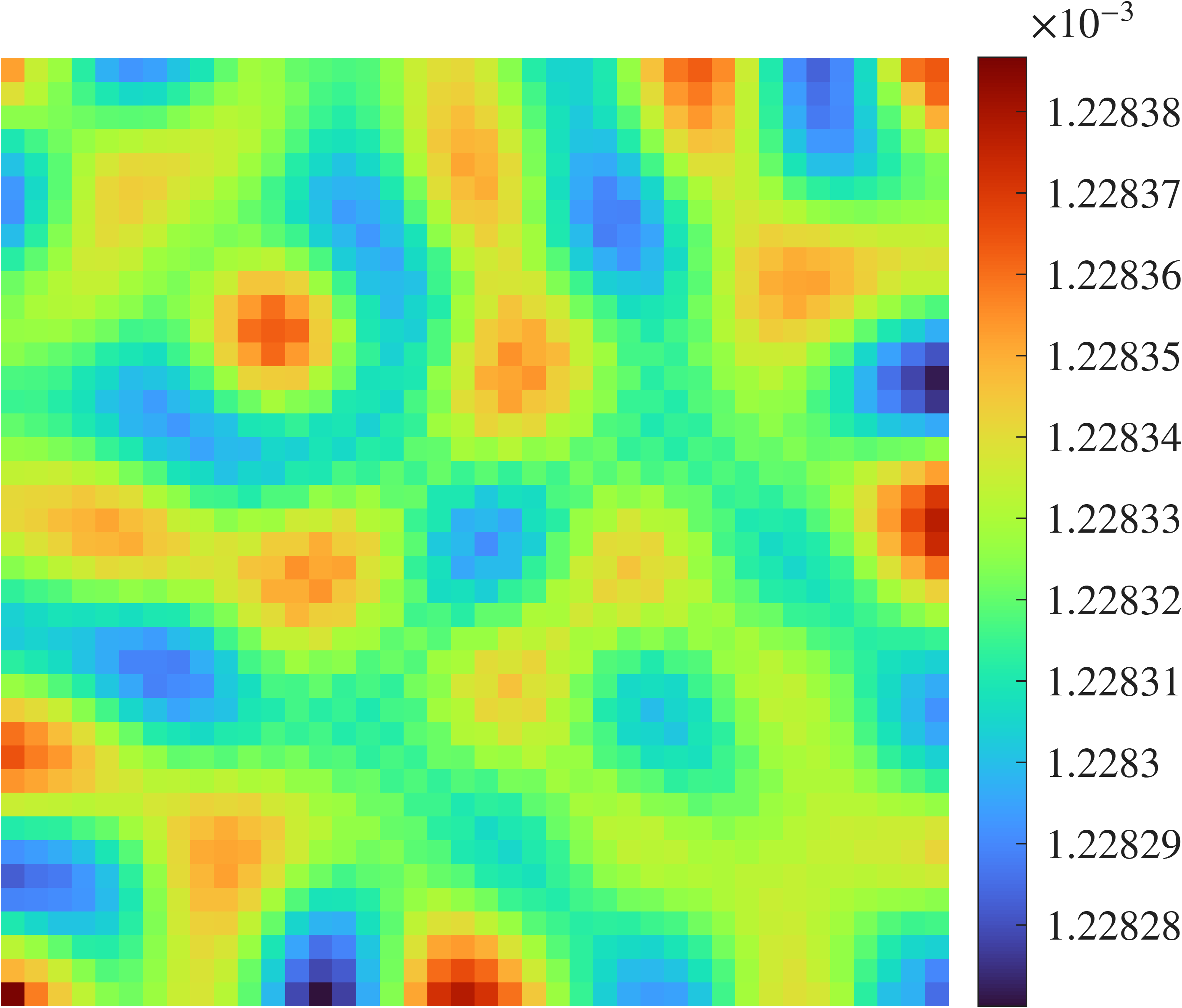}
            \label{}
        }
        \subfigure[]{%
            \includegraphics[width=0.31\textwidth]{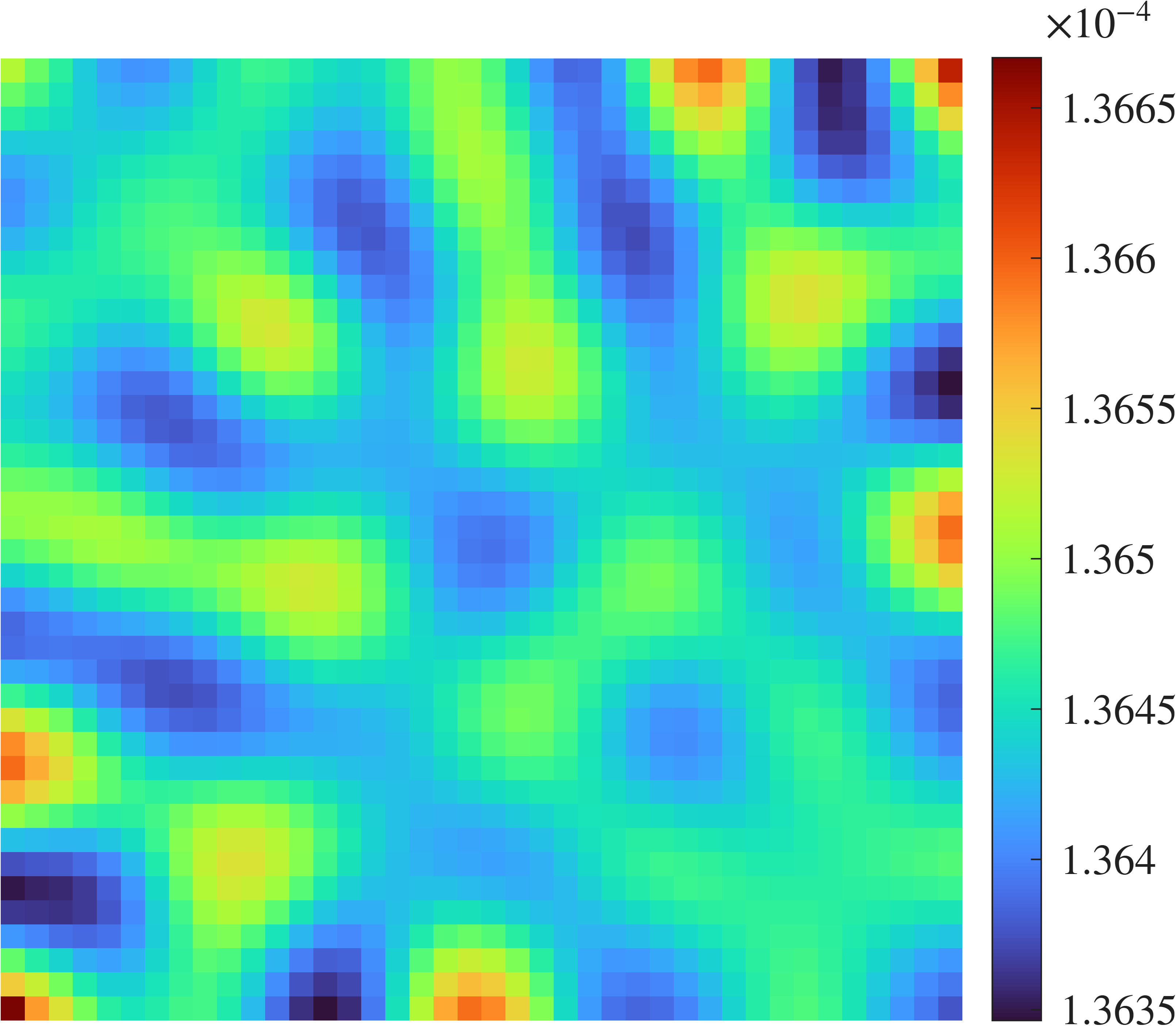}
            \label{}
        }
        \subfigure[]{%
            \includegraphics[width=0.305\textwidth]{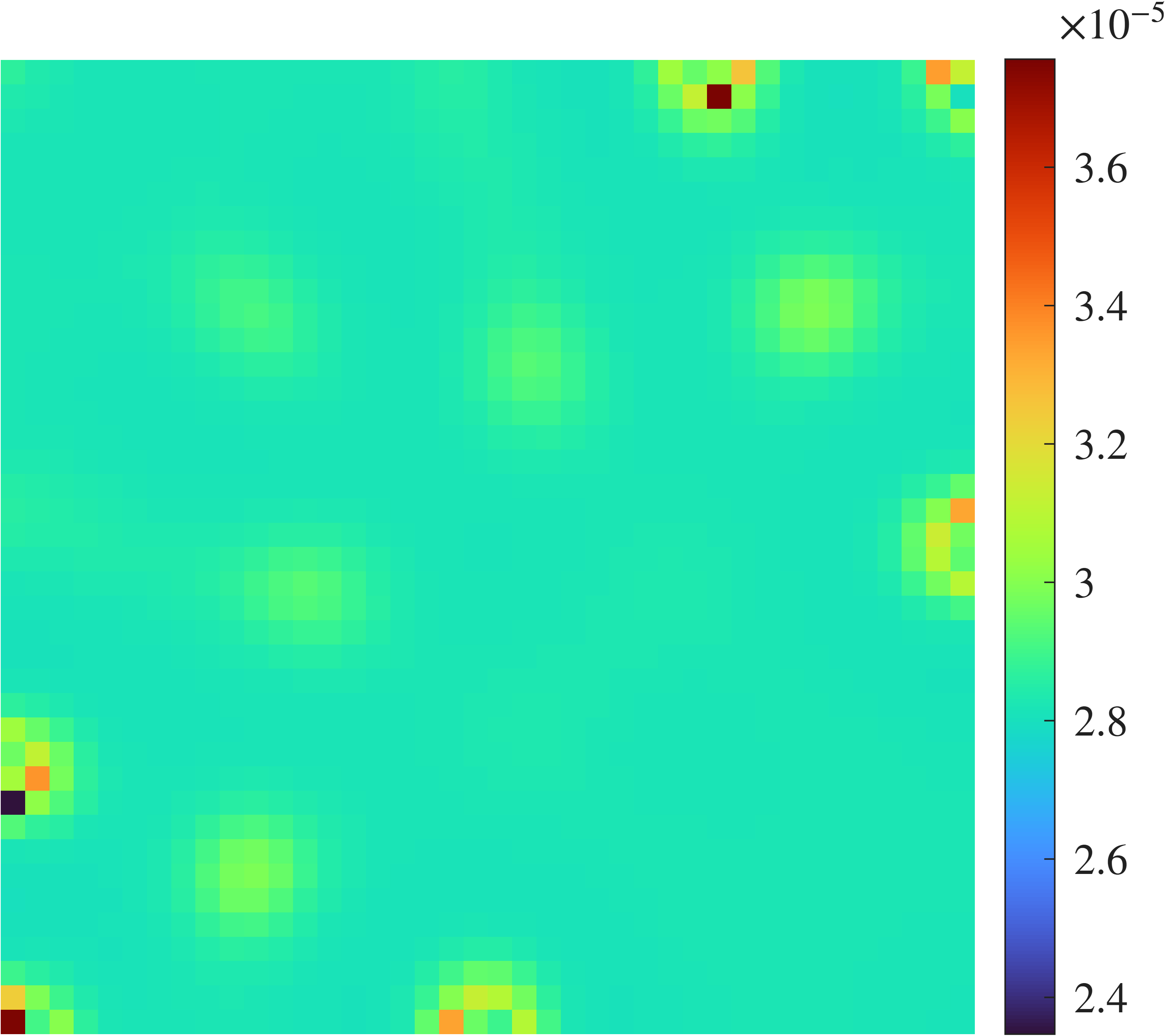}
            \label{}
        }
    }
    \caption{Spatial distribution of susceptible (first row), pathogen $1$ infection densities (second row), pathogen $2$ infection densities (third row), and co-infection densities (fourth row) on a LA12-LA12-LA4 network at $t=250$ (first column), $t=500$ (second column), and $t=700$ (third column).}
    \label{fig:coinfect_pattern}
\end{figure*}

 On a multiplex network with three identical LA12 networks for the $S$, $I$, and $J$ layers, the configuration in Example~\ref{ex:1} forms the patterns in \autoref{fig:superinfect_pattern}. This figure demonstrates the layers with $S$, $I$, and $J$ densities, in order of row, at times $t=750$, $1450$, and $1900$, respectively. As seen in the figure, we observe the emergence and growth of patterns with dots and stripes in the $S$ and $I$ layers. In those layers, the hotspots are still at the same locations between times $750$ and $1450$, which is a key feature of Turing patterns. On the other hand, pattern formation does not occur on the $J$ layer, demonstrating the possibility that Turing patterns occur only on some layers and not all of them. Finally, we notice from \autoref{fig:superinfect_pattern} that the $I$ layer densities are centered around approximately $0.52$ and the $J$ layer densities are centered around approximately $0.4$. This shows that it is possible for the $I$ layer to have a greater equilibrium even when $J$ steals hosts from $I$, and we believe this is largely because of $\beta_1$ being significantly larger than $\beta_2$. 

From our derivations, we expect that the type of network should not affect whether or not pattern formation occurs as long as the average degrees are constant, especially when the average degrees of each layer are large. We observe that this is indeed true for the {\bf MBRD-SI} dynamics described in Equation~(\ref{eq:superinfect-prelim}) and Example~\ref{ex:1}. The MBRD-SI model, with the parameters mentioned above, produces patterns on Watts-Strogatz networks where all layers have average degree $12$. \autoref{fig:superinfect_ws} displays the oscillating node densities at $t=700$ with the superinfection model parameters in Example~\ref{ex:1} on the WS multiplex network. 

We introduce a co-infection parameter setting below, which produces patterns with similar shapes to that of Example~\ref{ex:1}.

\begin{example}\label{ex:2}[{\bf MBRD-CI} model, Yu $\&$ Schaposnik \cite{yu2025spatial}]
Consider now the {\bf MBRD-CI} model  in Equation~(\ref{eq:coinfect-prelim}). We have the parameter setting below:
\begin{equation}\label{eq:p2-settings}
\begin{aligned}
\mu &= 0.005\quad & r&=0.1&\quad A&=0.1\quad& K&=1,\\
\beta_1&=0.3\quad & \beta_2&= 0.15\\ \beta_{10}&=0.1\quad & \beta_{02}&=0.1,\quad & \beta_{12}&=0.05\\
\gamma_1&=0.02\quad & \gamma_2&=0.05\\ \alpha_1&=0.02,\quad & \alpha_2&=0.15\quad & \alpha_{12}&=0.1\\
d_{11}&=0.4\quad & d_{12}&=-0.2\quad & d_{13}&=-0.2\\
d_{22}&=0.01\quad & d_{33}&=4.8.
\end{aligned}
\end{equation}
\end{example}

 On an LA12-LA12-LA4 multiplex network for the layers with $S$, $I$, $J$ densities, this configuration forms the patterns in  \autoref{fig:coinfect_pattern}. This figure presents the four layers, in order of row, at times $t=250$, $500$, and $700$, respectively. Again, as seen in the figure, we observe fine-grained Turing patterns with a maze-like structure, and the hotspots are still at the same locations between times $250$ and $500$. 

System collapse as a result of growing Turing patterns has been investigated in competition dynamics between pathogens~\cite{doumate2023competition}. We see this phenomenon occur in both {\bf MBRD-SI} and {\bf MBRD-CI} dynamics. First, as shown in \autoref{fig:superinfect_pattern}, the oscillations grow until a time at approximately $1600$ when system collapse begins. The third column of \autoref{fig:superinfect_pattern} displays the patterns at $t=1900$, when many susceptible and pathogen $1$-infected densities have reached $0$. In \autoref{fig:coinfect_pattern}, we observe similar system collapse at $t=700$, shown in the third column. This demonstrates that with the right parameter settings, it is possible for the uninfected population to disappear in most nodes. This shows how local oscillations can cause local system collapse, which propagates to other areas and leads to global collapse.

Overall, from observation, we find Turing-Hopf patterns are rarer than Turing patterns. With the presence of Turing patterns, location-targeted intervention will be easier as the hotspots formed are stationary. With Turing-Hopf patterns, the locations most in need of intervention will constantly be changing, making an effective response more difficult.

\subsection{Effects of Model Parameters}
\label{sec:pattern-parameters}

We consider here the effect of various model parameters on pattern formation and the growth of Turing instability-induced hotspots. Through our trials  and alalysis, 
we find that the following characteristics are present in most of the patterns we have studied and thus, we believe they help induce pattern formation.
\begin{itemize}
    \item There is a significant difference between the transmission rates $\beta_1$ and $\beta_2$ and the removal rates $\alpha_1$ and $\alpha_2$. This is especially true for the {\bf MBRD-SI} model.
    \item The cross-diffusion rates $d_{12}$ and $d_{13}$ are negative, meaning in the physical sense that susceptible individuals gravitate towards areas with a high number of infections of either pathogen.
    \item There is a substantial difference between at least two of the diffusion rates $d_{11}$, $d_{22}$, and $d_{33}$. This is also consistent with the parameter settings for the patterns described in~\cite{duan2019turing, zhao2025navigating}.
\end{itemize}

\begin{figure}[htbp]
    \centering
    \includegraphics[width=0.4\textwidth]{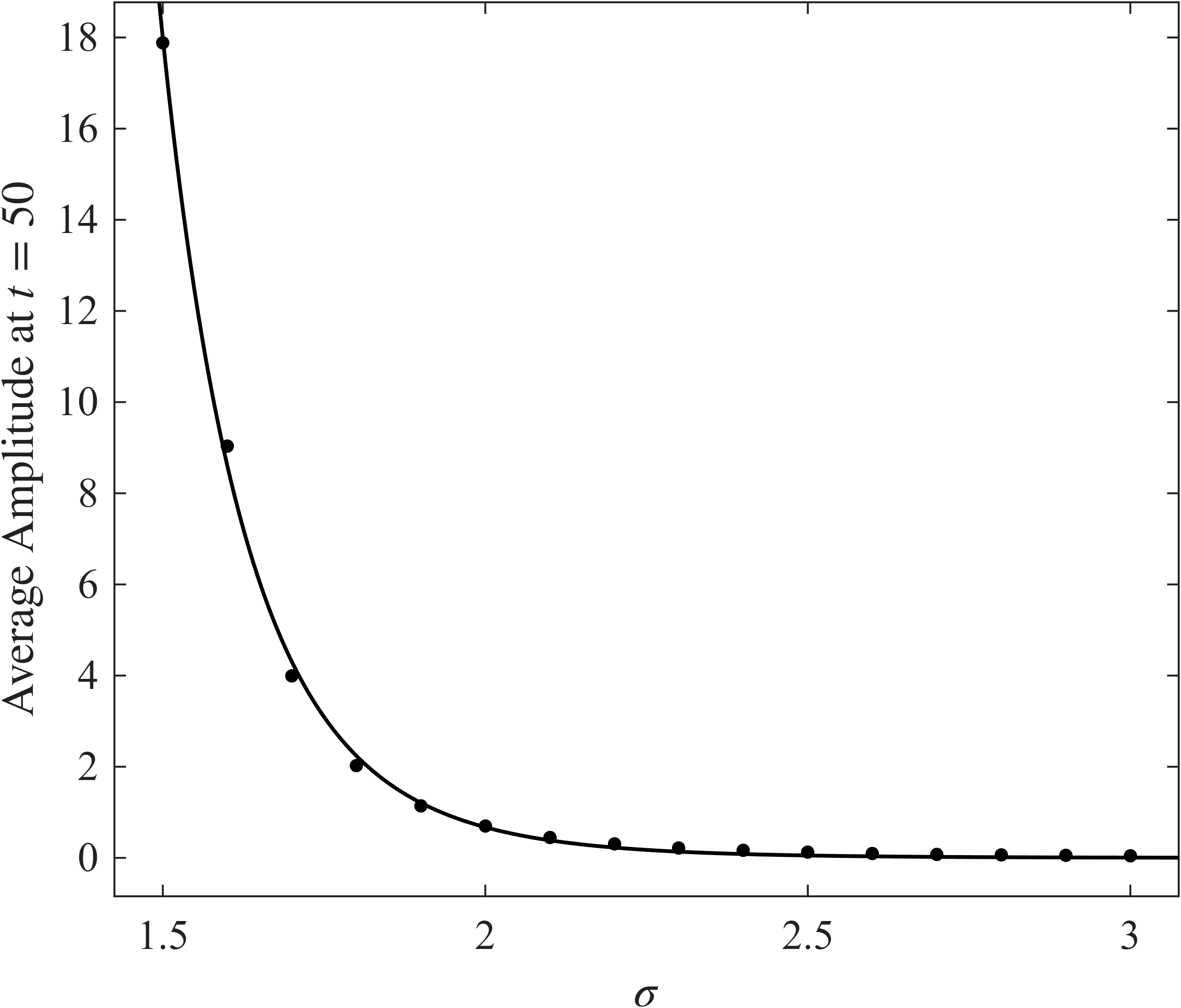}
    \caption{Amplitude at $t=50$ (average of $30$ trials) for varying $\sigma$ values on a LA4-LA4-LA4 network, based on Example~\ref{ex:1}. Fitted curve: $y=a\cdot x^b$ where $a=1847.5$ and $b=-11.4349$.}
    \label{fig:superinfect_sigma1}
\end{figure}

With the {\bf MBRD-SI} model in Equation~(\ref{eq:superinfect-prelim}), pathogen $2$ often dominates, inhibiting pattern formation. We observe through several scenarios that increasing $\sigma$ inhibits pattern growth; an example can be seen in \autoref{fig:superinfect_sigma1}, which is based on the parameter setting in Example~\ref{ex:1} and the superinfection model in Equation~(\ref{eq:superinfect-prelim}). 

We observe that the relationship between the average amplitude at $t=50$ can be described by a power curve. As a result, a greater $\sigma$ will make it less likely for the pattern to evolve back into the steady-state and system collapse to occur more quickly. Thus, in this case, a greater superinfection coefficient is most favorable in the long-term. 

\begin{figure}[htbp]
    \centering
    \includegraphics[width=0.4\textwidth]{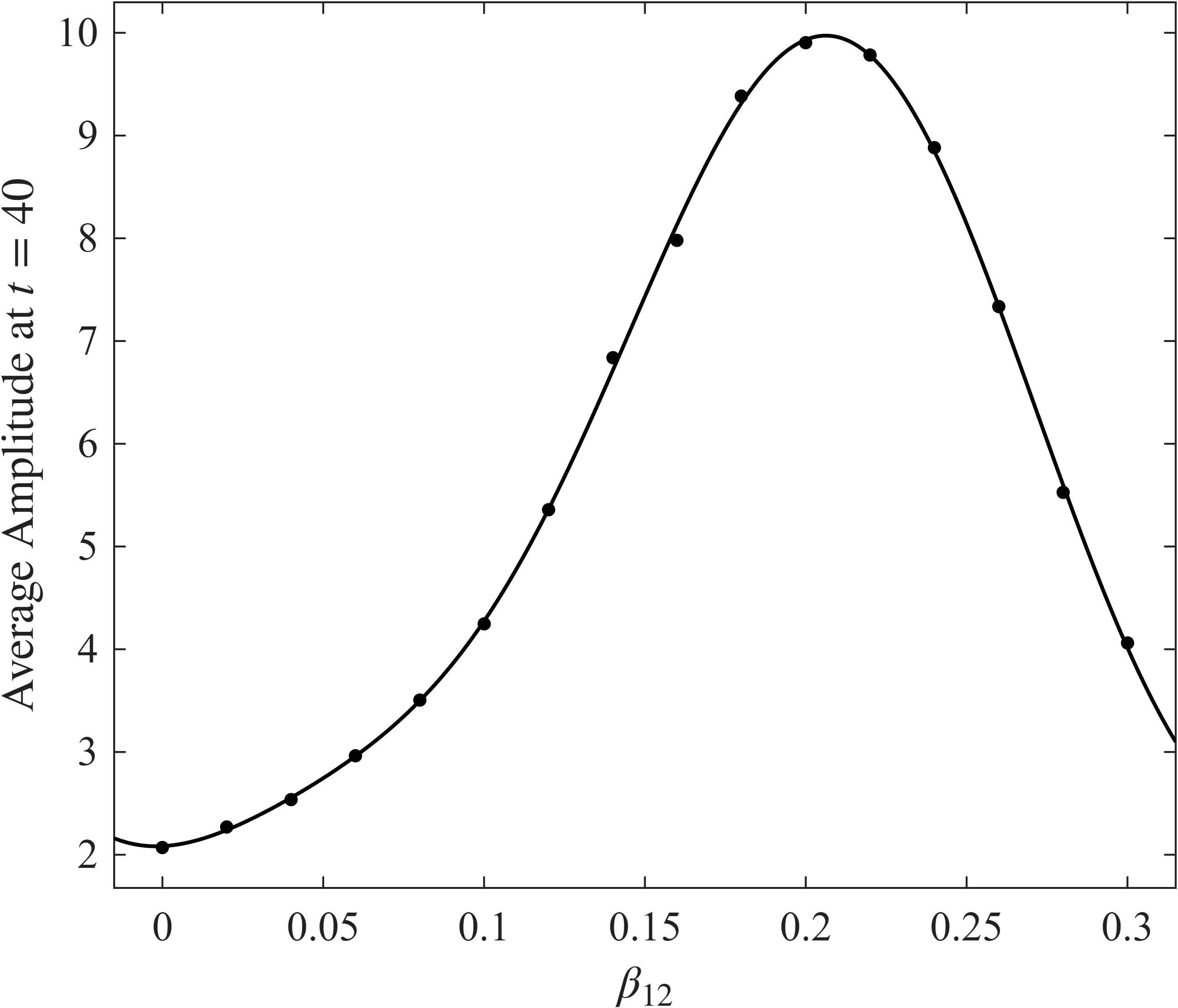}
    \caption{Amplitude at $t=40$ (average of $30$ trials) for varying $\beta_{12}$ values on a LA4-LA4-LA4 network, based on Example~\ref{ex:2}. Fitted curve:  $y=a_0+a_1\cos(xw)+b_1\sin(xw)+a_2\cos(2xw)+b_2\sin(2xw)$, where $a_0=5.4307$, $a_1=-3.6143$, $b_1=-1.3135$, $a_2=0.2676$, $0.7012$, $w=17.458$.}
    \label{fig:coinfect_b12}
\end{figure}

We consider the {\bf MBRD-CI} dynamics from the model in Equation~(\ref{eq:coinfect-prelim}). We note that in many scenarios, the majority of amplitude growth originates from fluctuations between nodes in mono-infections in co-infection dynamics. We use the configuration in Example~\ref{ex:2}, except for the co-transmission coefficient $\beta_{12}$, which we vary. From \autoref{fig:coinfect_b12}, we see that the average amplitude peaks approximately when $\beta_{12}=0.2$. We also fit the relationship to a Fourier series as described in the caption of  \autoref{fig:coinfect_b12}

This further analysis highlights the dual role of the co-transmission coefficient. Small values of $\beta_{12}$ are insufficient to sustain strong co-infection clusters, while very large values rapidly homogenize the system and suppress pattern formation. The intermediate regime (around $\beta_{12}=0.2$) maximizes oscillations, suggesting that there exists a critical threshold at which co-infections amplify spatial heterogeneity most strongly.  This is consistent with intuition from multi-strain epidemiology, where intermediate cross-immunity terms often reinforce chaotic and oscillatory behavior between competing strains~\cite{minayev2009improving}. We also note that our observations are also similar to prior results in oscillation quenching mechanisms~\cite{koseska2013oscillation} and animal synchrony~\cite{schaposnik2025modeling}, where moderate coupling parameters induce oscillations and large coupling parameters induce transitions to equilibria. In our setting, it implies that targeted interventions which reduce co-transmission could substantially weaken pattern growth and delay system collapse.

\begin{figure*}[htbp]
    \centering
    
    %\par\smallskip
    \makebox[\textwidth]{%
        \subfigure[]{%
            \includegraphics[width=0.32\textwidth]{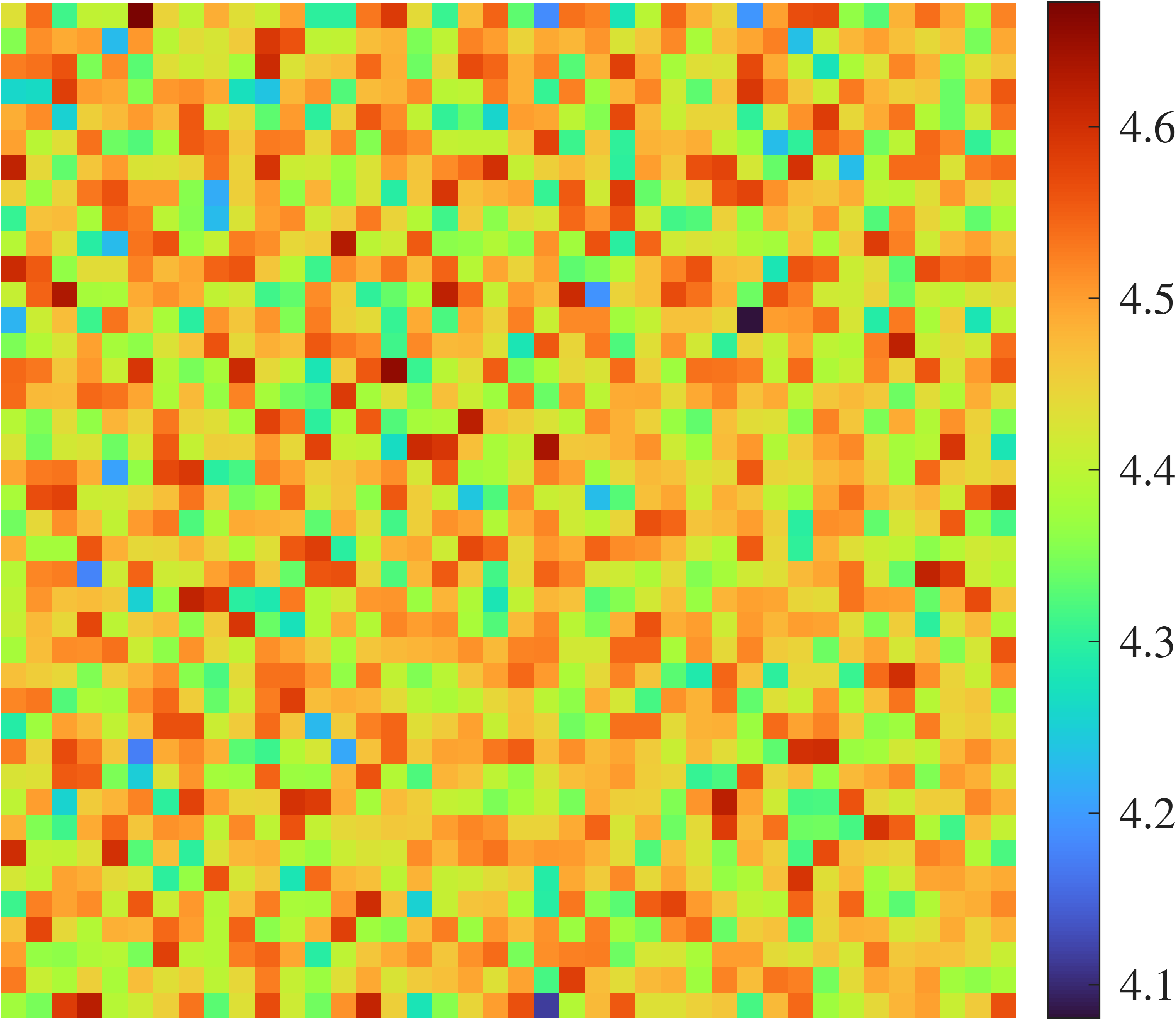}
            \label{}
        }
        % \hspace{0.04\textwidth}
        \subfigure[]{%
            \includegraphics[width=0.325\textwidth]{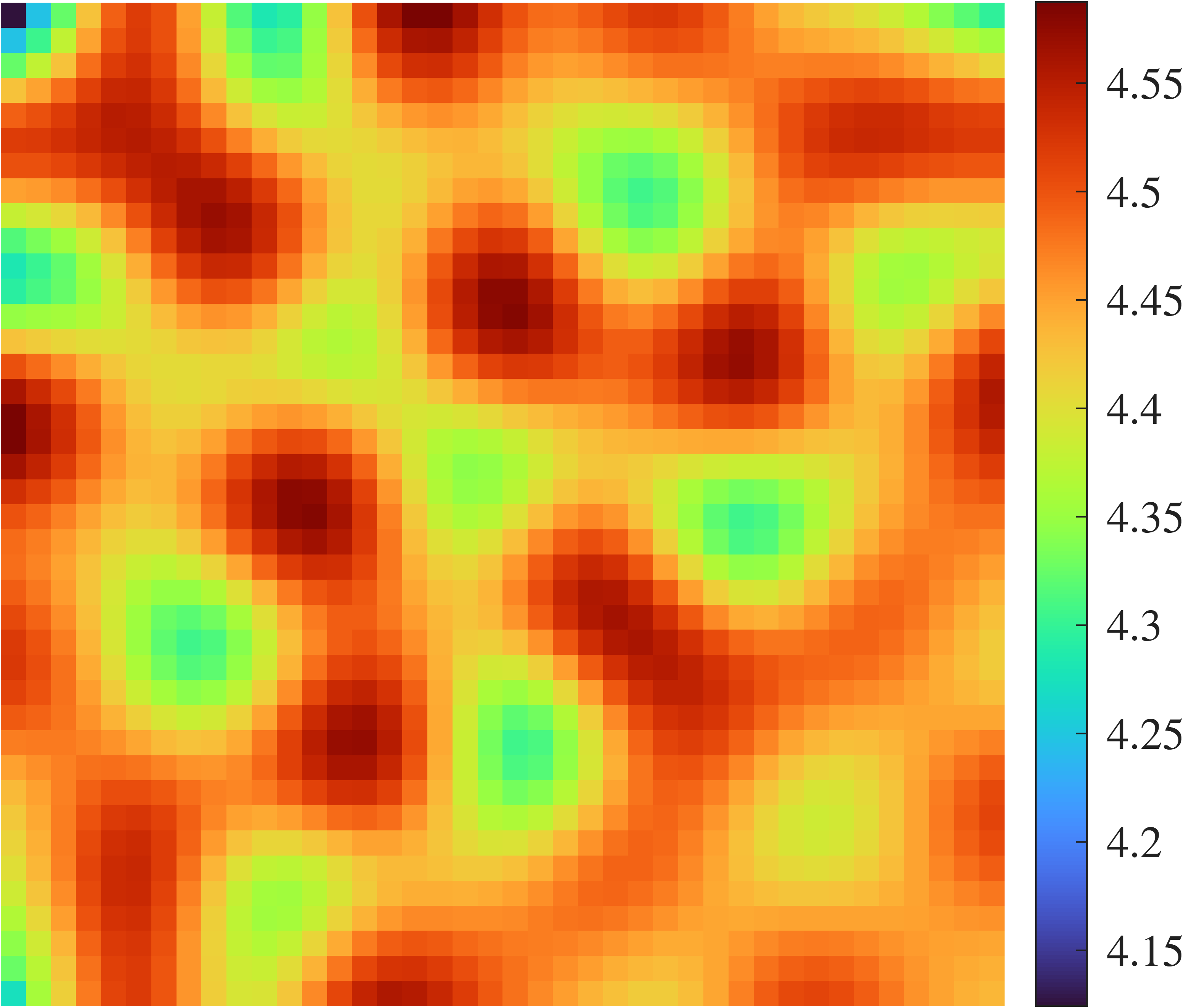}
            \label{}
        }
        % \hspace{0.04\textwidth}
        \subfigure[]{%
            \includegraphics[width=0.32\textwidth]{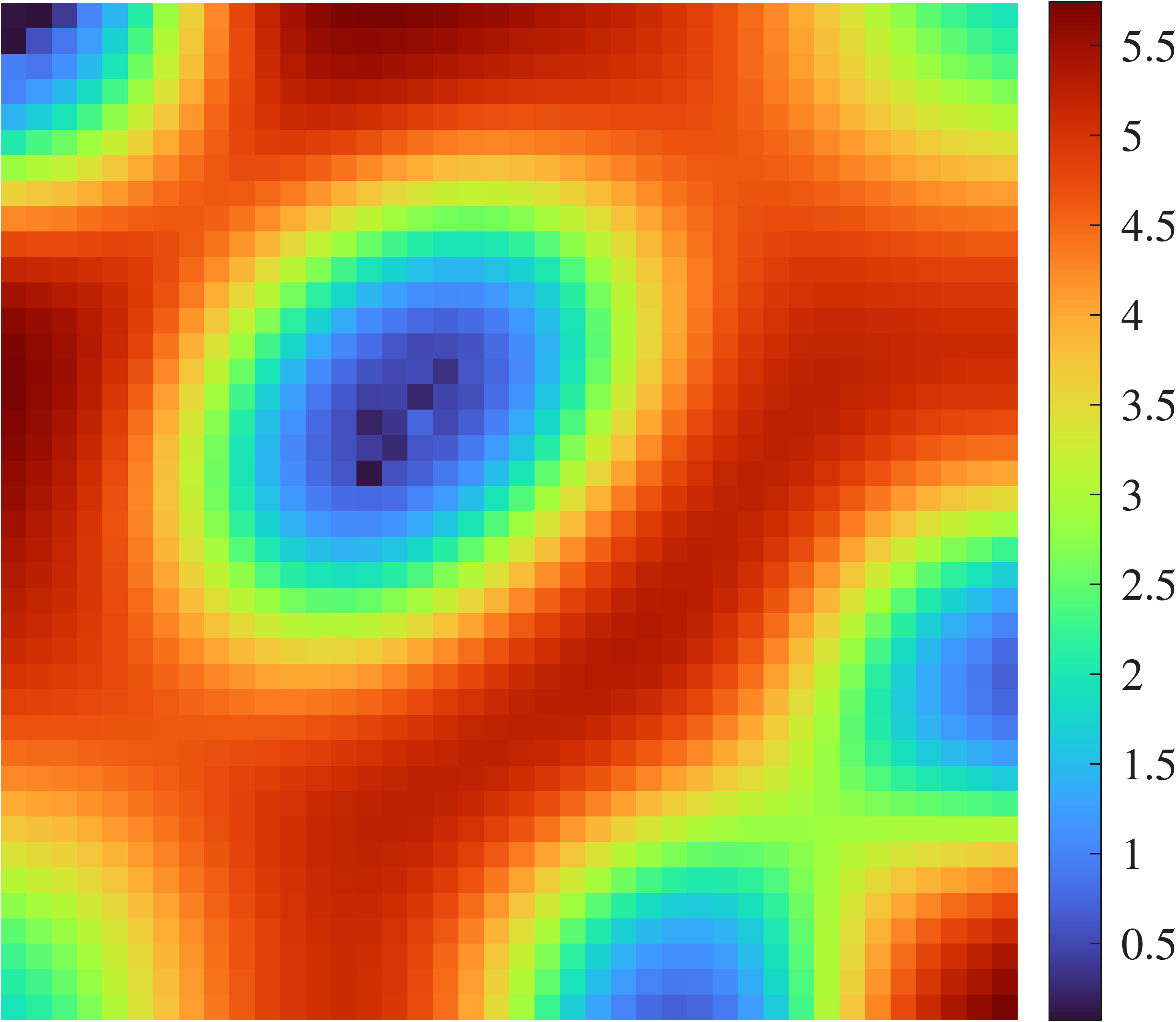}
            \label{}
        }
    }
    \caption{Layers with $I$ densities for three multiplex networks: LA14-LA4-LA4 (left, $t=450$), LA12-LA12-LA12 (center, $t=550$), and LA24-LA24-LA24 (right, $t=1000$).}
    \label{fig:coinfect_deg}
\end{figure*}

\subsection{Varying Network Degrees}\label{sec:simu-deg}

We shall dedicate this section to analyzing the effect of varying average degrees between layers on pattern formation and growth. We mainly analyze this from the standpoint of clustering, which influences how difficult a disease is to mitigate.

We analyzed the coinfection pattern with $8$ combinations of varying networks: (LA4, LA4, LA4), (LA12, LA12, LA4), (LA12, LA4, LA4), (LA12, LA4, LA12), (LA12, LA12, LA12), (LA24, LA12, LA4), (LA24, LA4, LA12), (LA24, LA24, LA24). \autoref{fig:coinfect_deg} presents patterns in the $I$ layer (first and second rows) for varying degree layers at time $700$ for the {\bf MBRD-CI} dynamics in Equation~(\ref{eq:coinfect-prelim}) and Example~\ref{ex:2}, for the other three pattern-forming layer combinations besides (LA12, LA12, LA4). From left to right, the figures depict pattern formation on (LA4, LA4, LA4), (LA12, LA12, LA12), and (LA24, LA24, LA24) networks. We make the following observations and illustrate them with \autoref{fig:coinfect_deg}.

\begin{itemize}
    \item For the MBRD-SI and MBRD-CI models with the parameter settings in Example~\ref{ex:1} and~\ref{ex:2}, respectively, pattern formation occurs in four combinations: (LA4, LA4, LA4), (LA12, LA12, LA4), (LA12, LA12, LA12), and (LA24, LA24, LA24). Thus, minimal variations between average layer degrees may induce pattern formation. Because pattern formation leads to infection hotspots that are difficult to mitigate, it is ideal for variation to occur between layers. Longer quarantine periods and limits on migration for infected populations would induce large average degree variations between the layers, preventing Turing patterns from forming.
    \item In single-pathogen dynamics, Zhao \emph{et al.} emphasized that increasing layerwise average degrees can lead to larger clusters~\cite{zhao2025navigating}. This also occurs with our MBRD-SI and MBRD-CI models and is illustrated in \autoref{fig:coinfect_deg}, where the clusters for (LA24, LA24, LA24) are the largest and clustering rarely occurs for the (LA4, LA4, LA4) configuration. Meanwhile, we also observe that larger average degrees overall also lead to slower amplitude growth. Thus, lower average degrees and minimal clustering may make migitation easier; however, hospitals and governments will have more time to prepare with overall higher average degrees.
\end{itemize}

\section{Point-Source Infections}\label{sec:point_source}

Many epidemics, such as the COVID pandemic, originate in a single region. In this section, we explore superinfection and co-infection dynamics after both are introduced at different nodes. In particular, we aim to understand the effects of the infection source locations, time difference of pathogen introductions, and the superinfection and co-infection parameters on infection spread for both pathogens. Finally, we consider the impact of different network topologies and varying average layer degrees on two-pathogen interactions.

\subsection{Methodology}

We let $I_0$ and $J_0$ be the initialized source nodes for pathogen $1$- and pathogen $2$-infections, respectively. We introduce the following terminology and metrics:
\begin{definition}
    We call a node $I_1$-\emph{active} (resp. $I_2$-\emph{active}) if its density of $I_1$-infections (resp. $I_2$-infections), including only mono-infections, is greater than $I_{thres}$ (resp. $J_{thres}$). We call a node $C$\emph{-active} if the density of co-infections is greater than $C_{thres}$.
\end{definition} 
\begin{definition}
    We define the $I_1$-\emph{spread index} (resp. $I_2$-\emph{spread index}) to be the fraction of nodes that are $I_1$-active (resp. $I_2$-active). Additionally, we define the $C$\emph{-spread index} to be the fraction of nodes that are $C$-active.
\end{definition}

For our analysis, we set $I_0=J_0=0.05$, and $I_{thres}=J_{thres}=0.01$ for both superinfection and co-infection simulations. We also define $C_{thres}=0.005$ for our co-infection analysis. We also introduce the following definitions, which will be useful in Subsection~\ref{sec:layers}.

\begin{definition}[Peak]
    Let $\delta(t)$ be the $I_1$-, $I_2$- or $C$-spread index as a function of the time $t$. For some integer time increment $\tau$, we define a \emph{peak time} $t^*>0$ to be integer multiple of $\tau$ that satisfies
    \begin{align*}
        \delta(t^*-\tau)&<\delta(t^*),
    \end{align*}
    under the condition that there exists an integer $u$ such that $u\equiv 0\bmod \tau$, $u\ge t^*$, $\delta(u)>\delta(u+\tau)$, and
    \begin{align*}
        \delta(t)=\delta(t^*) \mathrel{\forall}\, t\in \{(t^*,u]\mid t\equiv 0\bmod n\}.
    \end{align*}
    Additionally, for a \emph{peak time} $t^*$, we define the corresponding \emph{peak value} to be $\delta(t^*)$.
\end{definition}

\begin{definition}[Saturation time]
For time-increment $\tau$ and $\delta(t)$ as a spread index as a function of $t$, we define the \emph{saturation time} of a spread index to be an smallest integer multiple of $\tau$, $t^*$, such that $\delta(t^*)=1$.
\end{definition}

In our simulations, we use a time increment of $\tau=1$. Using a larger $\tau$ would provide smoother spread-index graphs. Finally, to illustrate our observations in the following subsections, we introduce these  superinfection parameters for Equation~(\ref{eq:superinfect-prelim}), and co-infection parameters for Equation~(\ref{eq:coinfect-prelim}).

\begin{example} \label{ex:4} (Superinfection model) 
\begin{equation}\label{eq:p4-settings}
\begin{aligned}
\mu &= 0.005,\quad &r&=0.1,\quad & A&=0.1,\quad & K&=1\\
\beta_1 &=0.5, \quad & \beta_2&=0.4,\quad & \sigma &=0.9,\\
\gamma_1&=0.2,\quad &\gamma_2&=0.1,\quad & \alpha_1&=0.01,\quad & \alpha_2&=0.05,\\
d_{11}&=0.3,\quad & d_{12}&=0.1,\quad & d_{13}&=0.1,\\
d_{22}&=0.3,\quad & d_{33}&=0.1.
\end{aligned}
\end{equation}
\end{example}

\pagebreak
\begin{example}\label{ex:5} (Co-infection model)\begin{small}
\begin{equation}\label{eq:p5-settings}
\begin{aligned}
\mu &= 0.005,\quad & r&=0.1,&\quad A&=0.1,\quad& K&=1,\\
\beta_1&=0.3,\quad & \beta_2&= 0.4,\\ \beta_{10}&=0.2,\quad & \beta_{02}&=0.3,\quad & \beta_{12}&=0.05,\\
\gamma_1&=0.1,\quad & \gamma_2&=0.05,\\ \alpha_1&=0.05,\quad & \alpha_2&=0.15,\quad & \alpha_{12}&=0.25,\\
d_{11}&=0.3,\quad & d_{12}&=0.1,\quad & d_{13}&=0.1,\\
d_{22}&=0.3,\quad & d_{33}&=0.1.
\end{aligned}
\end{equation}\end{small}
\end{example}

In what follows we shall study different aspects of the setting above:  we discuss the overall shape of the spread index evolutions in \ref{sec:nature},  we evaluate the impact of the distance and time difference, respectively, between the initialization of pathogen $1$ and $2$ in \ref{sec:source_loc} and~\ref{sec:time_diff};  we discuss the impact of superinfection or co-infection model-specific parameters  in \ref{sec:model_param} and  finally, in  \ref{sec:layers} we evaluate the impact of different network types and average layer degrees on the spread of infections.

\subsection{Nature of Infection Dynamics}
\label{sec:nature}

Consider the superinfection model in Equation~(\ref{eq:superinfect-prelim}). With the MBRD-SI model, we expect, in many scenarios, the eventual persistence of pathogen $2$ infections in every region and pathogen $1$ infections to die out as a result of pathogen $2$'s ability to steal hosts. We see that this is true for the parameter set in Example~\ref{ex:4} with the top row in \autoref{fig:superinfect_center}. We observe that in superinfection dynamics, the $I_1$-spread index often experiences a single peak and falls back to $0$. Meanwhile, the $I_2$-spread index eventually becomes $1$, as shown in Figures~\ref{fig:superinfect_center} and~\ref{fig:superinfect_corner}. 

In the co-infection dynamics shown in Equation~(\ref{eq:coinfect-prelim}), it is less likely for one pathogen to completely dominate the other. Thus, we expect that in some scenarios, both the $I_1$-, $I_2$-, and $C$-spread indexes will reach $1$ over time. We see in \autoref{fig:coinfect_center} that it is possible for that to occur. Moreover, it is possible for one pathogen to dominate over the other, causing one mono-spread index to peak and fall back to $0$ and the other mono-spread index to persist at $1$ as shown in \autoref{fig:coinfect_beta12} when $\beta_{12}=0.01$.

\subsection{Effect of Source Locations}
\label{sec:source_loc}

\begin{figure*}[htbp]
    \centering

    \makebox[\textwidth]{%
        \subfigure[]{%
            \includegraphics[width=0.29\textwidth]{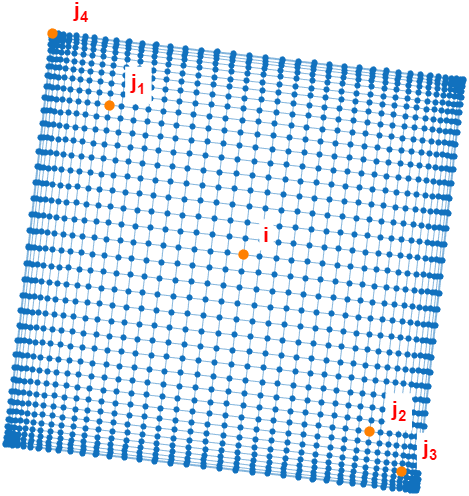}
            \label{fig:center_graph}
        }
        % \hspace{0.04\textwidth}
        \subfigure[]{%
            \includegraphics[width=0.36\textwidth]{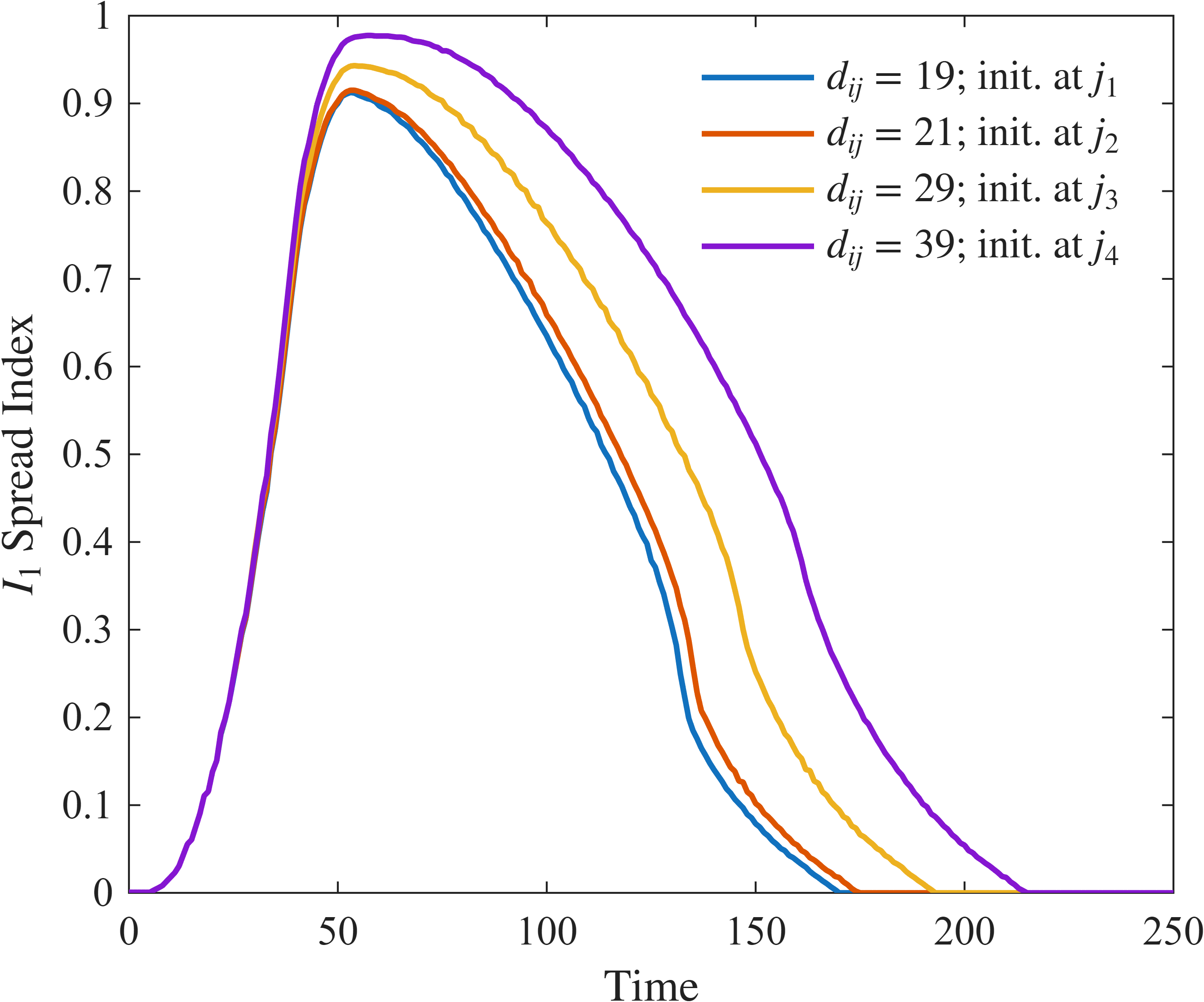}
            \label{fig:superinfect_center_I}
        }
        % \hspace{0.04\textwidth}
        \subfigure[]{%
            \includegraphics[width=0.36\textwidth]{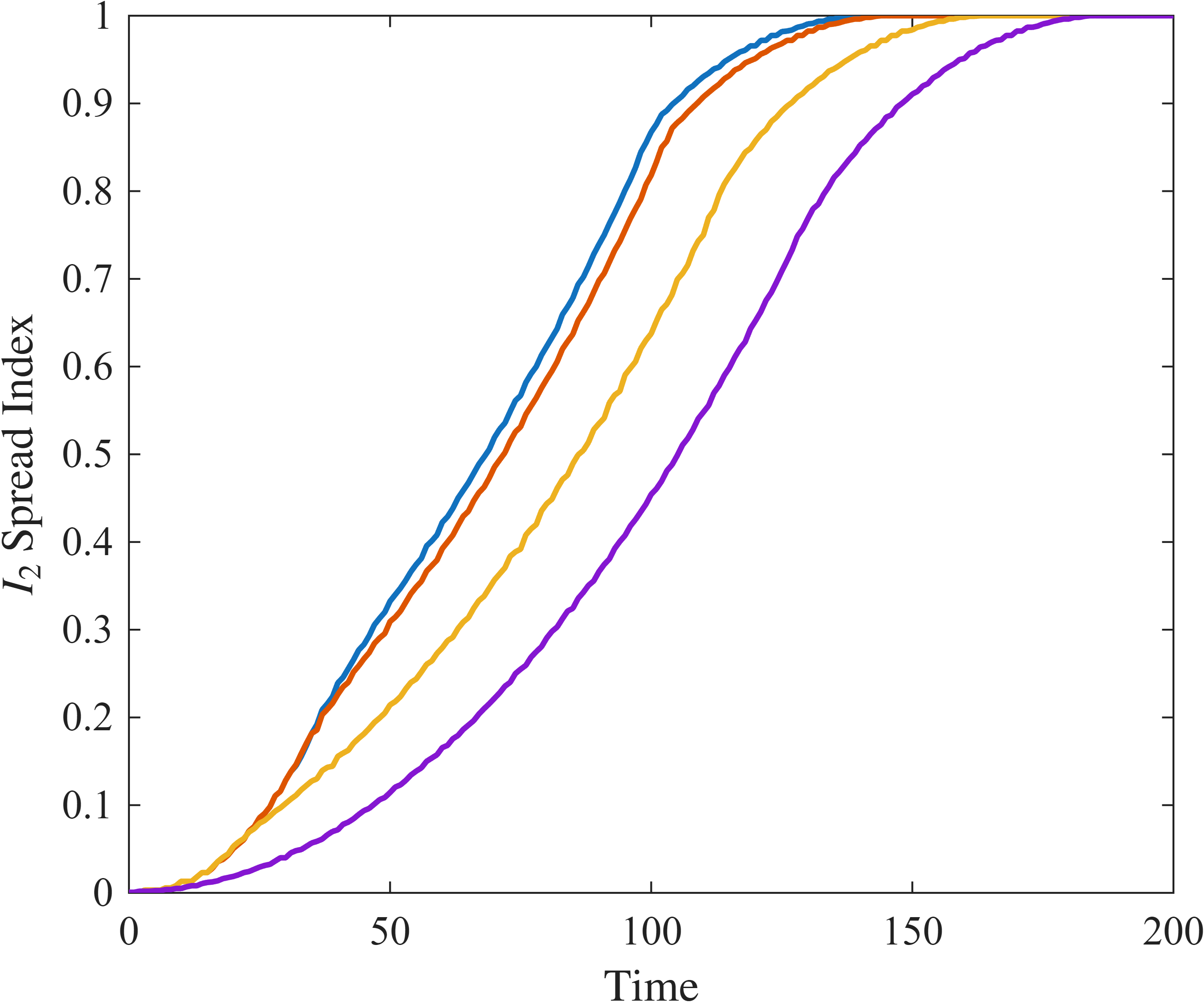}
            \label{fig:superinfect_center_J}
        }
    }
    \caption{$I_1$-spread index (center) and $I_2$-spread index (right) when $I$ is introduced in a central location and the starting location of $J$ varies (left). Based on Example~\ref{ex:4}.}
    \label{fig:superinfect_center}
\end{figure*}

We first focus on the effect that source locations for both pathogens have on the spread of both pathogens throughout networks. For simplicity, we perform our simulations with $40\times 40$ lattice networks. We denote $d_{ij}$ to be the starting path distance between the sources of pathogen $1$ and pathogen $2$.
% and $t_d$ to be the time difference in the introductions of pathogen $1$ and pathogen $2$ in the network.
With respect to location, we focus on two configurations. The first, as seen in Figure~\autoref{fig:center_graph}, features pathogen $1$ originating from the center of the network with initial density $0.05$. At the same time $t=0$, we let pathogen $2$ originate at $j_1$, $j_2$, $j_3$, or $j_4$, which are ordered based on distance from node $i$.  We analyze this scenario for both superinfection dynamics (see \autoref{fig:superinfect_center}) and co-infection dynamics (see \autoref{fig:coinfect_center}). 

\begin{figure*}
    %\par\smallskip
    
    \makebox[\textwidth]{%
        \subfigure[]{%
            \includegraphics[width=0.29\textwidth]{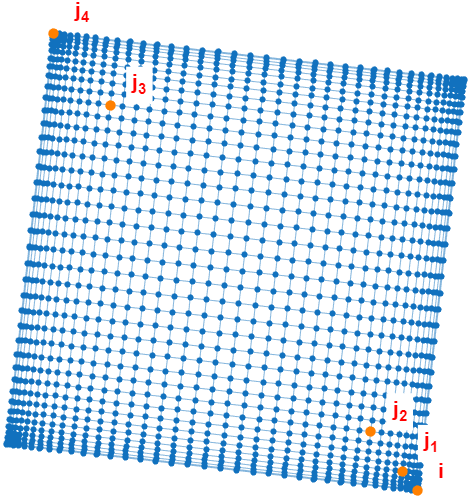}
            \label{fig:corner_graph}
        }
        % \hspace{0.04\textwidth}
        \subfigure[]{%
            \includegraphics[width=0.35\textwidth]{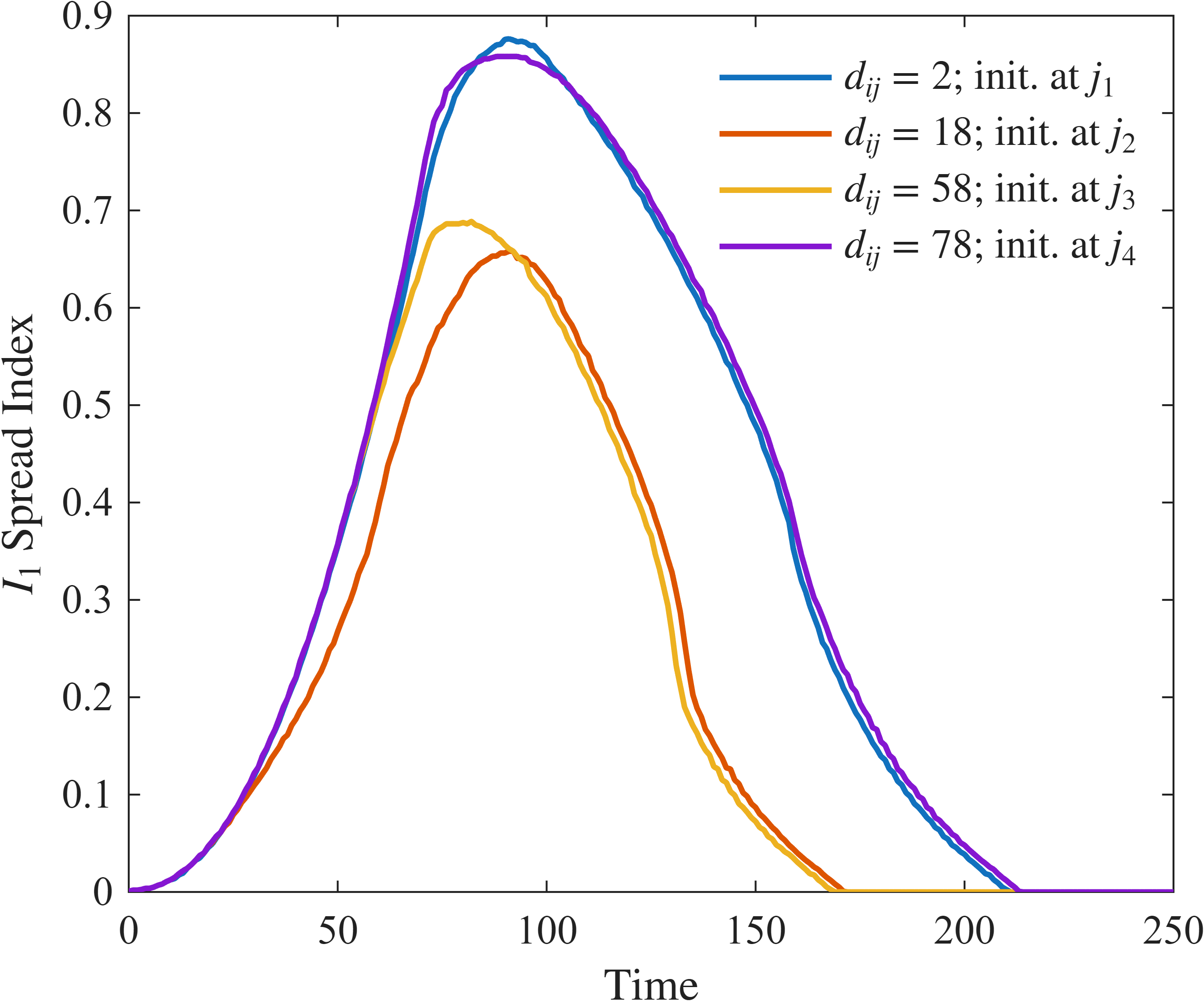}
            \label{fig:superinfect_corner_I}
        }
        % \hspace{0.04\textwidth}
        \subfigure[]{%
            \includegraphics[width=0.35\textwidth]{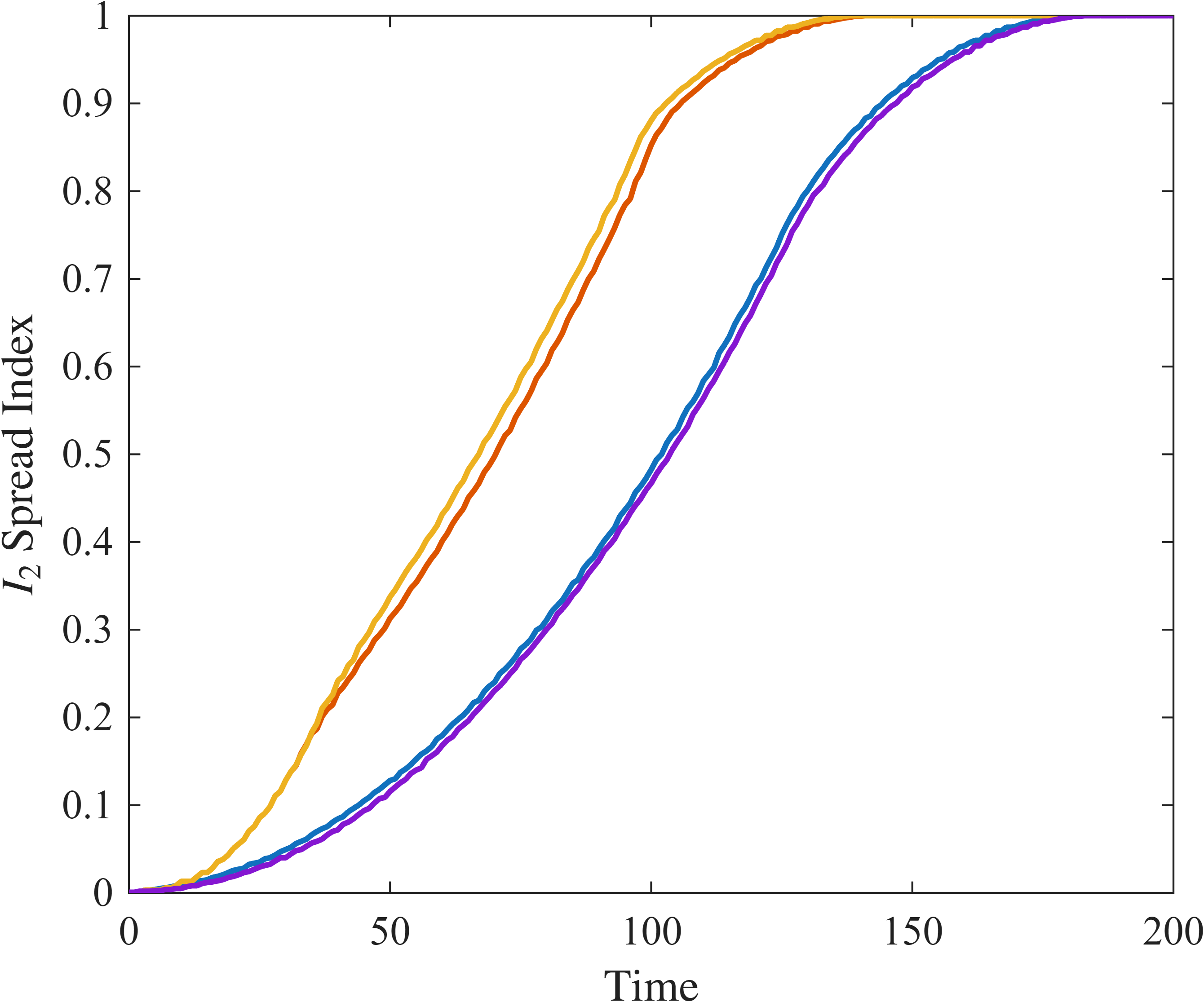}
            \label{fig:superinfect_corner_J}
        }
    }
    \caption{$I_1$-spread index (center) and $I_2$-spread index (right) when pathogen $1$ is introduced in a central location and the starting location of pathogen $2$ varies (left). Based on parameter configuration in Example~\ref{ex:4}.}
    \label{fig:superinfect_corner}
\end{figure*}

\begin{figure*}[htbp]
    \centering

    \makebox[\textwidth]{%
        \subfigure[]{%
            \includegraphics[width=0.33\textwidth]{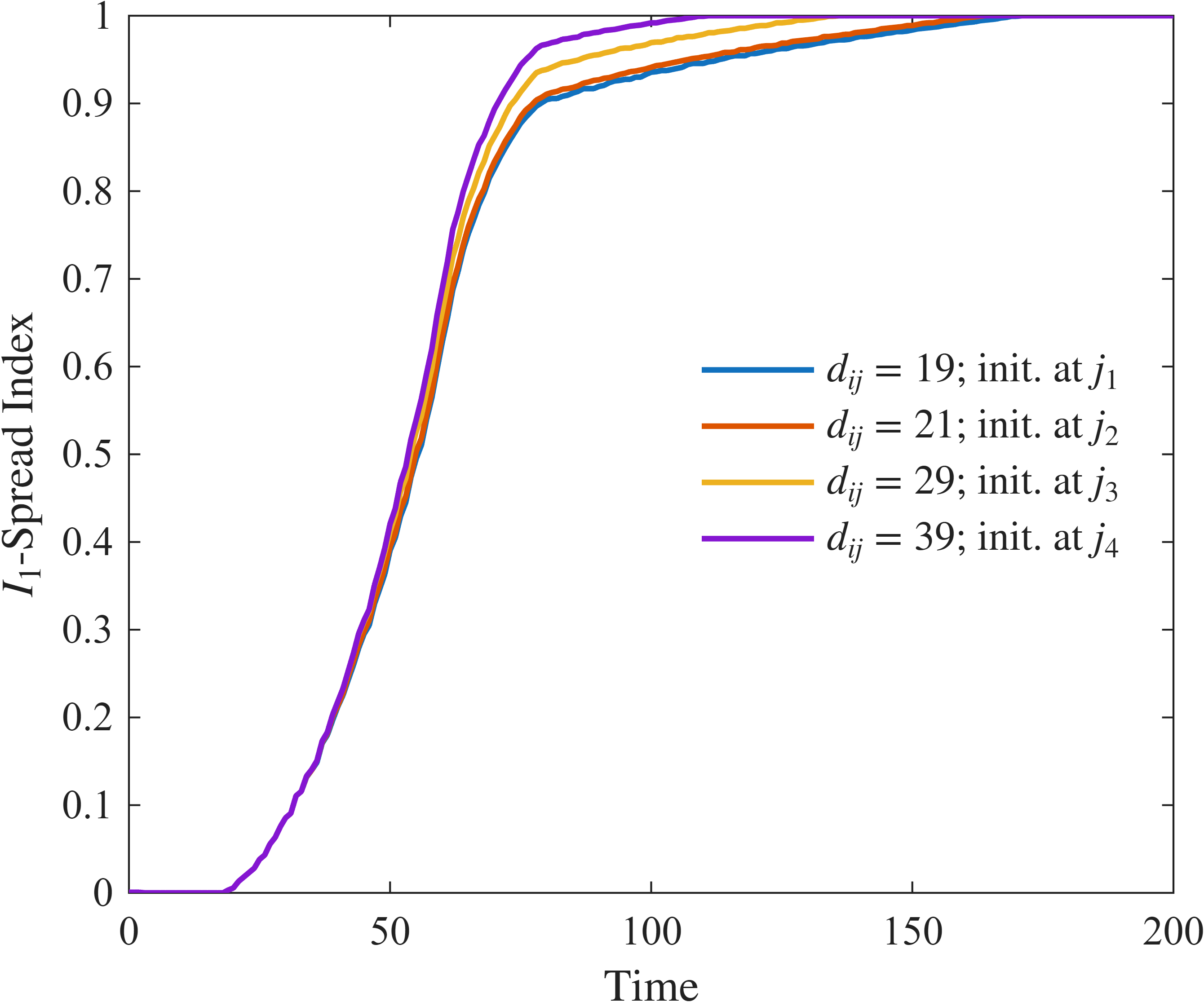}
            \label{coinfect_center_I}
        }
        \subfigure[]{%
            \includegraphics[width=0.33\textwidth]{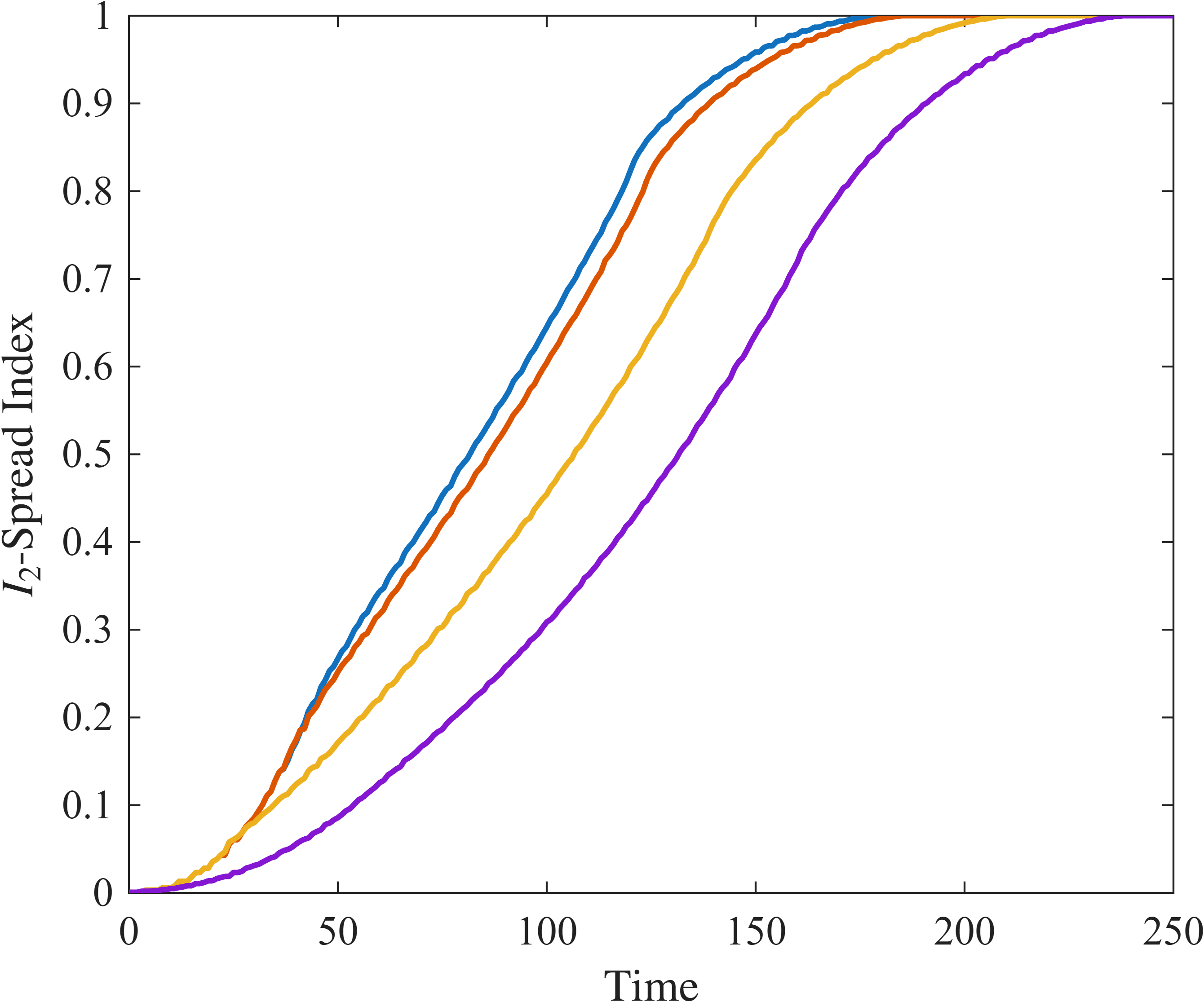}
            \label{coinfect_center_J}
        }
        \subfigure[]{%
            \includegraphics[width=0.33\textwidth]{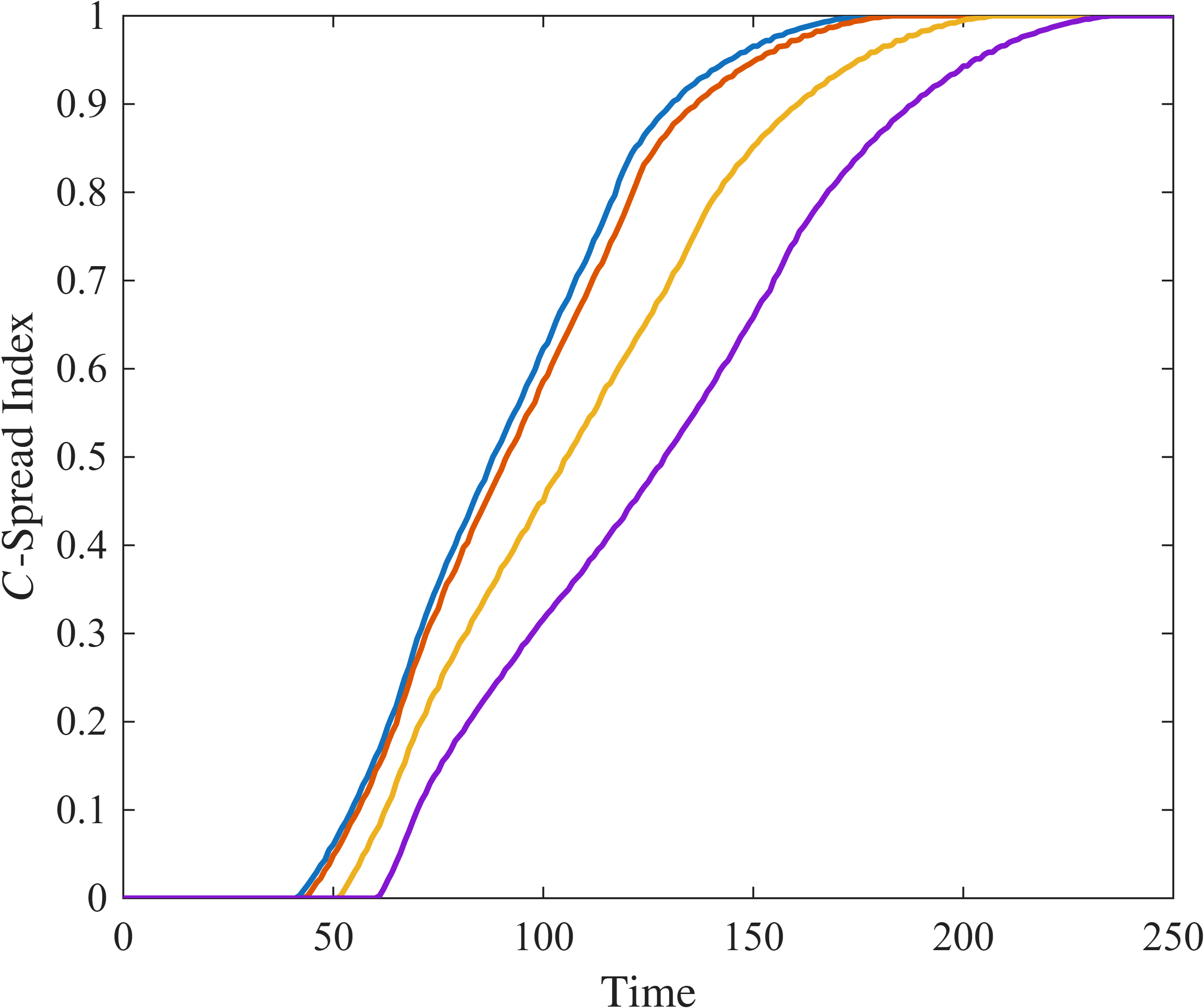}
            \label{coinfect_center_C}
        }
    }
    \caption{$I_1$-, $I_2$-, and $C$-spread indexes for both pathogens over time when pathogen $1$ is introduced in a central location and the starting location of pathogen $2$ varies. Based on \autoref{fig:center_graph} and Example~\ref{ex:5}.}
    \label{fig:coinfect_center}
\end{figure*}

From Example~\ref{ex:4} and the MBRD-SI model in Eq.~(\ref{eq:superinfect-prelim}), we expect that for both pathogen dynamics, the weaker pathogen's spread index peaks at a greater value or the time for that pathogen's spread index to reach $1$ decreases as the distance between the source locations of pathogens $1$ and $2$ increases. This is because pathogen $1$ would have more time to dominate a large area before both pathogens start to compete in the same locations. We observe from both figures that this is indeed true. However, in this specific configuration, we note interestingly that the initial distance between the pathogen sources has a minimal impact on the time at which the $I$-spread index peaks in the superinfection dynamics in Figure~    \autoref{fig:superinfect_center_I}. 

\begin{figure*}[htbp]
    \centering

    \makebox[\textwidth]{%
        \subfigure[]{%
            \includegraphics[width=0.45\textwidth]{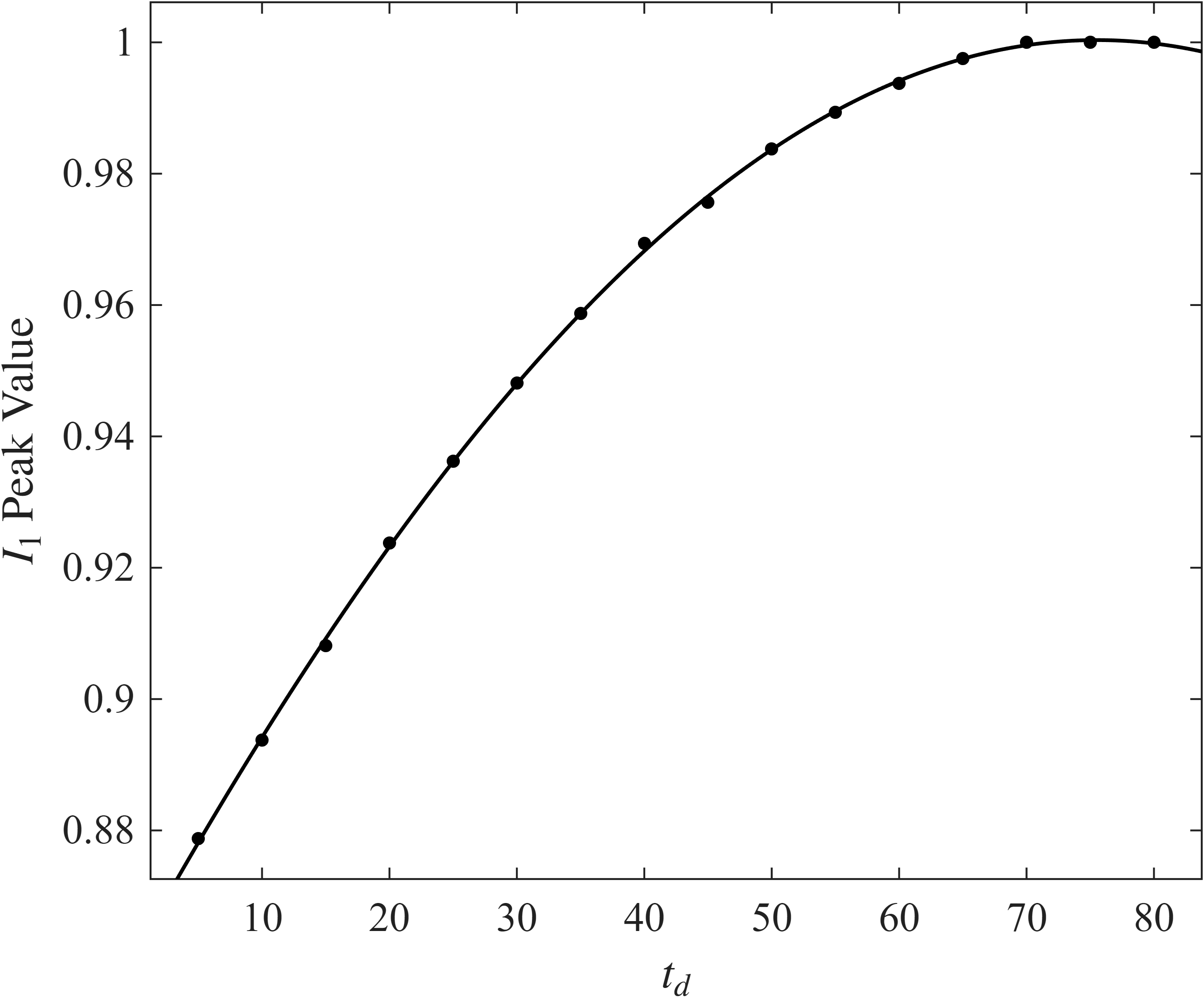}
            \label{fig:superinfect_time_I1}
        }
        \subfigure[]{%
            \includegraphics[width=0.45\textwidth]{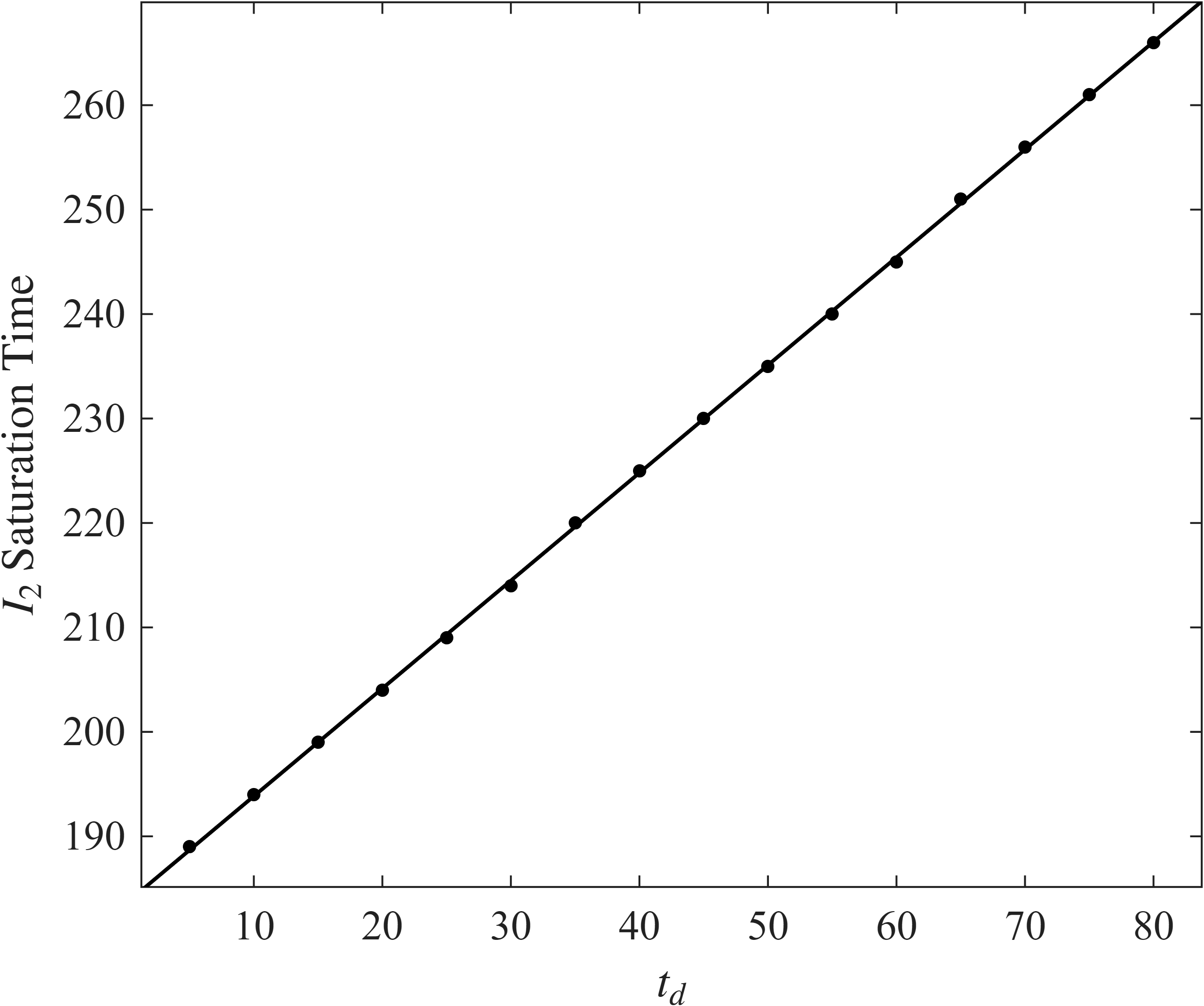}
            \label{fig:superinfect_time_I2}
        }
    }
    \caption{$I_1$-peak value (left) and $I_2$-saturation time (right) as $t_d$ varies. Left fitted curve: $y= a_0 + a_1\cos(xw) + b_1\sin(xw)$, where $a_0=0.548$, $a_1=0.3131$, $b_1=0.3264$, and $w=0.0107$. Right fitted curve: $y=ax+b$, where $a=1.0318$, $b=183.5250$.}
    \label{fig:superinfect_time}
\end{figure*}

\begin{figure*}[htbp]
    \centering

    \makebox[\textwidth]{%
        \subfigure[]{%
            \includegraphics[width=0.33\textwidth]{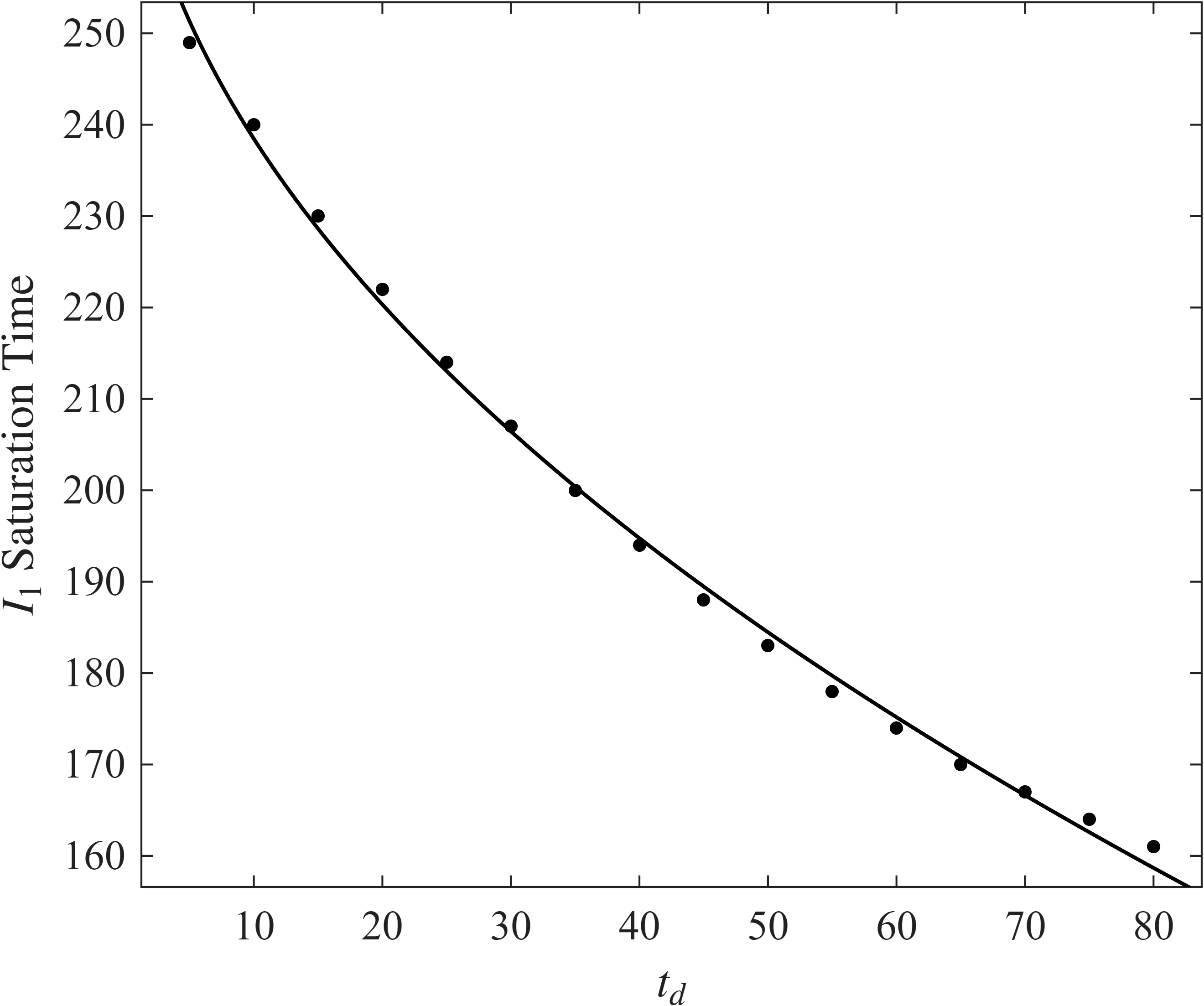}
            \label{}
        }
        \subfigure[]{%
            \includegraphics[width=0.33\textwidth]{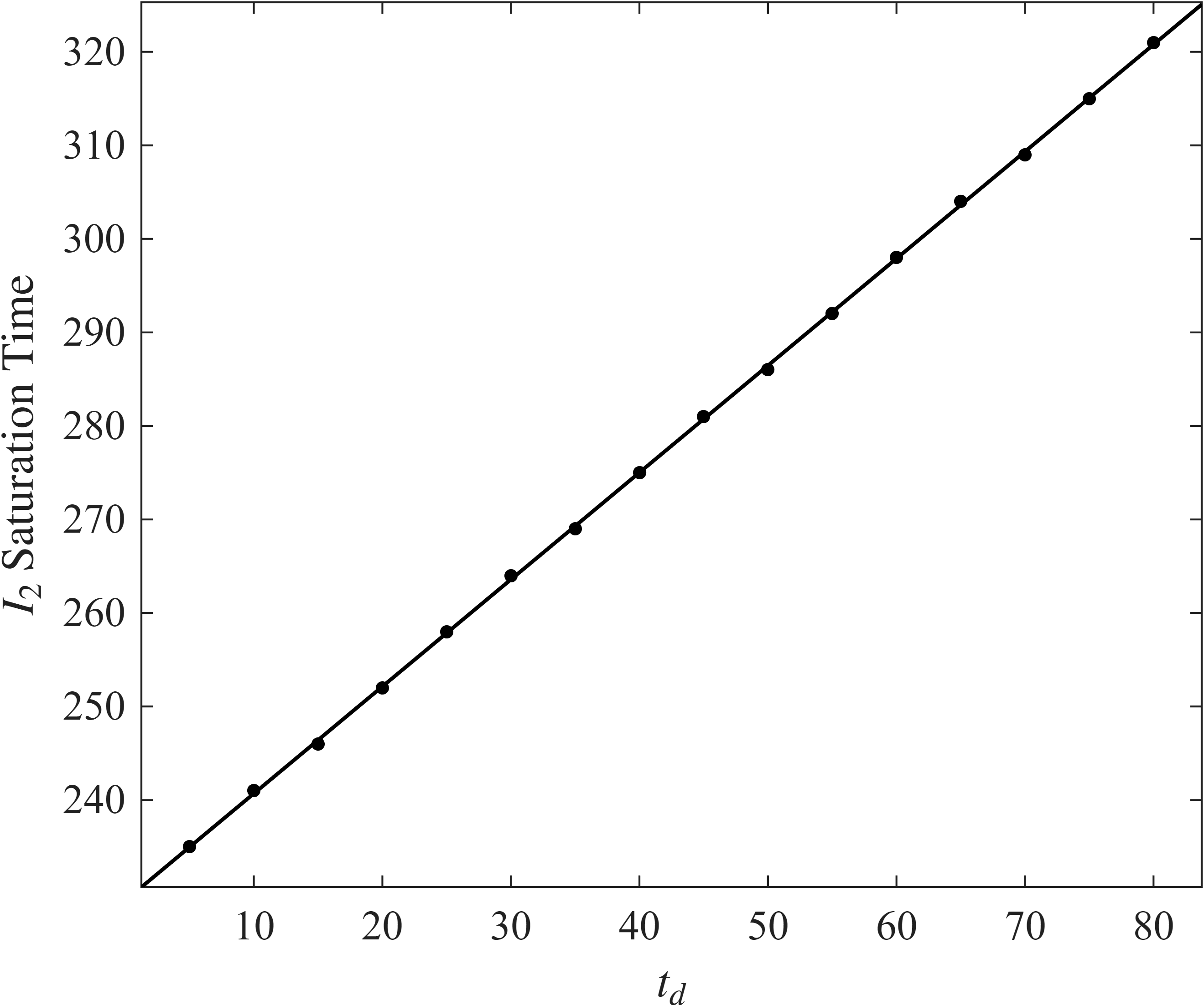}
            \label{}
        }
        \subfigure[]{%
            \includegraphics[width=0.33\textwidth]{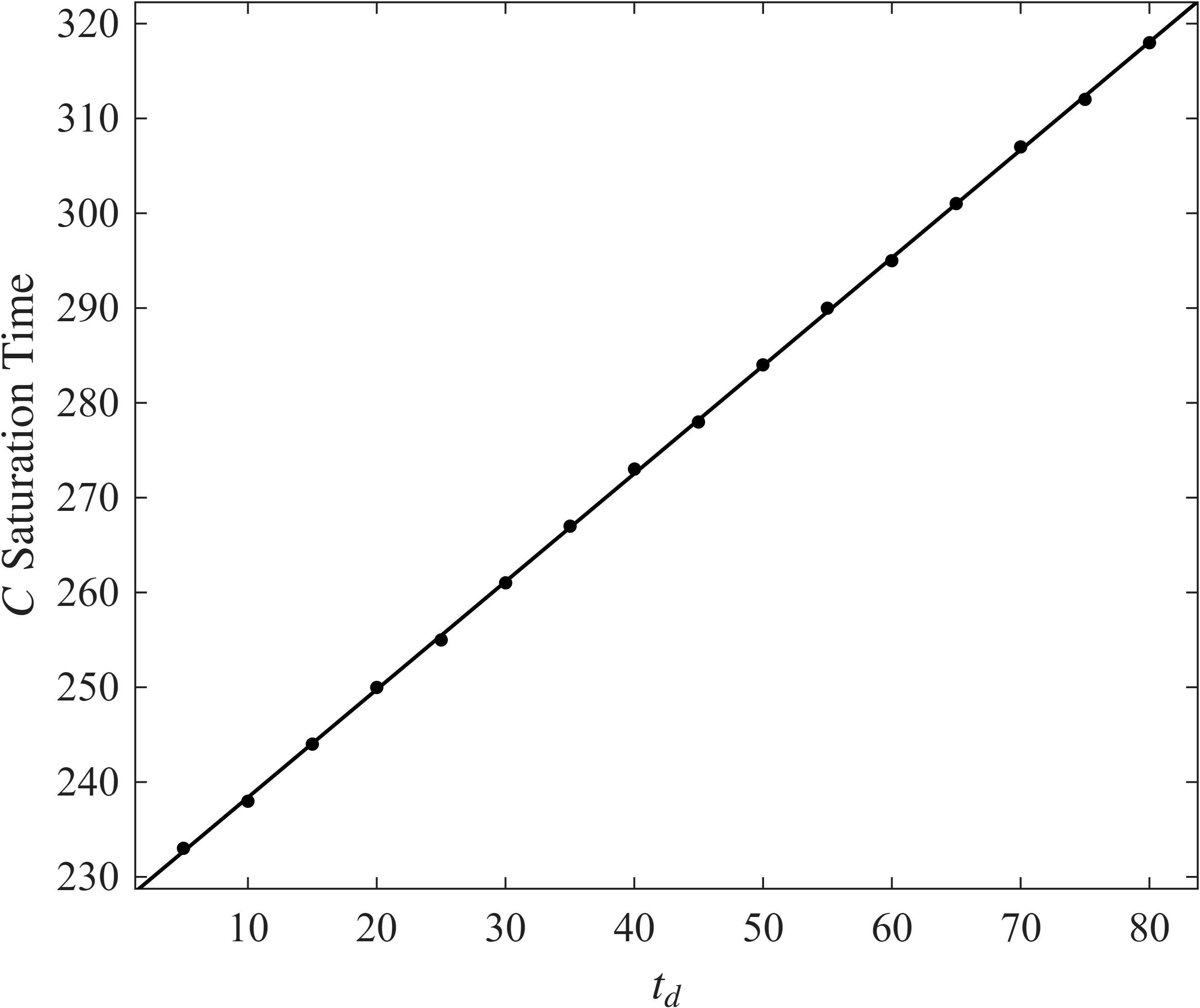}
            \label{}
        }
    }
    \caption{Saturation times of $I_1$ (left), $I_2$ (center), and $C$ (right) with varying $t_d$. Based on Example~\ref{ex:5}, $I$ is introduced at node $1$, and $J$ is introduced at node $1600$. Left fit curve: $y = ax^b+c$, where $a=-14.071$, $b=0.4964$, and $c=282.6$. Center fit curve: $y=ax+b$, where $a=1.1441$ and $b=229.25$. Right fit curve: $y=ax+b$, where $1.1382$ and $b=227$.}
    \label{fig:time_coinfect}
\end{figure*}

For the same initialization configuration but with the MBRD-CI model in Equation~(\ref{eq:coinfect-prelim}), we first note that as $d_{ij}$ increases, either the $I_1$-spread index or $I_2$-spread index reaches $1$ earlier while the other reaches $1$ later. For Example~\ref{ex:5}, pathogen $2$ has higher transmission rates than pathogen $1$. This may explain why $I_1$-spread index reaches $1$ at an earlier time and the $I_2$-spread index reaches $1$ at a later time as $d_{ij}$ increases; since pathogen $1$ is weaker, the $I_1$-spread index has more room to grow before the pathogens start interacting when the sources are far from each other. We also observe that the co-spread index reaches $1$ at a later time as $d_{ij}$ increases, mirroring the dynamics of the $I_2$-spread index.

The second initialization configuration, pictured in Figure~\autoref{fig:corner_graph} for the superinfection dynamics in Equation~(\ref{eq:superinfect-prelim}) with Example~\ref{ex:4}, features pathogen $1$-infections originating from a corner of the lattice, while the origin of pathogen $2$-infections varies. We observe that the source location for pathogen $2$ has a more irregular impact on the $I_1$-spread index's peak. The largest peaks in the $I_1$-spread index occur when pathogen $2$ is initialized at $j_1$ and $j_4$. This can be explained by the overall direction in movement of both infection spreads. When pathogen $2$ is initialized at $j_4$, the $I_1$-spread index has more room to grow before both pathogen infections start diffusing at the same locations. On the other hand, when the origins of pathogen $1$ and $2$ are close, pathogen $1$ more easily catches up to the places that pathogen $2$ is infecting already before pathogen $2$ infections can grow significantly. We find it surprising that the initializations at $j_2$ and $j_3$ produce the smallest peaks by a large margin, even though $j_1$ is fairly close to $j_2$ and $j_3$ is close to $j_4$.

Because pathogen $2$ is dominant and has a higher virulence $\alpha_2$, it is most ideal for the $I_2$-spread index to grow slowly. We see from the superinfection examples that the $I_2$-spread index grows relatively slowly when the pathogens originate at a far distance from each other, assuming the initialization in Figure~\autoref{fig:center_graph} or Figure~\autoref{fig:corner_graph}. For co-infection dynamics, if we seek to slow the spread of the infection that has the greatest virulence, different configurations are most beneficial depending on the nature of each infection. In this case, the co-infected group has the greatest virulence of $0.25$. Thus, it may be most ideal for the distances between the pathogen initializations to be as far as possible on the lattice, assuming the configuration in Figure~\autoref{fig:center_graph}.

\subsection{Effect of Source Time Differences}
\label{sec:time_diff}

In this subsection, we investigate the impact of time differences in the introduction of the two pathogens in the network, where pathogen $2$ infections are introduced some time after pathogen $1$ infections. We denote the time difference between the introductions of pathogens $1$ and $2$ in the network to be $t_d$. For superinfection dynamics, we expect that pathogen $1$ infections will have more time to spread to a greater area before superinfection becomes prominent. We see this is true when we incorporate the parameter values from Example~\ref{ex:4} with the model in Equation~(\ref{eq:superinfect-prelim}) and initialize pathogen $1$- and pathogen $2$-infections to originate at opposite corner nodes in a $1600$-node lattice network. In \autoref{fig:superinfect_time}, we observe that as the time difference $t_d$ of the introduction of the two pathogens-induced infections increases, both the peak $I_1$-spread index and the $I_2$-saturation time will increase. The progression of the $I_1$-peak value follows approximately a power curve as described in the captions of \autoref{fig:superinfect_time}, and the progression of the $I_2$-saturation time follows a linear relationship with the time $t_d$.

\begin{figure*}[htbp]
    \centering

    \makebox[\textwidth]{%
        \subfigure[]{%
            \includegraphics[width=0.34\textwidth]{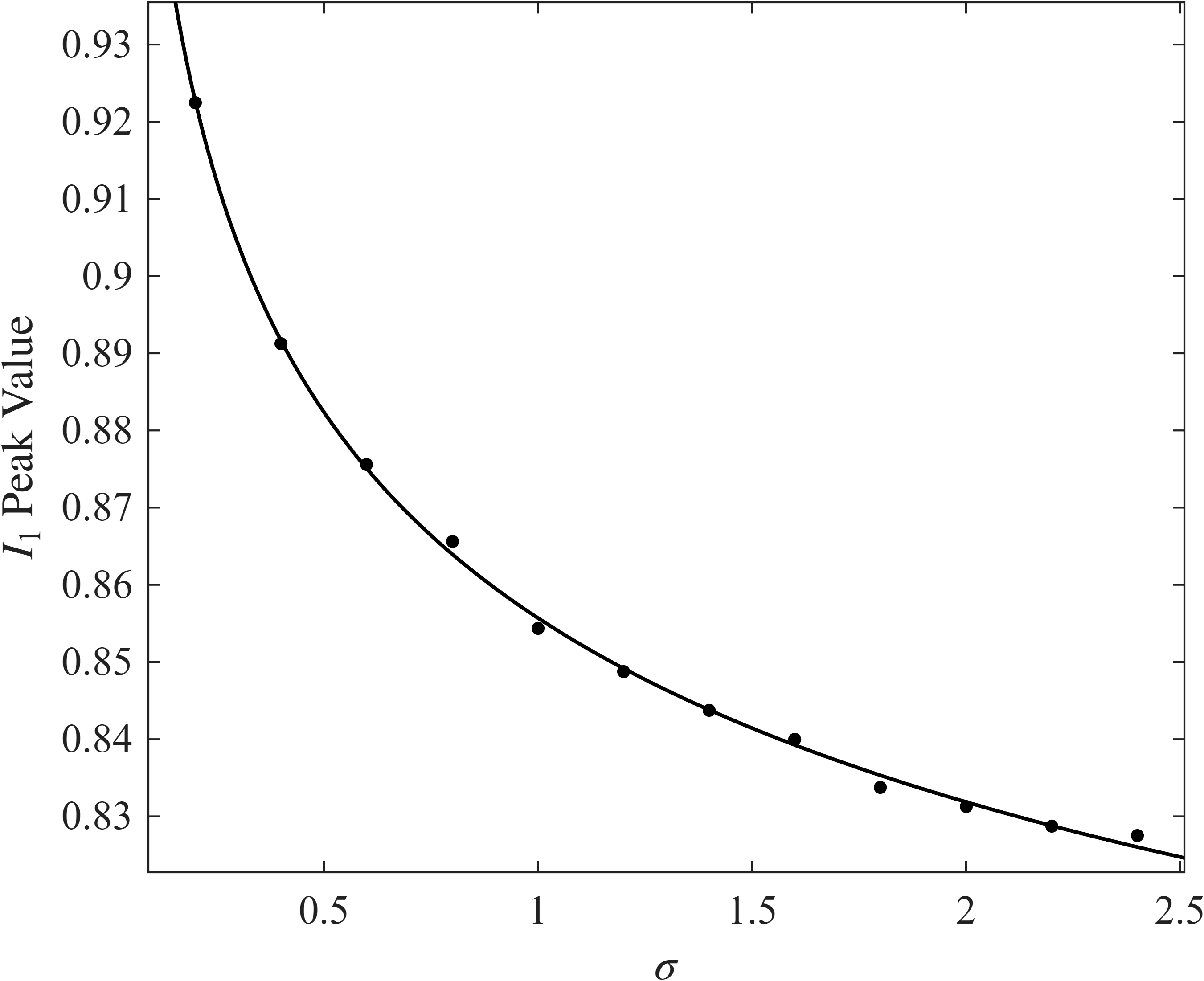}
            % \subcaption{Fit: $y=ax^b+c$, where $a=0.2239$, $b=-0.1624$, and $c=0.6318$}
            \label{fig:}
        }
        \subfigure[]{%
            \includegraphics[width=0.33\textwidth]{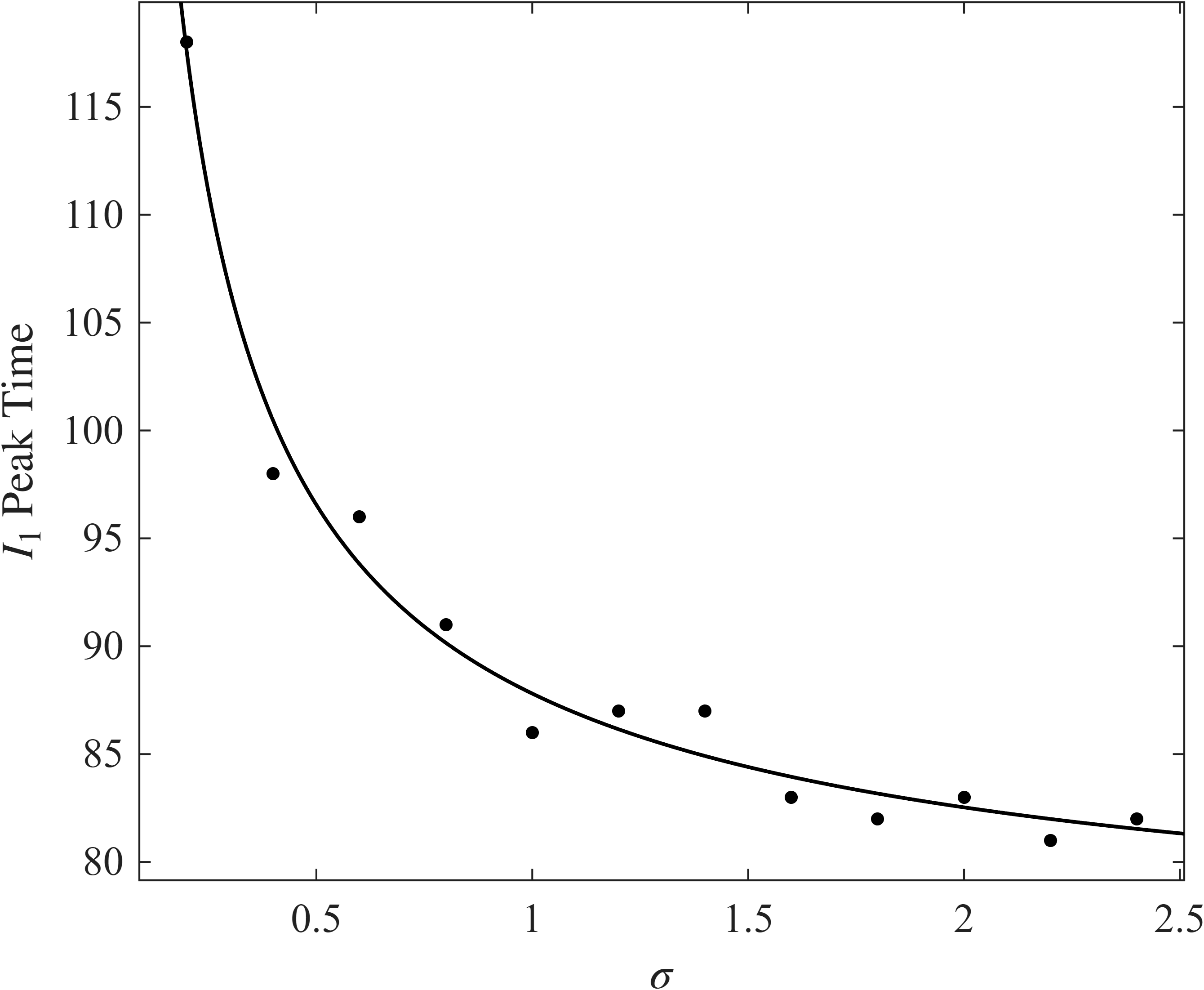}
            % \subcaption{Fit: $y=ax^b+c$, where $a=13.285$, $b=-0.73$, and $c=74.521$}
            \label{fig:}
        }
        \subfigure[]{%
            \includegraphics[width=0.32\textwidth]{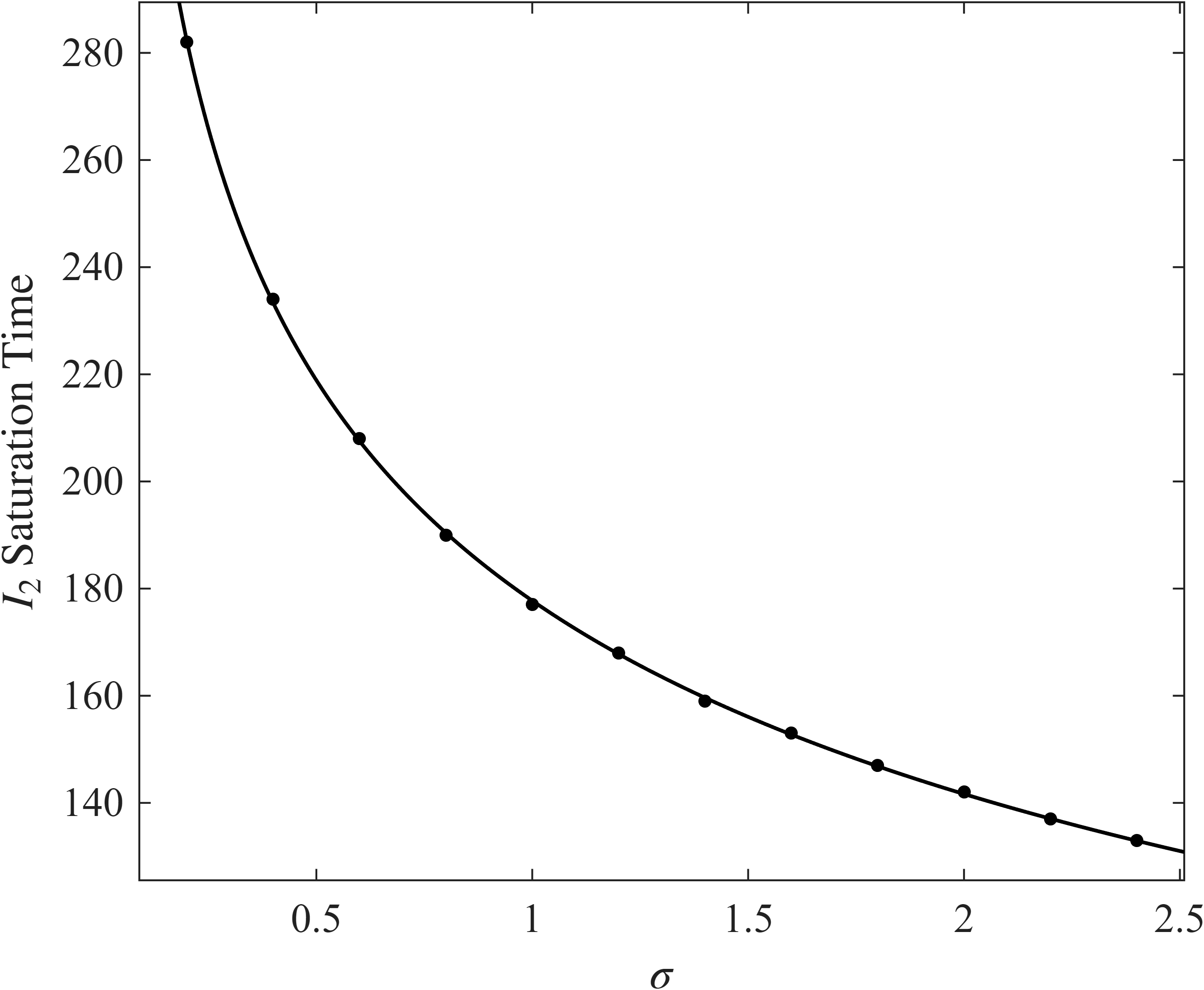}
            \label{fig:}
            % \subcaption{Fit: $y=ax^b+c$, where $a=293.11$, $b=-0.1895$, and $c=-115.36$}
        }
    }
    \caption{$I_1$-peak value (left), $I_1$-peak time (center), and $I_2$-saturation time (right) over varying $\sigma$. Left fitted curve: $y=ax^b+c$, where $a=0.2239$, $b=-0.1624$, and $c=0.6318$. Center fitted curve: $y=ax^b+c$, where $a=13.285$, $b=-0.73$, and $c=74.521$. Right fitted curve: $y=ax^b+c$, where $a=293.11$, $b=-0.1895$, and $c=-115.36$.}
    \label{fig:superinfect_sigma}
\end{figure*}

We keep the same settings as before, but incorporate the MBRD-CI model from Equation~(\ref{eq:coinfect-prelim}) and the co-infection parameters from Example~\ref{ex:5}. Again, we see that one mono-infection spread index reaches $1$ sooner, and the other spread index reaches $1$ at later times as $t_d$ increases, according to \autoref{fig:time_coinfect}. Here, the $I_1$-spread index reaches $1$ sooner and the $I_2$-spread index reaches $1$ later in response to that change. Moreover, we observe once again that the co-spread index follows the same trend in source time changes as the $I_2$-spread index. We believe this could be true in general co-infection dynamics and propose this as an interesting topic of future study.

\subsection{Effect of Superinfection and Co-Transmission Parameters}
\label{sec:model_param}

Because a larger superinfection coefficient $\sigma$ allows pathogen $2$ to dominate more easily, we expect that as $\sigma$ increases, the $I_1$-spread index will peak at a lower value and at an earlier time. This is because the $I_1$-spread index will have less room to grow before it becomes significantly dominated by pathogen $2$-infections and falls back to $0$ when pathogen $2$ is more dominant. We verify these statements with the model in Equation~(\ref{eq:superinfect-prelim}) and the parameter settings in Example~\ref{ex:4}, with varying values of $\sigma$. In \autoref{fig:superinfect_sigma}, we observe that as $\sigma$ increases, the $I_1$-peak value and $I_2$-peak time both decrease, and both relationships can be fitted to power curves. Because the correlations are all negative, it is most ideal for the superinfection coefficient to be small during the early stages of an epidemic.

\begin{figure*}[htbp]
    \centering
    \makebox[\textwidth]{%
        \subfigure[]{%
            \includegraphics[width=0.33\textwidth]{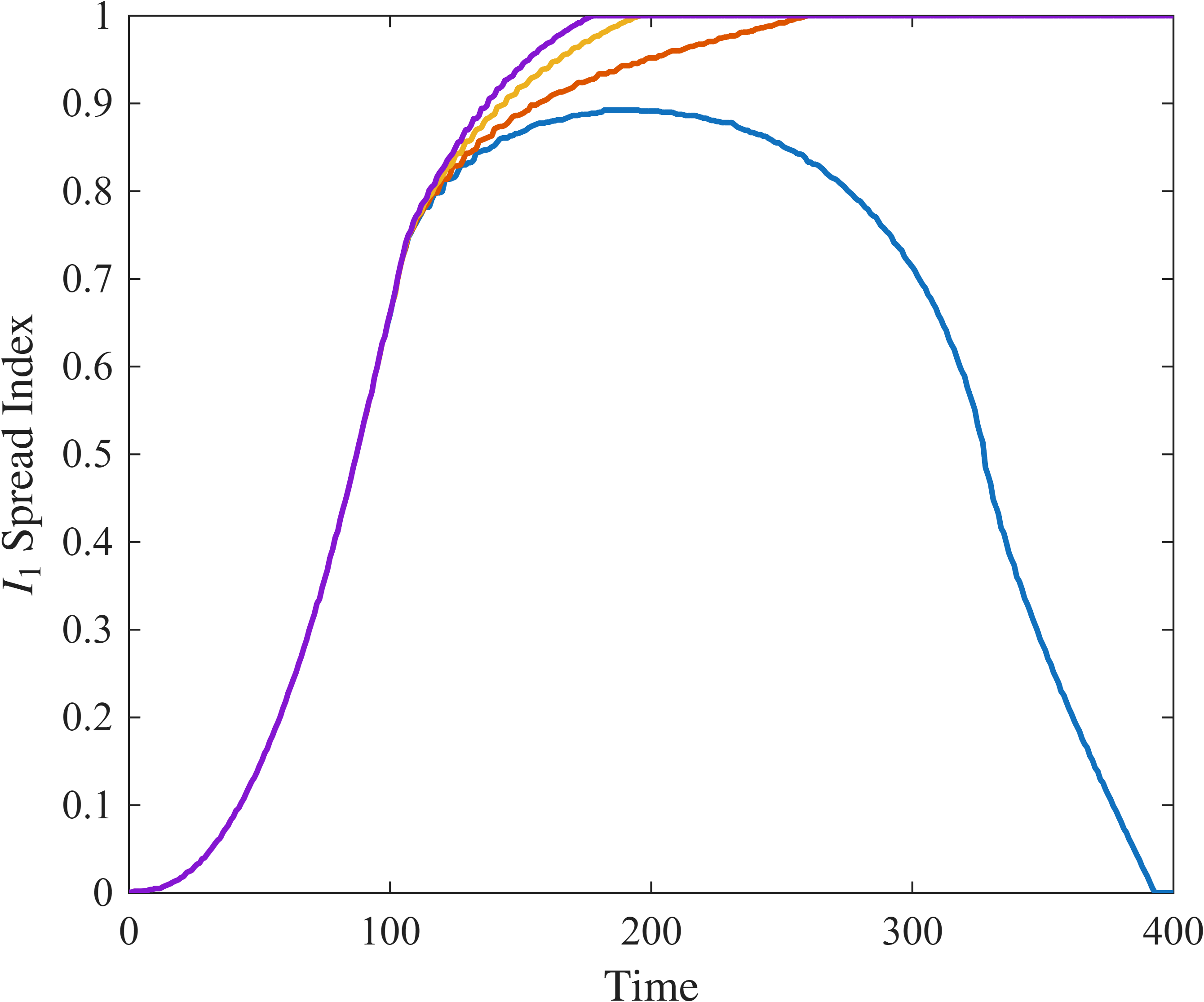}
            \label{fig:coinfect_I_beta12}
        }
        \subfigure[]{%
            \includegraphics[width=0.33\textwidth]{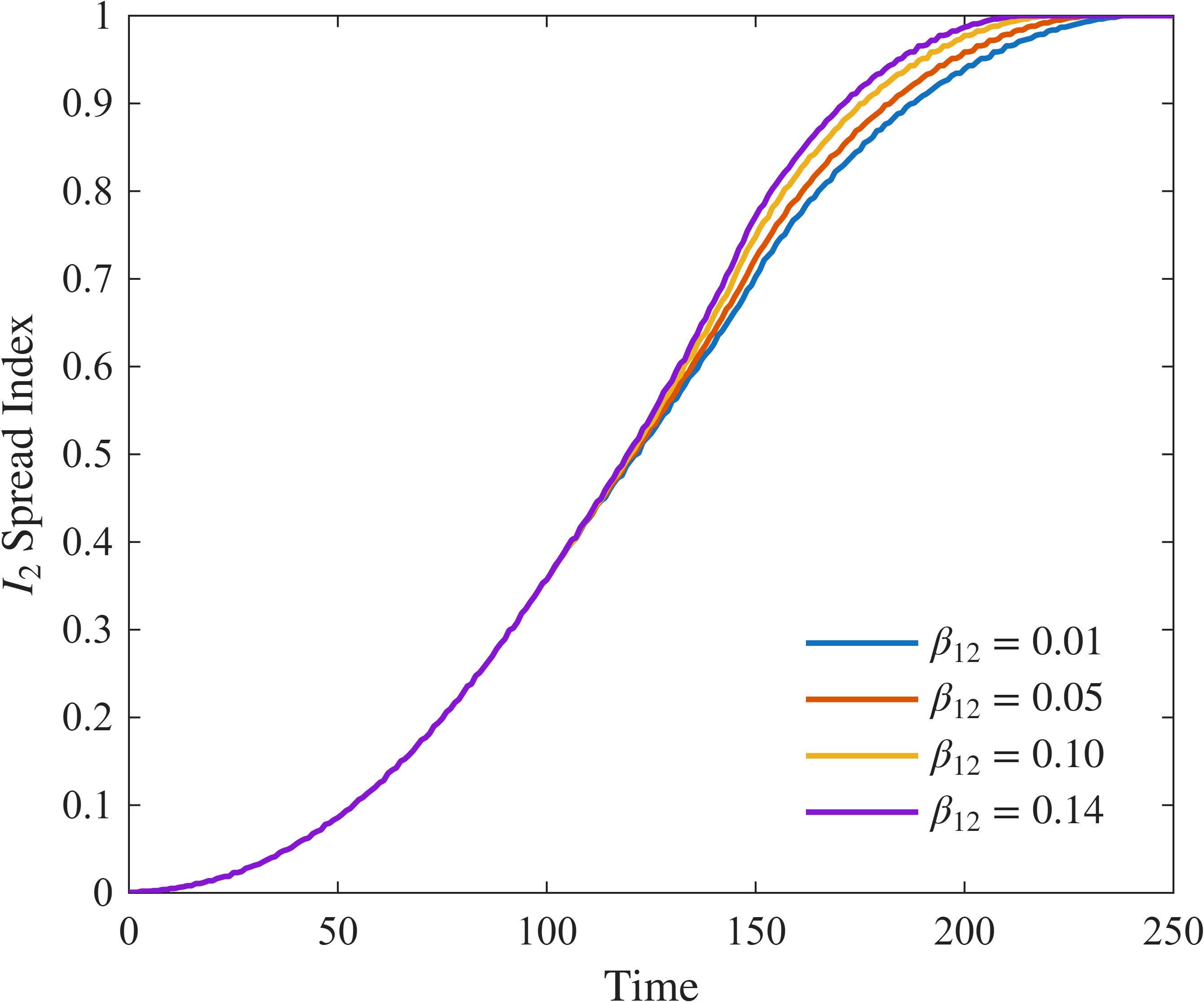}
            \label{fig:coinfect_J_beta12}
        }
        \subfigure[]{%
            \includegraphics[width=0.33\textwidth]{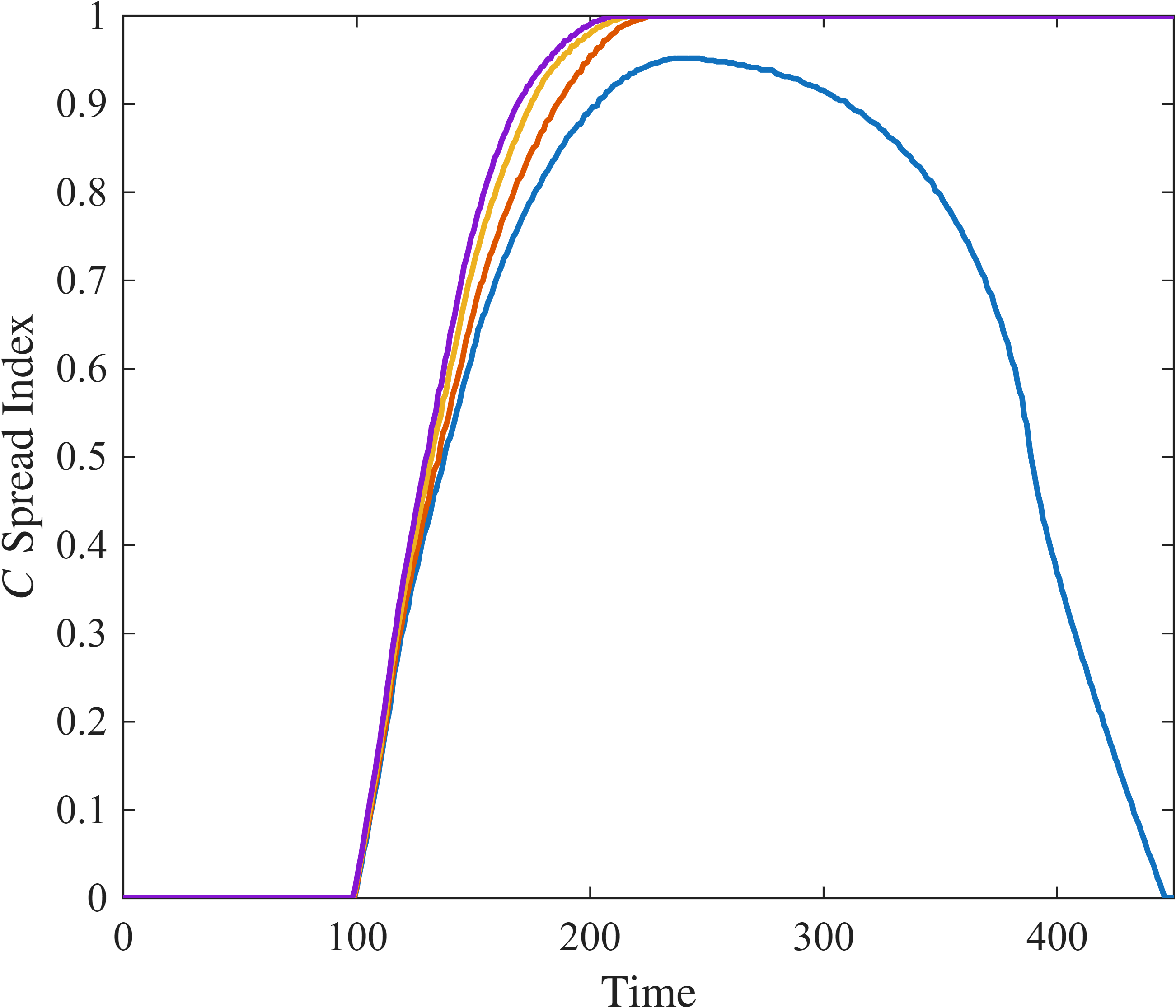}
            \label{fig:coinfect_C_beta12}
        }
    }
    \caption{$I_1$-spread index (left), $I_2$-spread index (center), and $C$-spread index (right) for four different values of $\beta_{12}$. Based on Example~\ref{ex:5}.}
    \label{fig:coinfect_beta12}
\end{figure*}

For the MBRD-CI model in Equation~(\ref{eq:coinfect-prelim}), we expect that as $\beta_{12}$ increases, the probability that an individual will be co-infected will increase, resulting in the $C$-spread index reaching $1$ sooner. We see this occurs in \autoref{fig:coinfect_beta12}, which was created using the parameter set from Example~\ref{ex:5}. From this figure, we also note that the dynamics when $\beta_{12}=0.01$, $0.05$ and $0.09$ can be described as mutual inhibition, while the dynamics when $\beta_{12}=0.14$ can be described as mutual enhancement. We observe that all three indexes reach $1$ at earlier times when mutual enhancement occurs as opposed to other conditions.

In this figure, we observe that $\beta_{12}=0.14$ allows all three indexes to reach $1$ at the earliest times, and $\beta_{12}=0.01$ enables the $I_1$-spread index and $C$-spread index to peak and then fall to $0$. This inspires the following definition:

\begin{definition}[$\beta_{12}$ threshold]
We define the $\beta_{12}$ threshold of a co-infection parameter configuration to be the largest value of $\beta_{12}$, with all other parameters fixed, such that the maximum $C$-spread index over time is less than $1$ and converges to $0$.
\end{definition}

\begin{figure}[htbp]
    \centering
    \includegraphics[width=0.45\textwidth]{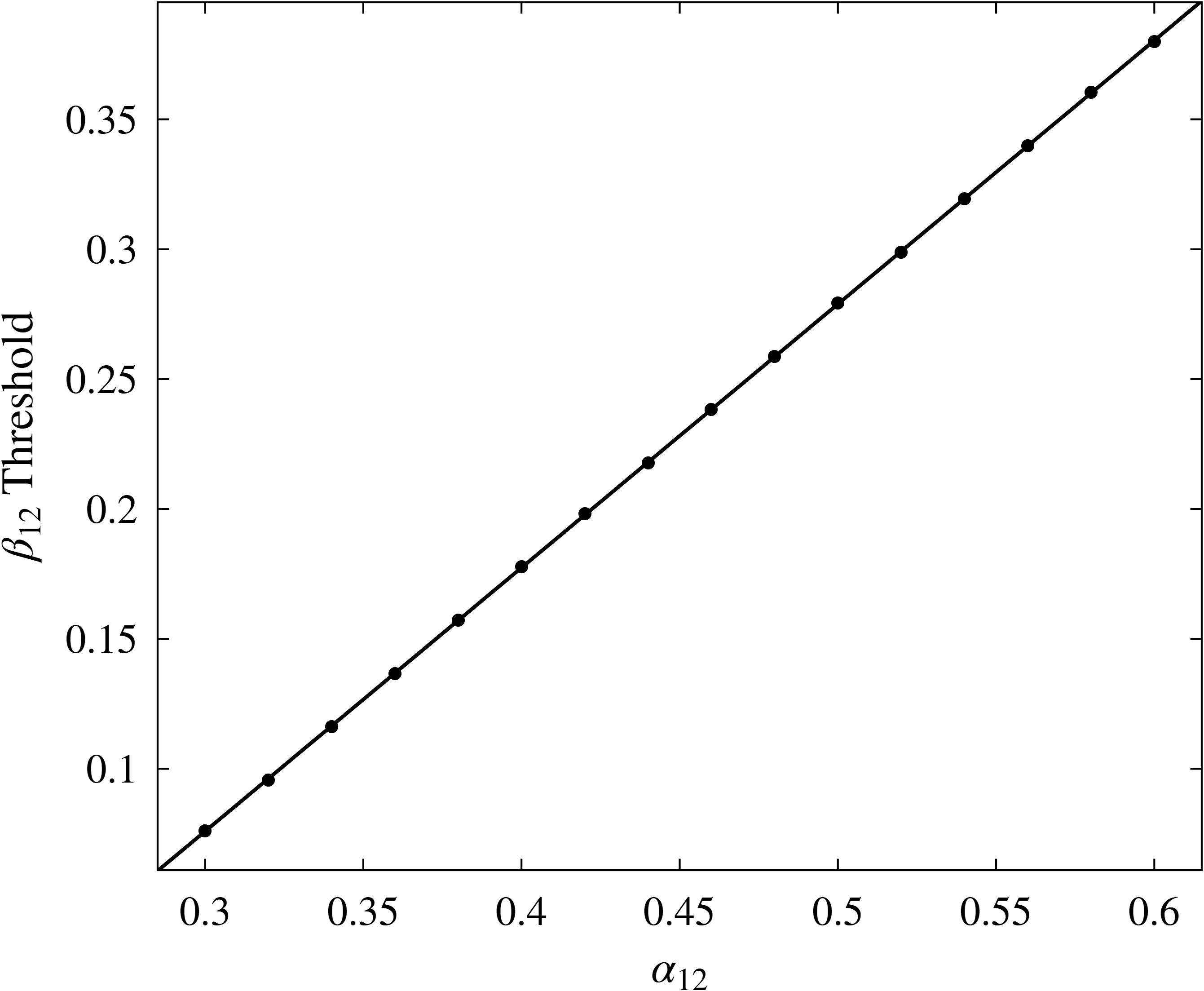}
    \caption{$\beta_{12}$-threshold for varying $\alpha_{12}$ values. Fitted curve: $y=ax+b$, where $a=1.0148$ and $b=-0.2285$.}
    \label{fig:coinfect_b12thres}
\end{figure}

When the co-infection removal rate $\alpha_{12}$ is large, we expect that the value of $\beta_{12}$ must make up for it, and the $\beta_{12}$ threshold must be larger. From \autoref{fig:coinfect_b12thres}, we observe that the relationship between the $\beta_{12}$ threshold and the value of $\alpha_{12}$. We believe that proving this relationship mathematically would be an interesting subject of future work.

\subsection{Effect of Network Type and Layer Degrees}
\label{sec:layers}

Investigating the impact of varying network types and layer average degrees is highly useful in understanding infection spread and effective public health policy. We analyze infection spreading in LA, WS and BA networks and with $8$ different combinations of average layer degrees. We analyze these factors with the following superinfection and co-infection parameter sets, as well as Examples~\ref{ex:4} and~\ref{ex:5}. 

\begin{example}[Superinfection model] \label{ex:6} 
\begin{equation}\label{eq:p6-settings}
\begin{aligned}
\mu &= 0.005\quad &r&=0.1\quad & A&=0.1\quad & K&=1\\
\beta_1 &=0.3, \quad & \beta_2&=0.5,\quad & \sigma &=1.2\\
\gamma_1&=0.2,\quad &\gamma_2&=0.1,\quad & \alpha_1&=0.01,\quad & \alpha_2&=0.1\\
d_{11}&=0.5,\quad & d_{12}&=0.1\quad & d_{13}&=0.1\\
d_{22}&=0.3,\quad & d_{33}&=0.1.
\end{aligned}
\end{equation}
\end{example}

\begin{example}[Superinfection model] \label{ex:7} 
\begin{equation}\label{eq:p7-settings}
\begin{aligned}
\mu &= 0.005\quad &r&=0.1\quad & A&=0.1\quad & K&=1\\
\beta_1 &=0.3, \quad & \beta_2&=0.2,\quad & \sigma &=0.3\\
\gamma_1&=0.2,\quad &\gamma_2&=0.05,\quad & \alpha_1&=0.01,\quad & \alpha_2&=0.02\\
d_{11}&=0.8,\quad & d_{12}&=-0.05\quad & d_{13}&=-0.07\\
d_{22}&=0.03,\quad & d_{33}&=0.5.
\end{aligned}
\end{equation}
\end{example}

\begin{example}\label{ex:8} (Co-infection model) 
\begin{equation}\label{eq:p8-settings}
\begin{aligned}
\mu &= 0.005,\quad & r&=0.1,&\quad A&=0.1,\quad& K&=1,\\
\beta_1&=0.5,\quad & \beta_2&= 0.4,\\ \beta_{10}&=0.4,\quad & \beta_{02}&=0.3,\quad & \beta_{12}&=0.15,\\
\gamma_1&=0.2,\quad & \gamma_2&=0.1,\\ \alpha_1&=0.05,\quad & \alpha_2&=0.2,\quad & \alpha_{12}&=0.15,\\
d_{11}&=0.5,\quad & d_{12}&=0.1,\quad & d_{13}&=0.1,\\
d_{22}&=0.2,\quad & d_{33}&=0.3.
\end{aligned}
\end{equation}
\end{example}

\begin{example}\label{ex:9} (Co-infection model) 
\begin{equation}\label{eq:p9-settings}
\begin{aligned}
\mu &= 0.005,\quad & r&=0.1,&\quad A&=0.1,\quad& K&=1,\\
\beta_1&=0.7,\quad & \beta_2&= 0.4,\\ \beta_{10}&=0.2,\quad & \beta_{02}&=0.05,\quad & \beta_{12}&=0.15,\\
\gamma_1&=0.2,\quad & \gamma_2&=0.1,\\ \alpha_1&=0.16,\quad & \alpha_2&=0.05,\quad & \alpha_{12}&=0.15,\\
d_{11}&=0.8,\quad & d_{12}&=0.05,\quad & d_{13}&=0.05,\\
d_{22}&=0.4,\quad & d_{33}&=0.1.
\end{aligned}
\end{equation}
\end{example}

\begin{table*}[ht]
\noindent
\begin{tabularx}{0.85\textwidth}{|X|X|X|X|X|X|X|X|X|X|X|X|}
\hline
\multicolumn{12}{|c|}{\textbf{Layerwise Degree Variations, using Examples~\ref{ex:4},~\ref{ex:6},~\ref{ex:7} and Lattice Networks}} \\
\hline
\multicolumn{3}{|c|}{\textbf{Network Type}} & 
\multicolumn{3}{|c|}{\textbf{Example ~\ref{ex:4}}} &
\multicolumn{3}{|c|}{\textbf{Example~\ref{ex:6}}} &
\multicolumn{3}{|c|}{\textbf{Example~\ref{ex:7}}} \\
\hline
\textbf{$S$ layer} & \textbf{$I$ layer} & \textbf{$J$ layer} & \textbf{$I_1$-peak value} & \textbf{$I_1$-peak time} & \textbf{$I_2$-sat. time} & \textbf{$I_1$-peak value} & \textbf{$I_1$-peak time} & \textbf{$I_2$-sat. time}& \textbf{$I_1$-peak value} & \textbf{$I_1$-peak time} & \textbf{$I_2$-sat. time}\\
\hline
LA24 & LA24 & LA24 & 1 & 28 & 43 & 0.0006 & 0 & 38 & 0.0006 & 0 & 48\\
LA24 & LA4 & LA12 & 0.5056 & 71 & 74 & 0.0881 & 62 & 66 & 0.1581 & 84 & 60\\
LA24& LA12 & LA4 & 1 & 49 & 184 & 0.6756 & 84 & 158 & 0.7963 & 99 & 128\\
LA12 & LA12 & LA12 & 0.9719 & 54 & 75 & 0.0969 & 60 & 66 & 0.3844 & 80 & 60\\
LA12 & LA4 & LA12 & 0.5056 & 71 & 74 & 0.0881 & 62 & 66 & 0.1581 & 84 & 60\\
LA12 & LA4 & LA4 & 0.8581 & 88 & 183 & 0.3444 & 111 & 158 & 0.3525 & 116 & 128\\
LA12 & LA12 & LA4 & 1 & 49 & 184 & 0.6756 & 84 & 158 & 0.7963 & 100 & 128\\
LA4 & LA4 & LA4 & 0.8581 & 88 & 183 & 0.3444 & 111 & 158 & 0.3525 & 116 & 128\\
\hline
\end{tabularx}

\begin{tabularx}{0.85\textwidth}{|X|X|X|X|X|X|X|X|}
\hline
\multicolumn{8}{|c|}{\textbf{Layerwise Degree Variations, using Example~\ref{ex:4}}} \\
\hline
\multicolumn{3}{|c|}{\textbf{Average Degrees}} & 
\multicolumn{3}{|c|}{\textbf{WS}} &
\multicolumn{2}{|c|}{\textbf{BA}} \\
\hline
\textbf{$S$ layer} & \textbf{$I$ layer} & \textbf{$J$ layer} & \textbf{$I_1$-peak value} & \textbf{$I_1$-peak time} & \textbf{$I_2$-sat. time} & \textbf{$I_1$-sat. time} & \textbf{$I_2$-sat. time}\\
\hline
24 & 24 & 24 & 1 & 21 & 25.01 & 21 & 24\\
24 & 4 & 12 & 0.9458 & 45.573 & 29.649 & 21 & 24\\
24& 12 & 4 & 1 & 22.061 & 72.345 & 21 & 29.341\\
12 & 12 & 12 & 1 & 22.049 & 29.622 & 21 & 24\\
12 & 4 & 12 & 0.95034 & 45.721 & 29.555 & 21 & 24\\
12 & 4 & 4 & 0.99901 & 46.391 & 71.883 & 21.912  & 29.358\\
12 & 12 & 4 & 1 & 22.059 & 72.611 & 21 & 29.321\\
4 & 4 & 4 & 0.99874 & 46.353 & 72.434 & 21.913   & 29.348\\
\hline
\end{tabularx}

\begin{tabularx}{0.85\textwidth}{|X|X|X|X|X|X|X|X|}
\hline
\multicolumn{8}{|c|}{\textbf{Layerwise Degree Variations, using Example~\ref{ex:6}}} \\
\hline
\multicolumn{3}{|c|}{\textbf{Average Degrees}} & 
\multicolumn{3}{|c|}{\textbf{WS}} &
\multicolumn{2}{|c|}{\textbf{BA}} \\
\hline
\textbf{$S$ layer} & \textbf{$I$ layer} & \textbf{$J$ layer} & \textbf{$I_1$-peak value} & \textbf{$I_1$-peak time} & \textbf{$I_2$-sat. time} & \textbf{$I_1$-peak value} & \textbf{$I_2$-sat. time}\\
\hline
24 & 24 & 24 & 0.0006 & 0 & 21.041 & 0.000625 & 20\\
24 & 4 & 12 & 0.00063062 & 0.019 & 25.637 & 0.000625 & 20\\
24& 12 & 4 & 0.000625 & 0 & 63.19 & 0.000625 & 25.464\\
12 & 12 & 12 & 0.000625 & 0 & 25.591 & 0.000625 & 20\\
12 & 4 & 12 & 0.00063062 & 0.02 & 25.644 & 0.000625 & 20\\
12 & 4 & 4 & 0.0013719 & 6.899 & 63.026 & 0.000625 & 25.494\\
12 & 12 & 4 & 0.000625 & 0 & 63.265 & 0.000625 & 25.482\\
4 & 4 & 4 & 0.0014444 & 6.631 & 62.924 & 0.000625 & 25.495\\
\hline
\end{tabularx}

\begin{tabularx}{0.85\textwidth}{|X|X|X|X|X|X|X|X|}
\hline
\multicolumn{8}{|c|}{\textbf{Layerwise Degree Variations, using Example~\ref{ex:7}}} \\
\hline
\multicolumn{3}{|c|}{\textbf{Average Degrees}} & 
\multicolumn{3}{|c|}{\textbf{WS}} &
\multicolumn{2}{|c|}{\textbf{BA}} \\
\hline
\textbf{$S$ layer} & \textbf{$I$ layer} & \textbf{$J$ layer} & \textbf{$I_1$-peak value} & \textbf{$I_1$-peak time} & \textbf{$I_2$-peak time} & \textbf{$I_1$-peak value} & \textbf{$I_2$-sat. time}\\
\hline
24 & 24 & 24 & 0.000625 & 0 & 47 & 0.0006 & 47\\
24 & 4 & 12 & 0.17353 & 77.012 & 47 & 0.0006 & 47\\
24& 12 & 4 & 0.43153 & 78.423 & 64.423 & 0.0006 & 47\\
12 & 12 & 12 & 0.000625 & 0 & 47 & 0.0006 & 47\\
12 & 4 & 12 & 0.17657 & 77.039 & 47 & 0.0006 & 47\\
12 & 4 & 4 & 0.21016 & 79.89 & 64.475 & 0.0006 & 47\\
12 & 12 & 4 & 0.43003 & 77.688 & 64.422 & 0.0006 & 47\\
4 & 4 & 4 & 0.20097 & 79.75 & 64.365 & 0.0006 & 47\\
\hline
\end{tabularx}
\caption{Variation of layerwise degrees with superinfection dynamics, each an average of $1000$ trials.}
\label{table:superinfect_deg}
\end{table*}

\begin{table*}[ht]
\noindent
\begin{tabularx}{0.85\textwidth}{|X|X|X|X|X|X|X|X|X|X|X|X|}
\hline
\multicolumn{12}{|c|}{\textbf{Layerwise Degree Variations, using Examples~\ref{ex:5},~\ref{ex:8},~\ref{ex:9} and Lattice Networks}} \\
\hline
\multicolumn{3}{|c|}{\textbf{Network Type}} & 
\multicolumn{3}{|c|}{\textbf{Example ~\ref{ex:5}}} &
\multicolumn{3}{|c|}{\textbf{Example~\ref{ex:8}}} &
\multicolumn{3}{|c|}{\textbf{Example~\ref{ex:9}}} \\
\hline
\textbf{$S$ layer} & \textbf{$I$ layer} & \textbf{$J$ layer} & \textbf{$I_1$-sat. time} & \textbf{$I_2$-sat. time} & \textbf{$C$-sat. time} & \textbf{$I_1$-sat. time} & \textbf{$I_2$-sat. time} & \textbf{$C$-sat. time}& \textbf{$I_1$-sat. time} & \textbf{$I_2$-sat. time} & \textbf{$C$-sat. time}\\
\hline
LA24 & LA24 & LA24 & 53 & 50 & 56 & 34 & 61 & 55 & 22 & 58 & 52\\
LA24 & LA4 & LA12 & 586 & 89 & 353 & 132 & 81 & 133 & 103 & 95 & 97\\
LA24 & LA12 & LA4 & 73 & 234 & 232 & 55 & 155 & 152 & 36 & 301 & 292\\
LA12 & LA12 & LA12 & 84 & 90 & 89 & 55 & 78 & 73 & 37 & 108 & 101\\
LA12 & LA4 & LA12 & 586 & 89 & 353 & 132 & 81 & 133 & 103 & 95 & 97\\
LA12 & LA4 & LA4 & 260 & 229 & 227 & 132 & 161 & 157 & 93 & 279 & 270\\
LA12 & LA12 & LA4 & 73 & 234 & 232 & 55 & 155 & 152 & 36 & 301 & 292\\
LA4 & LA4 & LA4 & 260 & 229 & 227 & 132 & 161 & 157 & 93 & 279 & 270\\
\hline
\end{tabularx}

\begin{tabularx}{0.85\textwidth}{|X|X|X|X|X|X|X|X|X|}
\hline
\multicolumn{9}{|c|}{\textbf{Layerwise Degree Variations, using Example~\ref{ex:5}}} \\
\hline
\multicolumn{3}{|c|}{\textbf{Average Degrees}} & 
\multicolumn{3}{|c|}{\textbf{WS}} &
\multicolumn{3}{|c|}{\textbf{BA}} \\
\hline
\textbf{$S$ layer} & \textbf{$I$ layer} & \textbf{$J$ layer} & \textbf{$I_1$-sat. time} & \textbf{$I_2$-sat. time} & \textbf{$C$-sat. time} & \textbf{$I_1$-sat. time} & \textbf{$I_2$-sat. time} & \textbf{$C$-sat. time}\\
\hline
24 & 24 & 24 & 42 & 30.997 & 49 & 42 & 30 & 49\\
24 & 4 & 12 & 206.54 & 35.534 & 109.15 & 42 & 30 & 49\\
24& 12 & 4 & 47.186 & 86.964 & 87.678 & 43.798 & 30.154 & 50.412\\
12 & 12 & 12 & 44.741 & 35.522 & 50.677 & 42 & 30 & 49\\
12 & 4 & 12 & 205.06 & 35.519 & 108.87 & 42 & 30 & 49\\
12 & 4 & 4 & 79.151 & 85.998 & 88.342 & 44.495 & 35.139 & 50.394\\
12 & 12 & 4 & 47.269 & 87.004 & 87.816 & 43.823 & 35.156 & 50.462\\
4 & 4 & 4 & 80.926 & 85.362 & 88.37 & 44.548 & 35.12 & 50.396\\
\hline
\end{tabularx}

\begin{tabularx}{0.85\textwidth}{|X|X|X|X|X|X|X|X|X|}
\hline
\multicolumn{9}{|c|}{\textbf{Layerwise Degree Variations, using Example~\ref{ex:8}}} \\
\hline
\multicolumn{3}{|c|}{\textbf{Average Degrees}} & 
\multicolumn{3}{|c|}{\textbf{WS}} &
\multicolumn{3}{|c|}{\textbf{BA}} \\
\hline
\textbf{$S$ layer} & \textbf{$I$ layer} & \textbf{$J$ layer} & \textbf{$I_1$-sat. time} & \textbf{$I_2$-sat. time} & \textbf{$C$-sat. time} & \textbf{$I_1$-sat. time} & \textbf{$I_2$-sat. time} & \textbf{$C$-sat. time}\\
\hline
24 & 24 & 24 & 24 & 59 & 53 & 24 & 59 & 53\\
24 & 4 & 12 & 55.093 & 62.9 & 65.06 & 25 & 59 & 53\\
24 & 12 & 4 & 26.305 & 80.404 & 75.829 & 24 & 59 & 53\\
12 & 12 & 12 & 26.285 & 59 & 53.001 & 24 & 59 & 53\\
12 & 4 & 12 & 55.406 & 62.943 & 65.275 & 25 & 59 & 53\\
12 & 4 & 4 & 54.962 & 83.433 & 79.132 & 25 & 59 & 53\\
12 & 12 & 4 & 26.274 & 80.755 & 76.213 & 24 & 59 & 53\\
4 & 4 & 4 & 55.104 & 83.323 & 79.063 & 25 & 59 & 53\\
\hline
\end{tabularx}

\begin{tabularx}{0.85\textwidth}{|X|X|X|X|X|X|X|X|X|}
\hline
\multicolumn{9}{|c|}{\textbf{Layerwise Degree Variations, using Example~\ref{ex:9}}} \\
\hline
\multicolumn{3}{|c|}{\textbf{Average Degrees}} & 
\multicolumn{3}{|c|}{\textbf{WS}} &
\multicolumn{3}{|c|}{\textbf{BA}} \\
\hline
\textbf{$S$ layer} & \textbf{$I$ layer} & \textbf{$J$ layer} & \textbf{$I_1$-sat. time} & \textbf{$I_2$-sat. time} & \textbf{$C$-sat. time} & \textbf{$I_1$-sat. time} & \textbf{$I_2$-sat. time} & \textbf{$C$-sat. time}\\
\hline
24 & 24 & 24 & 18 & 26.059 & 29 & 18 & 25 & 29\\
24 & 4 & 12 & 35.974 & 33.44 & 38.772 & 18 & 25 & 29\\
24 & 12 & 4 & 18.599 & 121.13 & 111.84 & 18.676 & 34.168 & 32.378\\
12 & 12 & 12 & 18.536 & 34.862 & 32.733 & 18 & 25 & 29\\
12 & 4 & 12 & 36.073 & 33.423 & 38.907 & 18 & 25 & 29\\
12 & 4 & 4 & 35.407 & 116.22 & 106.95 & 18.915 & 34.313 & 32.459\\
12 & 12 & 4 & 18.602 & 121.28 & 111.97 & 18.668 & 34.155 & 32.389\\
4 & 4 & 4 & 35.654 & 116.21 & 106.86 & 18.923 & 34.227 & 32.409\\
\hline
\end{tabularx}
\caption{Variation of layerwise degrees with coinfection dynamics, each an average of $1000$ trials.}
\label{table:coinfect_deg}
\end{table*}

We first analyze infection diffusion for different network topology and layerwise average degrees in superinfection dynamics. It is expected that lowering the migration of infected individuals will help mitigate the diffusion of infections to other areas. We confirm this with Table~\ref{table:superinfect_deg}, which compares different combinations of networks and the $I_1$-peak value, $I_1$-peak time, and $I_2$-saturation time that occur. We start by focusing specifically on lattice networks because they are deterministic. We note that because pathogen $2$ has a larger removal rate, it is often best to  maximize the $I_2$-saturation time to lessen the number of deaths. From all of the subtables in Table~\ref{table:superinfect_deg}, we observe the following:
\begin{itemize}
    \item We find that in all the subtables in Table~\ref{table:superinfect_deg}, the degree combinations that give the highest $I_2$ saturation times are LA24-LA12-LA4, LA12-LA4-LA4, LA12-LA12-LA4, and LA4-LA4-LA4. Thus, in superinfection dynamics, a low average degree of the $J$-density layer of the multiplex network leads to a larger $I_2$ saturation time. We also note that out of the four combinations mentioned, LA12-LA4-LA4 and LA4-LA4-LA4 produce the largest $I_1$-saturation times. Thus, to slow the propagation of both infections, it is important for the average degrees of both the $I$ and $J$ layers to be minimal. However, because pathogen $2$ has the highest virulence, limiting the migration of pathogen $2$-infected individuals should take priority.
    \item Recent research has found that diffusion generally occurs faster in BA networks than WS networks~\cite{zhu2025pattern,ju2022online}. As a result, we expect that in BA networks, the number of instances of superinfection will grow earlier and faster, leading to smaller peaks in the $I_1$-spread index and larger $I_2$-saturation times. We verify this is true in Tables~\ref{table:superinfect_deg}, where the time that the $I_2$-spread index reaches $1$ is consistently lower for BA networks than WS networks. Moreover, degree variations have a smaller impact on the values on the spread indexes over time for both infections. The BA network topology may also explain why outbreaks often occur during holidays and policies become less effective in those time periods.
\end{itemize}

Moving to co-infection dynamics, we similarly expect that it is most beneficial for the graph corresponding to the more dominant pathogen to have a lower average degree to slow the spread of that pathogen across the network. Table~\ref{table:coinfect_deg} incorporates the MBRD-CI model in Equation~\ref{eq:coinfect-prelim} and the parameters from Example~\ref{ex:5},~\ref{ex:8}, and~\ref{ex:9}. Here, we analyze each of them separately, as follows:
\begin{itemize}
    \item \textbf{Example~\ref{ex:5}}. This parameter configuration satisfies $\alpha_{12}>\alpha_2>\alpha_1$. Thus, it is most beneficial to limit the infections of pathogen $2$, including co-infections. Overall, the combinations that produce the largest saturation time for $I_2$ and $I_{12}$ are the LA12-LA4-LA4 and LA4-LA4-LA4. Moreover, for the LA24-LA4-LA12 and LA12-LA4-LA12, there is also a large saturation time for $I_1$.
    \item \textbf{Example~\ref{ex:8}}. This parameter configuration satisfies $\alpha_2>\alpha_{12}>\alpha_1$. Again, it is most beneficial to limit the infections of pathogen $2$, including co-infections. In this case, the multiplex network combinations LA24-LA12-LA4, LA12-LA4-LA4, LA12-LA12-LA4, and LA4-LA4-LA4. Out of these, the combinations LA12-LA4-LA4 and LA4-LA4-LA4 produce the highest $I_1$-saturation times in addition to the high $I_2$ and and $C$-saturation time, and are thus most beneficial.
    \item \textbf{Example~\ref{ex:9}}. This parameter configuration satisfies $\alpha_1>\alpha_{12}>\alpha_2$, where $\alpha_1$ and $\alpha_{12}$ are only $0.01$ apart. Thus, it may be most beneficial to limit the infections of pathogen $1$, including co-infections. It is the most beneficial to maximize the saturation times for pathogen $1$ and co-infections. Thus, the most beneficial layer combinations are LA12-LA4-LA4 and LA4-LA4-LA4. 
\end{itemize}

As a result, to minimize the spread of both infections, it is most important to reduce the movement of individuals infected by either, and limiting the movement of susceptible individuals is less important. This is consistent with our findings from the MBRD-SI model.

In human metapopulations, migration patterns can be best described with BA network topologies. A key feature of scale-free networks such as BA networks is large hubs that dominate the network. This accurately reflects how during holiday season, major cities serve as large tourism hubs, as~\cite{wu2021study} shows is true for Chinese tourism on the May Day holiday. During other seasons, human metapopulations are better represented with small-world networks such as the WS network. The WS network has a high clustering coefficient, representing how at work and local events, humans form many close social circles. Small-world networks are also characterized by short average path lengths, which represent how communities in human metapopulations are highly interconnected~\cite{zhao2025navigating,salathe2010high}.
Studies have shown that infections often spike during holidays~\cite{arnarson2021school,qiao2024public}. From the results discussed above, we observe for both superinfection and co-infection dynamics that BA network topologies allow both pathogens to spread more quickly throughout the network. As a result, we believe that network topology may be a reason for infection spikes during holiday seasons. 

\section{Comparison with Real-World Infections}\label{sec:realworld}

In this section, we explore some of the similarities between our co-infection simulations and real-world data. First, we present two observations from our simulations, which are illustrated with Example~\ref{ex:10}. Then, we show that these phenomena can be found in real-world data.

\begin{figure*}[htbp]
    \centering

    \makebox[\textwidth]{%
        \subfigure[]{%
            \includegraphics[width=0.45\textwidth]{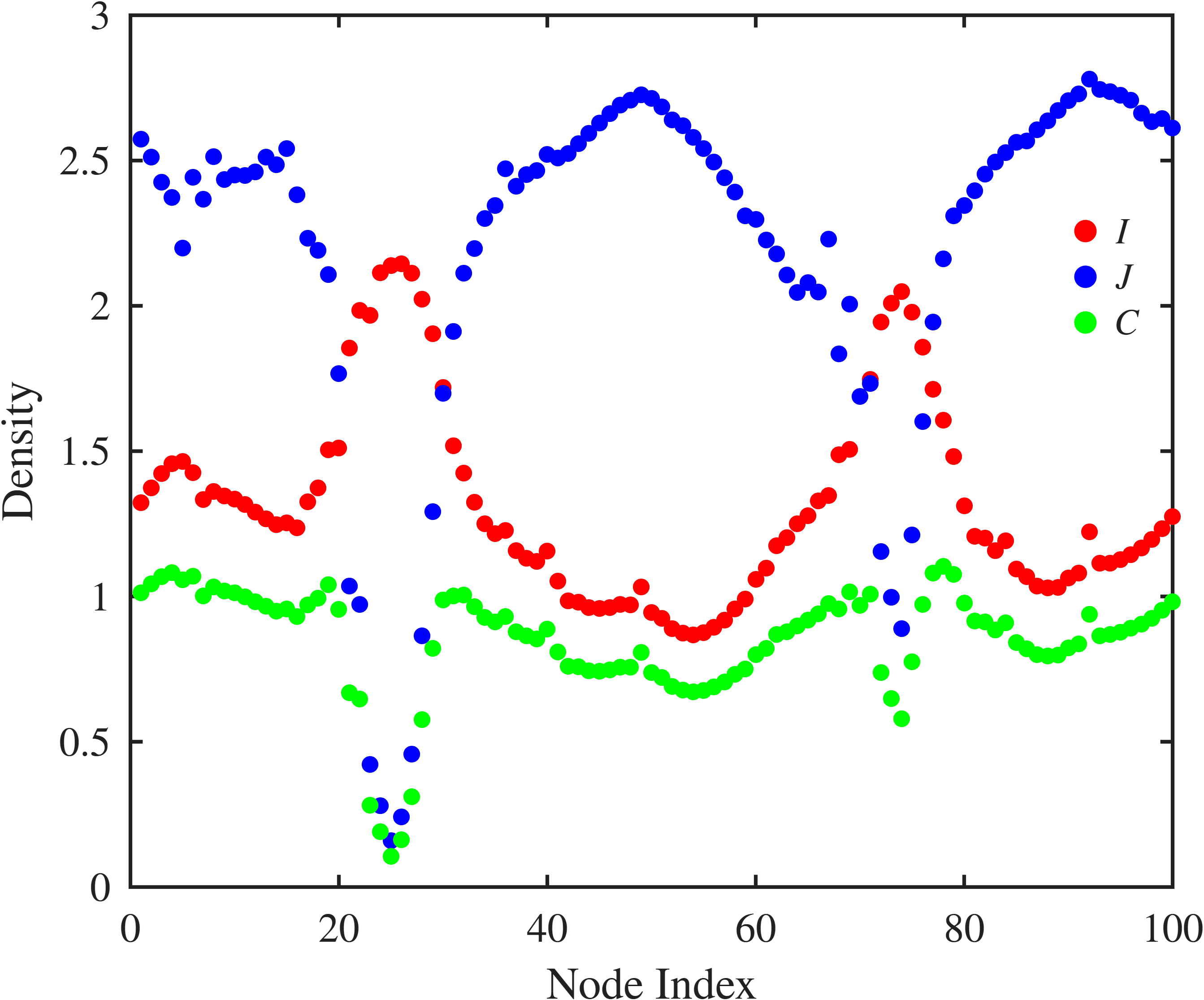}
            \label{fig:compare_ws_far}
        }
        % \hspace{0.04\textwidth}
        \subfigure[]{%
            \includegraphics[width=0.45\textwidth]{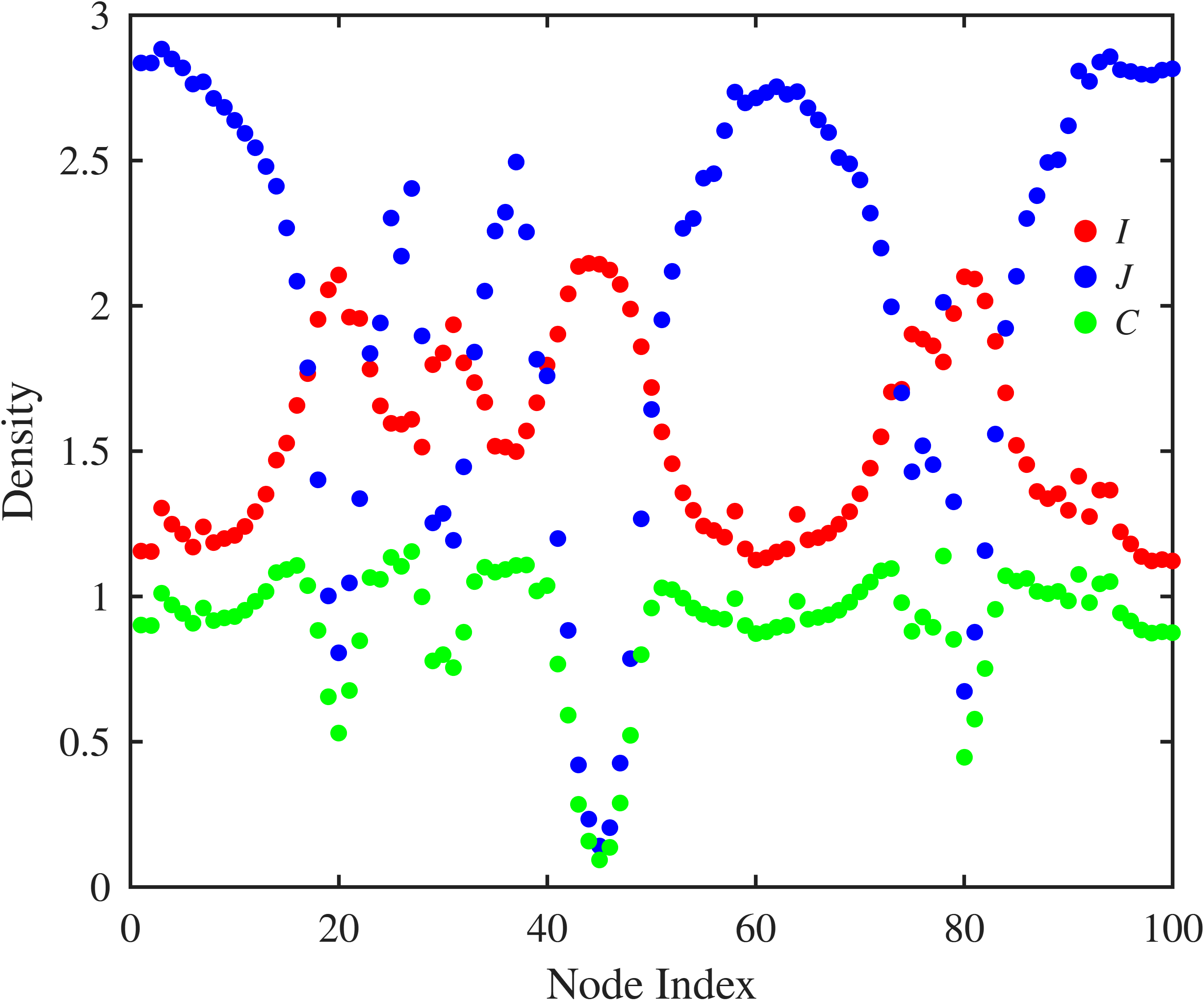}
            \label{fig:compare_ws_close}
        }
    }

    %\par\smallskip
    
    \makebox[\textwidth]{%
        \subfigure[]{%
            \includegraphics[width=0.45\textwidth]{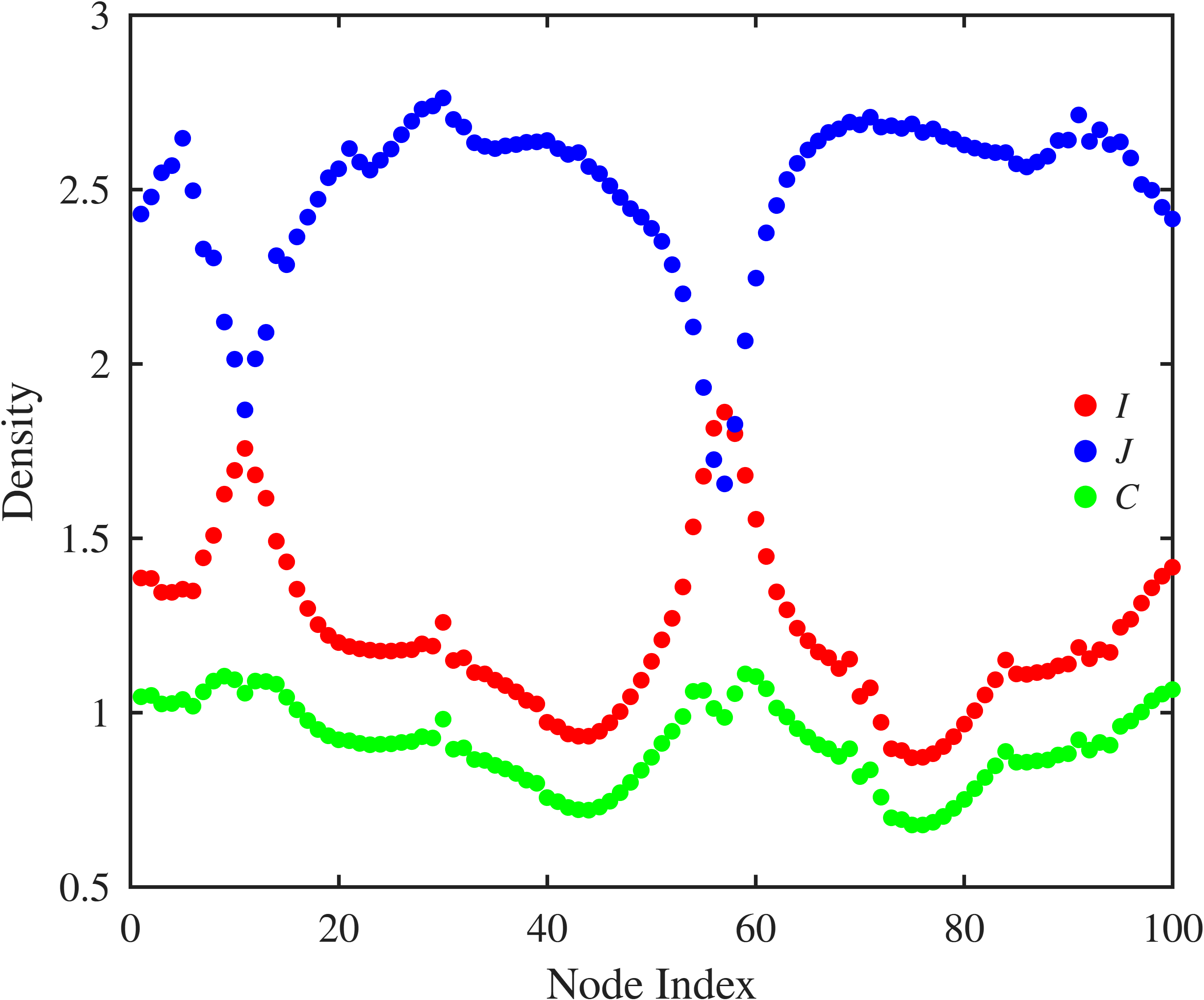}
            \label{fig:compare_ba_far}
        }
        \subfigure[]{%
            \includegraphics[width=0.45\textwidth]{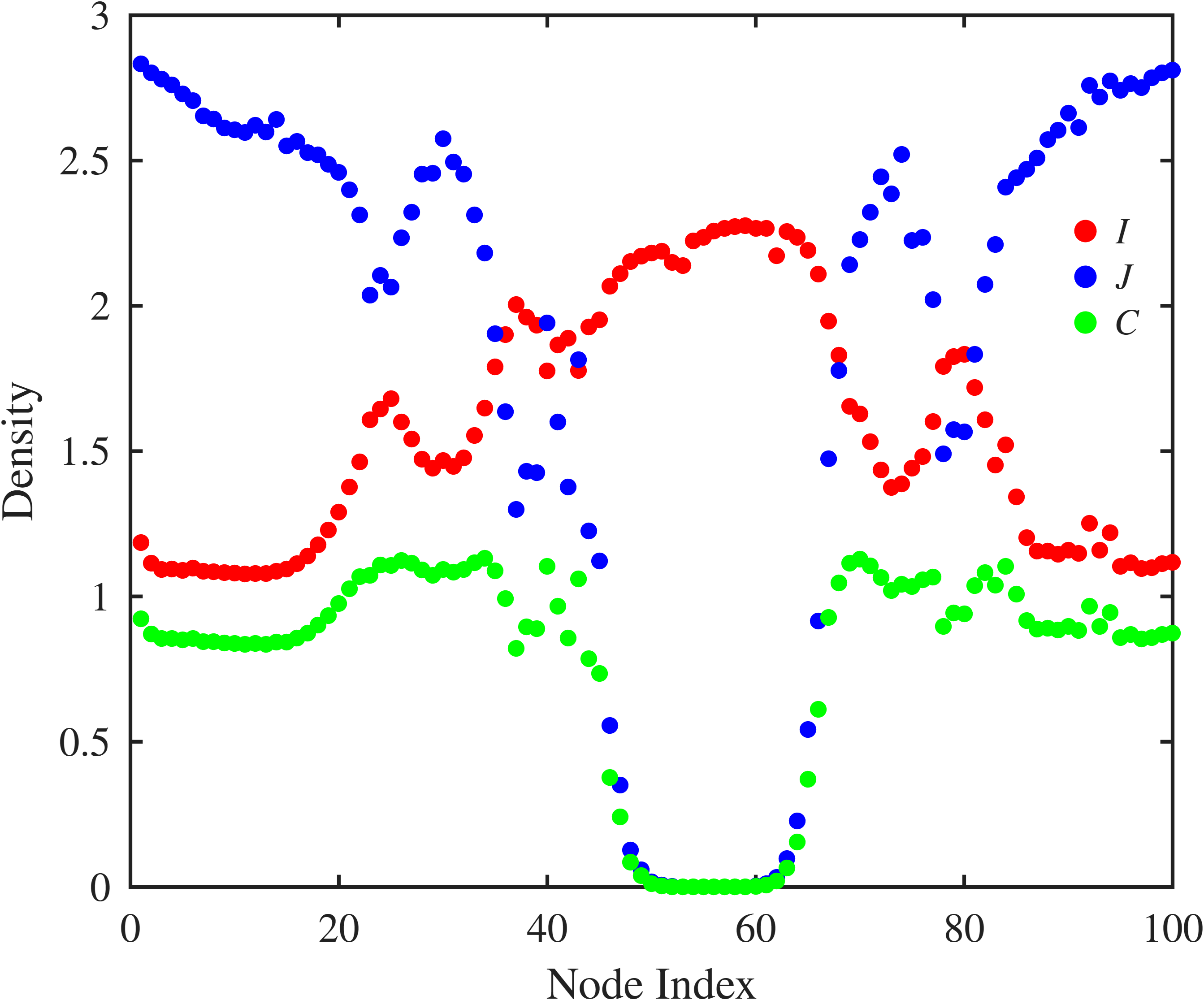}
            \label{fig:compare_ba_close}
        }
    }

    \caption{At $t=9$ with different infection initializations on WS (top row) and BA (bottom row) networks with $100$ nodes. Red represents combined pathogen $1$ and co-infection densities, blue represents combined pathogen $2$ and co-infection densities, and green represents co-infection densities only.}
    \label{fig:compare_simu}
\end{figure*}

We conduct simulations with the co-infection model in Equation~(\ref{eq:coinfect-prelim}) and the parameters in Example~\ref{ex:10}. The results are shown in \autoref{fig:compare_simu}.

\begin{example}\label{ex:10} (Co-infection model) 
\begin{equation}\label{eq:p10-settings}
\begin{aligned}
\mu &= 0.005,\quad & r&=0.1,&\quad A&=0.1,\quad& K&=1,\\
\beta_1&=0.4,\quad & \beta_2&= 0.3,\\ \beta_{10}&=0.1,\quad & \beta_{02}&=0.3,\quad & \beta_{12}&=0.05,\\
\gamma_1&=0.1,\quad & \gamma_2&=0.05,\\ \alpha_1&=0.01,\quad & \alpha_2&=0.02,\quad & \alpha_{12}&=0.05,\\
d_{11}&=0.3,\quad & d_{12}&=0.001,\quad & d_{13}&=0.001,\\
d_{22}&=0.03,\quad & d_{33}&=0.01.
\end{aligned}
\end{equation}
\end{example}

Note that these figures show the density on each layer as a function of the node index. Considering our simulations in \autoref{fig:compare_simu}, we make the following two observations:
\begin{itemize}
\item Oftentimes, the peaks of one infection and a valley of the other infection occur at the same location. We see this occur for different types of initializations, including when both initializations are far from each other and when they are close. Specifically, we observe in \autoref{fig:compare_simu} that most of the time, co-infection distributions follow pathogen $1$-infection distributions. 
\item When the source locations for the infections are close, we also see from \autoref{fig:compare_simu} that the density of pathogen $2$ infections and the overall severity tends to be higher in locations that also have a greater density of co-infections, and generally co-infection distributions tend to mirror the density distribution of one pathogen more than the other. This aligns with the conclusions in~\cite{susi2015co}, which conclude this from investigating co-infections of different strains of the pathogen \emph{Podosphaera plantaginis} in \emph{Plantago lanceolata} gardens. 
\end{itemize}

We also analyze the validity of our observations in the context of the co-circulation of COVID-19 and tuberculosis. The city of Recife, Brazil, can be broken down into $94$ neighborhoods. Silva \emph{et al.} examined the incidences of COVID-19, tuberculosis, and co-infections in this region in 2020~\cite{silva2025spatial}. We assume that the incidences of COVID-19 and tuberculosis include co-infections. Figures 2-4 in~\cite{silva2025spatial} categorize the incidences in four categories for each type of infection. We record each region with ``1'', ``2'', ``3'', or ``4'', based on the category. We denote ``1'' to represent the category with the lowest incidences and ``4'' to represent the category with the highest incidences, and call these the $i$-th COVID, tuberculosis, and coinfection category numbers for region $i$. We see that for more than $35$\% of the regions, the absolute difference between the COVID and tuberculosis category numbers are greater than $2$. Moreover, around $7$\% of the regions have an exact absolute difference of $3$ between the COVID and tuberculosis category numbers, showing that peak and valleys can occur in the same region at the same time for different infections. On the other hand, we note that around 87 percent of the time, the absolute differences between the tuberculosis and coinfection categories take on the values $0$ or $1$, showing that the spatial trends of coinfections closely resemble that of tuberculosis in this scenario.

We also look at Wu \emph{et al.}'s analysis of tuberculosis and HIV co-infections from 2011-2019 in Jiangsu province~\cite{wu2023spatio-temporally}. We see that from 2011-2014 in particular, the northern side of Jiangsu province is filled with tuberculosis infection hotspots but is almost entirely an HIV cold spot. Moreover, many areas in the south are HIV hotspots are tuberculosis cold spots in the same years. Finally, we see that spatial co-infection distributions closely follow that of HIV infections, where the north is almost entirely a coldspot but there are some hotspots in the south, supporting our simulation-based observations.

\section{Final Remarks}\label{sec:conclusion}

The propagation of a single infection has been well-studied, especially during and after the COVID-19 pandemic~\cite{zhao2025navigating}. However, spatial-temporal interactions between two pathogens have not been thoroughly analyzed. Our paper \cite{yu2025spatial} proposed two reaction-diffusion models taking into account superinfection and co-infection, respectively, and provided a theoretical analysis of these models. In this paper, we provide a simulation-based approach of analyzing both models to understand the spatial spread of both infections over time, observing  the following relationships throughout our simulations:
\begin{itemize}
    \item \textbf{Source separation.} We analyze how the initialization locations of both infections impact the dynamics across networks. When one infection is initialized at a lattice's center, a small source separation produces rapid overlap of infection fronts and faster saturation of co-infections across the network (e.g. see \autoref{fig:superinfect_center}-\ref{fig:coinfect_center}, which shows infection front overlaps and spread indexes when pathogens start from different locations). Additionally, delayed interaction between spreading fronts and slower emergence of co-infected clusters occur for large source separations in the same configurations.
    \item \textbf{Pathogen strengths.} The stronger pathogen, which is typically accompanied by higher $\beta$ or lower $\alpha$ values, tends to dominate in overlap regions and suppresses the weaker pathogen, as can be seen in \autoref{fig:superinfect_sigma}, where we analyze the effect of the superinfection parameter on infection propagation. With respect to pattern formation, the dominance of one pathogen, including high superinfection rates, may suppress pattern formation (see \autoref{fig:superinfect_sigma1}, which illustrates how changes in $\sigma$ affects amplitude and oscillations). 
    
    \item \textbf{Transmission.} When transmission rates are high, infection waves reach overlap earlier, and increased mixing leads to stronger co-infection presence across networks. On the other hand, low transmission promotes lower propagation and weaker overlap, and pathogens remain localized longer.
    Moreover, co-transmission parameters can control how endemic both mono- and co-infections become (see \autoref{fig:coinfect_beta12}-\ref{fig:coinfect_b12thres}, which describes the $\beta_{12}$ vs. $\alpha_{12}$ tradeoff), and in many scenarios intermediate values of co-transmission are best for inducing pattern formation, as shown in \autoref{fig:coinfect_b12}, which illustrates how changes in $\beta_{12}$ affect amplitude and oscillations.
    \item \textbf{Population Migration.} Variations in the migration of susceptible and infected populations prevent pattern formation and increases in average degrees overall allow for larger clusters during pattern formation. Moreover, we find that reducing migration of all infected populations is important for slowing the spread of both pathogens in early stages of epidemics. Thus, we show that mitigation efforts should focus on limiting the movement of infectious individuals, with less priority on limiting the movement of susceptible individuals. This is 
supported by the systematic comparisons in 
\autoref{fig:coinfect_deg}, \autoref{table:superinfect_deg}, and 
\autoref{table:coinfect_deg}, which demonstrate how different degree 
configurations influence clustering and saturation times.

    \item \textbf{Network topology.} From comparing saturation times for different types of networks, we find that infections spread slower and are less severe in Watts-Strogatz (WS) network topologies in comparison to Barab{\'a}si-Albert (BA) topologies. This finding, 
illustrated in \autoref{table:superinfect_deg} and 
\autoref{table:coinfect_deg}, provides a possible explanation for why 
outbreaks often spike during holiday seasons, when human mobility patterns 
resemble scale-free networks with large hubs.
\end{itemize}

We have illustrated that early containment of both types of infections is crucial to slowing the reach of these pathogens. Many previously implemented containment methods align with our findings. For example, quarantining of infection individuals was implemented during the COVID-19 pandemic~\cite{memon2021assessing} and the 2002-2004 SARS outbreak~\cite{hsieh2007impact}. Moreover, border control policies for infected people prevent migration between regions, which correspond to reducing edges in our networks~\cite{nishiura2009quarantine}. According to our findings, this is beneficial not only for the respective countries or regions, but also for the entire metapopulation.

 As discussed in Section~\ref{sec:realworld}, our observations can apply to different co-infection spread in the physical world. We shall mention here a few possible lines of research that could arise from our work:  
 The spread of infections is often influenced by environmental factors, such as temperature, humidity, and air quality~\cite{issarow2018environmental}. The impact of these conditions on infection spread could be investigated through our model,  and point-source infections could be studied with stochastic noise factored in.
 The impact of various vaccination strategies and other specific mitigation strategies can be further investigated through simulations, or through formulating our model into an optimal control problem, similar to how~\cite{lemecha2020optimal} analyzes COVID-19 spread.

  Applications of our model in \cite{yu2025spatial} include understanding the popularity of political parties and predicting elections. With similarly constructed simulations to those in this paper, we can investigate how newly created political parties gain popularity in countries and the factors that influence it.
Finally,   higher-dimensional multiplex networks can be used to investigate the effects of information propagation, socio-economic, and demographic factors on bi-pathogen interactions through adding new layers to the models introduced in~\cite{yu2025spatial} and a similar simulation-based analysis.
\smallbreak
\noindent {\bf Acknowledgments.}\\
The authors are thankful to the MIT PRIMES-USA program for their support and the opportunity to conduct this research together. The research of LPS was partially supported by  NSF FRG Award DMS- 2152107, and  an NSF CAREER Award DMS 1749013.\
\smallbreak
\noindent {\bf Affiliations.}\\
(a) Poolesville High School, Maryland, USA.\\ 
(b)  University of Illinois at Chicago,  USA. \\
\bibliographystyle{unsrt}
\bibliography{PRIMES_2025}
\end{document}